\documentclass[12pt]{article}
\usepackage{graphicx}

\usepackage{amsfonts}
\usepackage{url}

\usepackage{caption}
\usepackage{subcaption}

\def\acos{{\rm acos}}
\def\acosh{{\rm acosh}}

\begin{document}

\def\bbH{\mathbb{H}}
\def\bbE{\mathbb{E}}
\def\bbS{\mathbb{S}}
\def\bbR{\mathbb{R}}
\def\bbG{\mathbb{G}}
\def\bbA{\mathbb{A}}
\def\bbB{\mathbb{B}}
\def\bbZ{\mathbb{Z}}
\def\ra{\rightarrow}

\title{Modelling brain connectomes networks: Solv is a worthy competitor to hyperbolic geometry!}
\author{Dorota Celi\'nska-Kopczy\'nska, Eryk Kopczy\'nski \\
Institute of Informatics, University of Warsaw, Warsaw, Poland}
\maketitle

\begin{abstract}
  Finding suitable embeddings for connectomes (spatially embedded complex networks that map neural connections in the brain) is crucial for analyzing and understanding cognitive processes.
  Recent studies have found two-dimensional hyperbolic embeddings superior to Euclidean embeddings in modeling connectomes across species, especially human connectomes.
  However, those studies had limitations: geometries other than Euclidean, hyperbolic, or spherical were not considered.
  Following William Thurston's suggestion that the networks of neurons in the brain could be successfully represented in Solv geometry, we study the goodness-of-fit of the embeddings for
  21 connectome networks (8 species). To this end, we suggest an embedding algorithm based on Simulating Annealing that allows us to embed connectomes to Euclidean, Spherical, Hyperbolic,
  Solv, Nil, and product geometries. Our algorithm tends to find better embeddings than the state-of-the-art, even in the hyperbolic case. Our findings suggest that while three-dimensional hyperbolic embeddings yield the best results in many cases, Solv embeddings perform reasonably well.
\end{abstract}


\section{Introduction}

Connectomes are comprehensive maps of the neural connections in the brain. Understanding the interactions they shape is a key to understanding cognitive processes. Given their spatially embedded complexity, shaped by physical constraints and communication imperatives, connectomes exhibit properties inherent to non-Euclidean geometries. Therefore, a vast amount of recent research has been devoted to finding the appropriate embeddings for connectome networks.
Recent studies (e.g., \cite{whkl,brainserrano}) have advocated for the superiority of two-dimensional hyperbolic embeddings over Euclidean embeddings in modeling connectomes across species, especially human connectomes. However, those studies had limitations: they restricted the focus to Euclidean, hyperbolic, or spherical geometries, neglecting to explore other potential embedding spaces.

Our study expands the perspectives for suitable embeddings.
We analyze the goodness of fit (measured with widely used quality measures) of the embeddings for 21 connectome networks (8 species) to 15 unique tessellations (Euclidean, Spherical, Hyperbolic, Solv, Nil, and also product geometries). We consider both two-dimensional and three-dimensional manifolds. Following William Thurston's suggestion that the networks of neurons in the brain could be successfully represented in Solv geometry (one of eight so-called Thurston geometries), we stipulate that this geometry would outperform hyperbolic geometry.

Against this background, our contribution in this paper can be summarized as follows:

\begin{itemize}
\item We present a novel embedding method based on Simulated Annealing (SA). Experiments show that our algorithm outperforms the state-of-the-art, even for the hyperbolic embeddings,
as evaluated by standard measures (mAP, MeanRank, greedy routing success, and stretch).
\item To the best of our knowledge, we are the first to compare embeddings of connectomes to all Thurston geometries. As a result, we expand the horizons of connectome modeling and open up new possibilities for analysis. We show that connectome modeling is more nuanced than previously presented.
\item We find that while three-dimensional hyperbolic geometry yields the best results in many cases, other geometries, such as Solv, are worth considering. Supported by an extensive simulation scheme, our results bring confidence and reliability beyond previous studies.
\end{itemize} 

\begin{figure}
\begin{center}
\includegraphics[width=.32\textwidth]{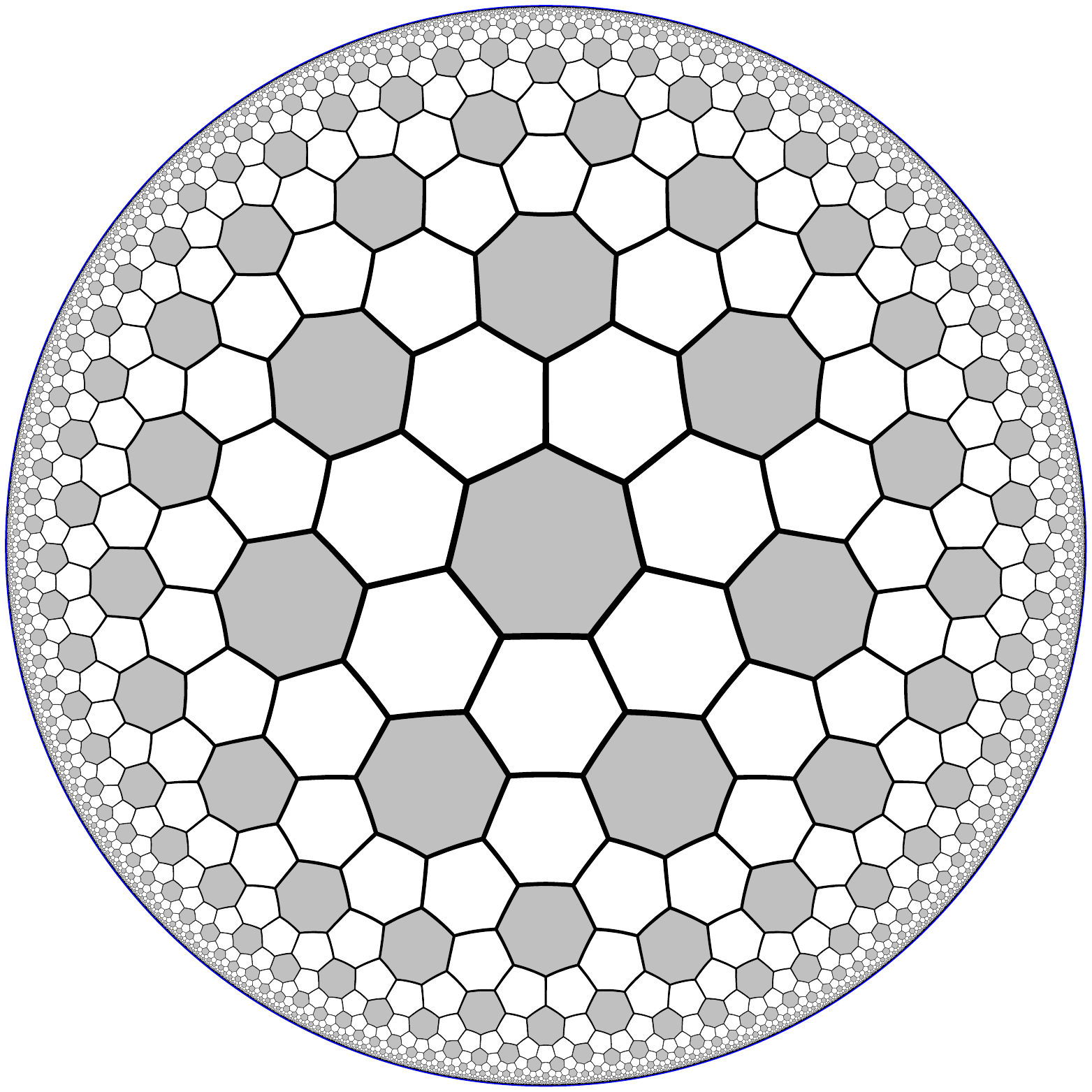}
\includegraphics[width=.32\textwidth]{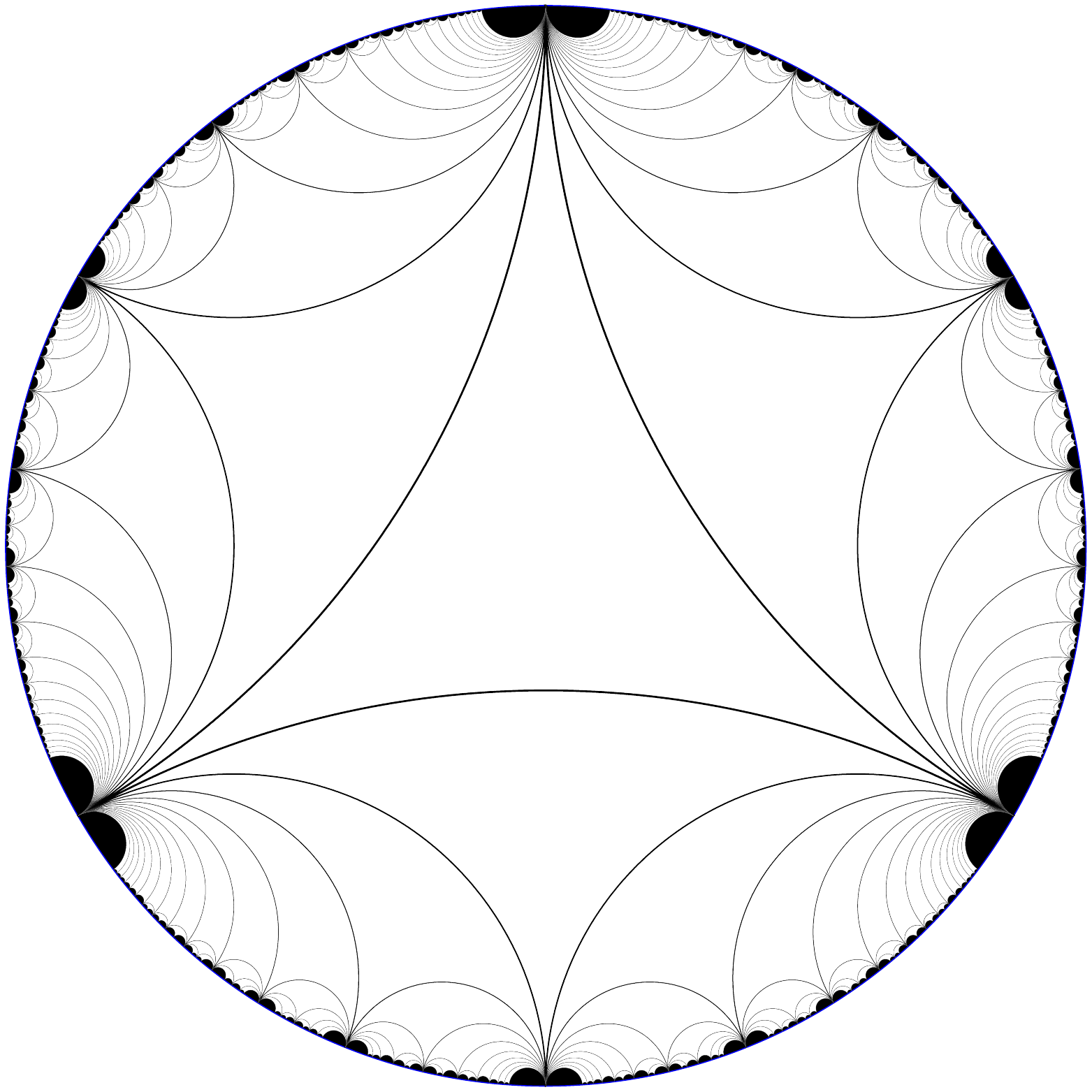}
\includegraphics[width=.32\textwidth]{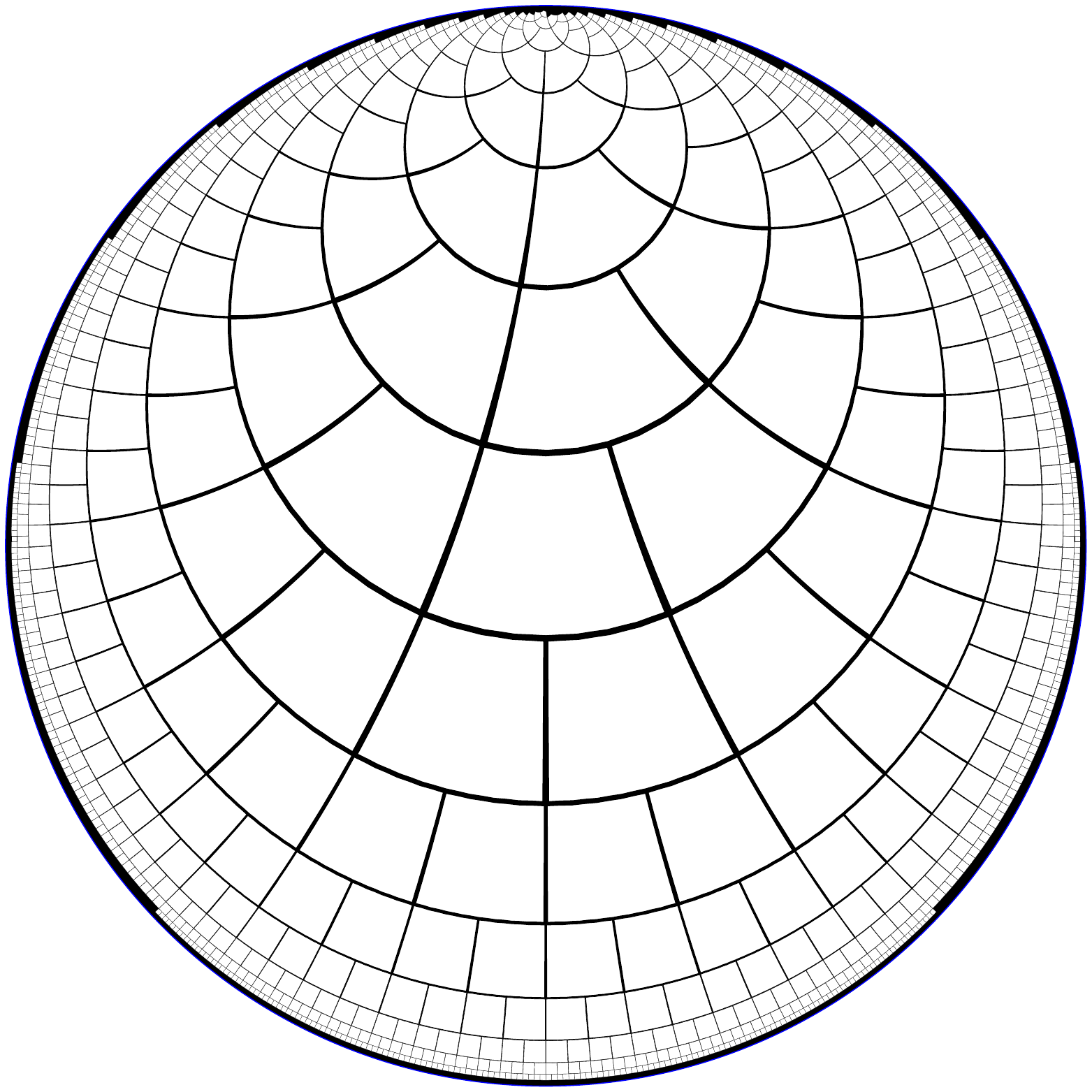}
\end{center}
\caption{Tessellations of the hyperbolic plane. 
From left to right: (a) bitruncated order-3 heptagonal tiling ($\{7,3\}$),
(b) infinite-order triangular tiling ($\{3,\infty\}$), 
(c) binary tiling.\label{fig:tess}}
\end{figure}

This paper is accompanied with supplementary material containing our implementation, data and results \cite{brainsupplement} and a video containing our 3D visualizations \cite{brainvideo}.

\section{Prerequisities}

\subsection{Thurston geometries}
By the uniformization theorem, every closed two-dimensional topological surface can be given spherical ($\bbS^2$), Euclidean ($\bbE^2$), or hyperbolic ($\bbH^2$) geometry, that is, 
there exists a Riemannian manifold with the same topology as $M$ and locally isometric to a sphere, Euclidean plane, or hyperbolic plane.
William Thurston conjectured \cite{thurston1982} that three-dimensional topological manifolds can be similarly decomposed into fragments, each of which can be given one
of eight \emph{Thurston geometries}, which are homogeneous Riemannian manifolds. The eight Thurston geometries include:

\begin{itemize}
\item isotropic geometries: spherical ($\bbS^3$), Euclidean ($\bbE^3$), and hyperbolic ($\bbH^3$).
\item product geometries: $\bbS^2\times\bbR$ and $\bbH^2\times\bbR$, In geometry $\bbA \times \bbB$, the distance $d_{\bbA\times\bbB}$ between $(a_1,b_1), (a_2,b_2)\in \bbA\times\bbB$
is defined using the Pythagorean formula: \[ d_{\bbA\times\bbB}((a_1,b_1), (a_2,b_2)) = \sqrt{d_\bbA(a_1,a_2)^2 + d_\bbB(b_1,b_2)^2}. \]
Intuitively, using the Pythagorean formula here means that the third dimension is added to $\bbS^2$ or $\bbH^2$ in the Euclidean way. 
\item Twisted product geometries: twisted $\bbE^2\times\bbR$, also known as Nil, and twisted $\bbH^2\times\bbR$, referred to as Twist in this paper, also known as the universal cover of $SL(2,\bbR)$. \cite{rtviz}
\item Solv geometry, also known as Solve or Sol, which is fully anisotropic.
\end{itemize}

The more exotic Thurston geometries have been successfully visualized only very recently \cite{rtviz,segmarching}, and thus are much less known than isotropic geometries. 
We refer to these papers and explanatory videos \cite{yt_solv,yt_nil} and demos \cite{segdemo} for detailed explanations of Solv and Nil geometries. In the rest of this section, 
we include a brief description of Solv and an intuitive explanation of twisted product geometries.

The $n$-dimensional sphere is $\bbS^n = \{x \in \bbR^{n+1}: g(x,x)=1\}$, where $g$ is the Euclidean inner product,
$g(x,y)=x_1y_1+x_2y_2+\ldots+x_{n+1}y_{n+1}$. The distance between two points
$a$, $b$ on the sphere is the length of the arc connecting $a$ and $b$, which can be computed as $d(a,b) = \acos\ g(a,b)$.
Similarly, we can define $n$ dimensional hyperbolic geometry using the Minkowski hyperboloid model. In this model, $\bbH^n = \{x \in \bbR^{d+1}: x_{d+1}>0, g^-(x,x)=-1$,
where $g^-$ is the Minkowski inner product, $g^-(x,y)=x_1y_1+x_2y_2+\ldots+x_{n}y_{n}-x_{n+1}y_{n+1}$. The distance between two points can be computed as $d(a,b) = \acosh\ g^-(a,b)$.

Typically, tessellations of the hyperbolic plane $\bbH^2$ are visualized using the Poincar\'e disk model, which is a projection of $\bbH^2$ to the Euclidean plane
that distorts the distances (Figure \ref{fig:tess}). In each of these tessellations, all the shapes (of the same color) have the same hyperbolic size, even though ones closer to the boundary 
look smaller in the projection.

To explain Solv, we should start with the horocyclic coordinate system of $\bbH^2$. Horocycles are represented in the Poincar\'e disk model as circles tangent to the boundary;
these can be seen as hyperbolic analogs of circles with infinite radius and circumference, centered in an ideal point (point on the boundary of the Poincar\'e disk).
 Figure \ref{fig:tess}c depicts concentric horocycles; the distance between two adjacent horocycles in this picture is $\log(2)$, and if two points $A$ and $B$ on given horocycle are in the distance
$x$, then the distance between their projections on the next (outer) horocycle is $2x$. For a point $P\in\bbH^2$,
we project $P$ orthogonally to $Q$ on the horocycle going through the center $C$ of the Poincar\'e model. The $x$ coordinate is the (signed) length of the horocyclic arc $CQ$, and $y$ is the 
(signed) length of the segment $PQ$. (This is similar to the upper half-plane
model \cite{cannon}, except that we take the logarithm of the $y$ coordinate.) In this coordinate system, the length of the curve $((x(t), y(t)): t \in [a,b])$ is defined as $\int_a^b \sqrt{ (x'(t) \exp{y{t}})^2 + y'(t)^2} dt$.

A similar coordinate system for $\bbH^3$ defines the length of the curve $((x(t), y(t), z(t)): t \in [a,b])$ as $\int_a^b \sqrt{ (x'(t) \exp{z(t)})^2 + (y'(t) \exp{z(t)})^2 + z'(t)^2} dt$.
The surfaces of constant $z$ are called \emph{horospheres}; the geometry on a horosphere is Euclidean.
We obtain Solv geometry by switching the sign in this formula. That is, each point also has three coordinates ($x$, $y$, and $z$), but the length of a curve is now equal to
$\int_a^b \sqrt{ (x'(t) \exp{z(t)})^2 + (y'(t) \exp{-z(t)})^2 + z'(t)^2} dt$. The distance between two points is the length of the shortest curve connecting them; this length is difficult to compute \cite{segmarching,rtvizfinal}.

In Nil, we have well-defined directions at every point, which we can intuitively call North, East, South, West, Up and Down.
However, while in Euclidean geometry, after moving 1 unit to the North, East, South, and West,
we return to the starting point; in Nil, such a loop results in a move by 1 unit in the Up direction. In general, the vertical movement is equal to the signed area of the projection of
the loop on the horizontal plane. Twist is based on the same idea, but the horizontal plane is now hyperbolic.

\begin{figure*}[t!]
  \begin{subfigure}[b]{0.3\textwidth}
   \centering
   \includegraphics[width=\textwidth]{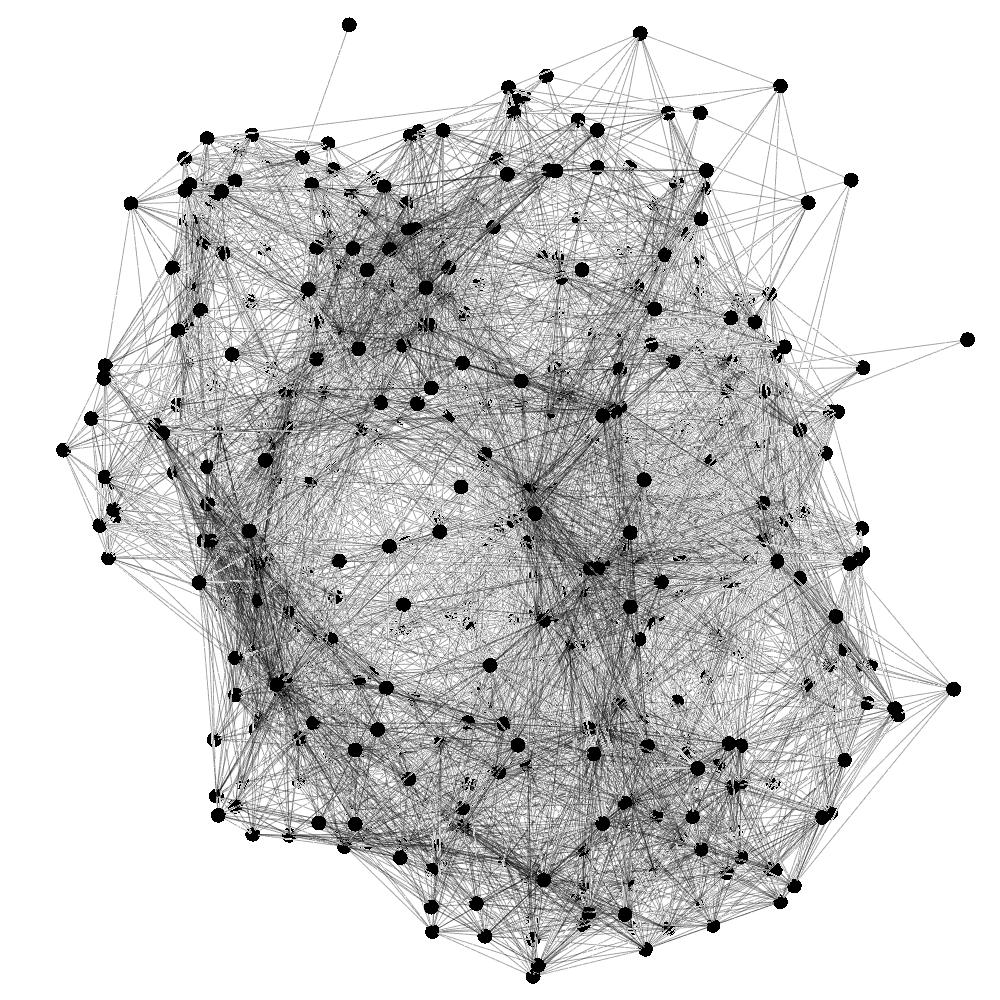}
   \caption{Human2 in $\bbE^3$}
  \end{subfigure}
  \begin{subfigure}[b]{0.3\textwidth}
   \centering
   \includegraphics[width=\textwidth]{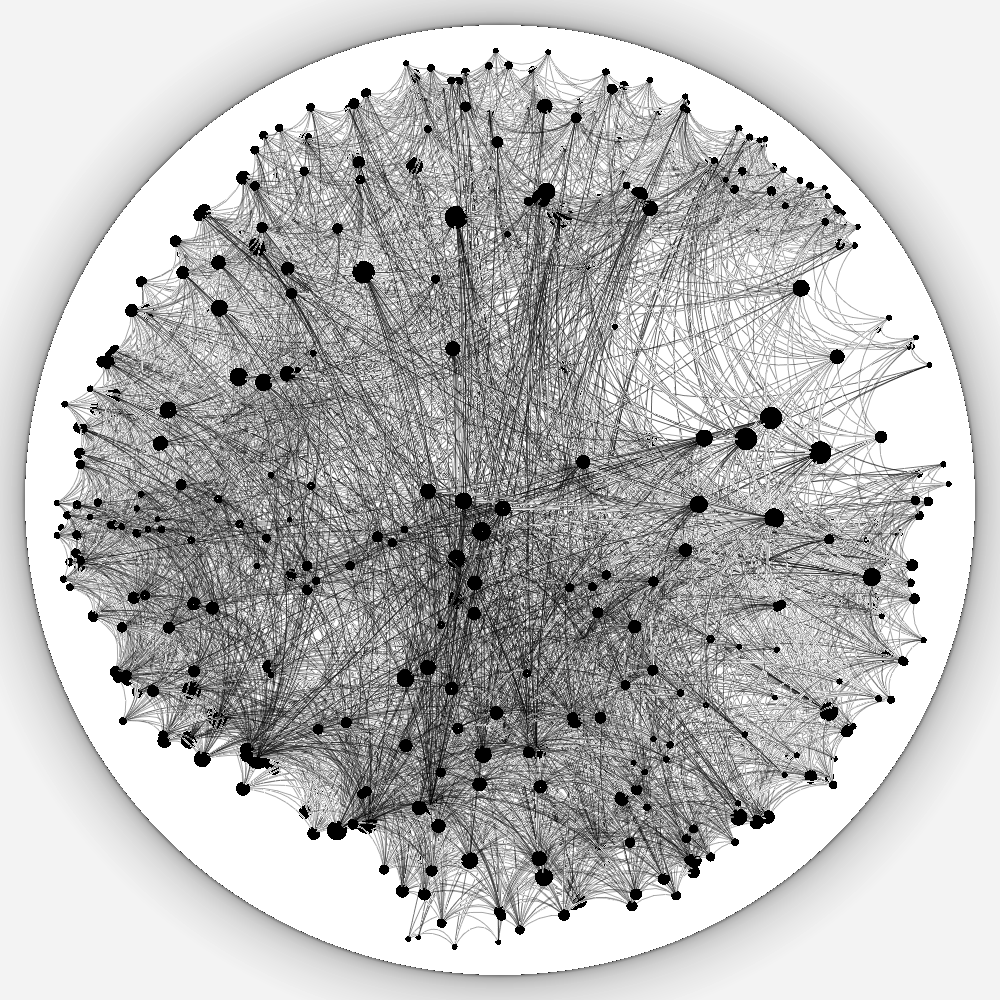}
   \caption{Human2 in $\bbH^3$}
  \end{subfigure}
  \begin{subfigure}[b]{0.3\textwidth}
   \centering
   \includegraphics[width=\textwidth]{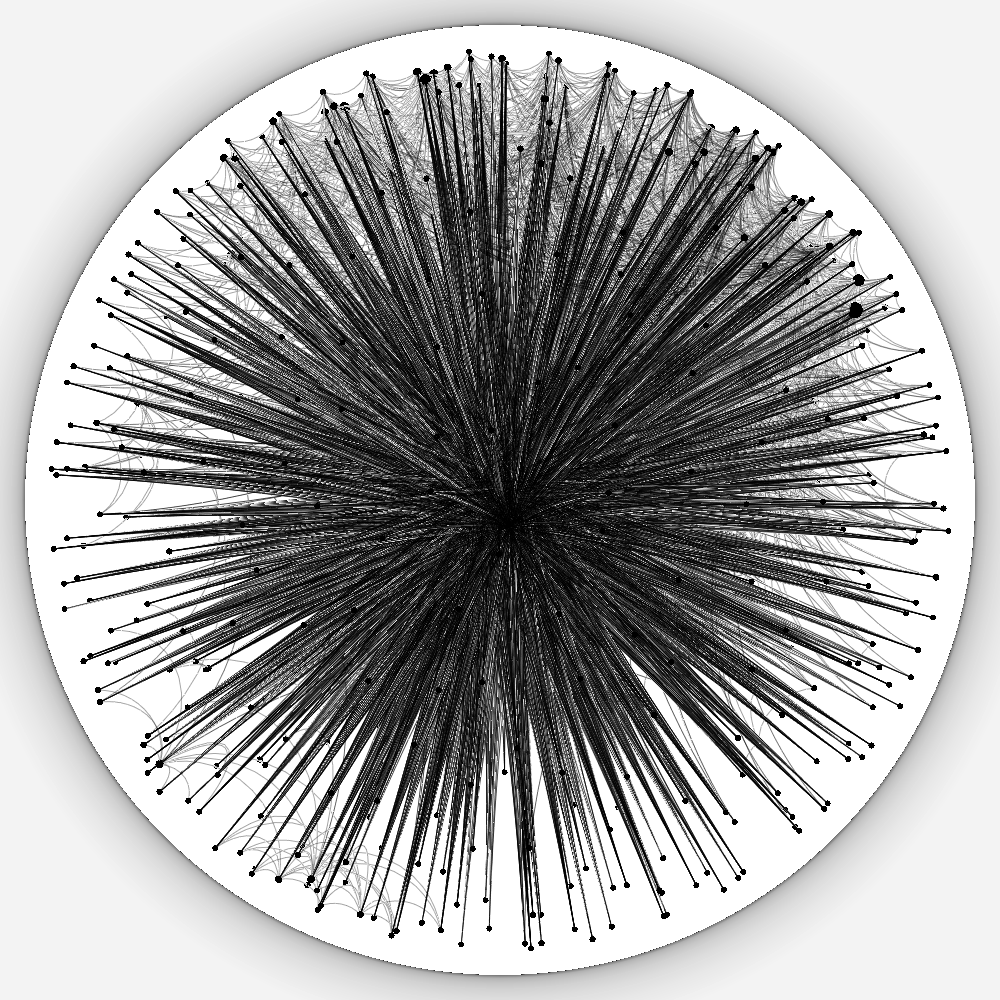}
   \caption{Rat1 in $\bbH^3$}
  \end{subfigure}\\
  \begin{subfigure}[b]{0.3\textwidth}
   \centering
   \includegraphics[width=\textwidth]{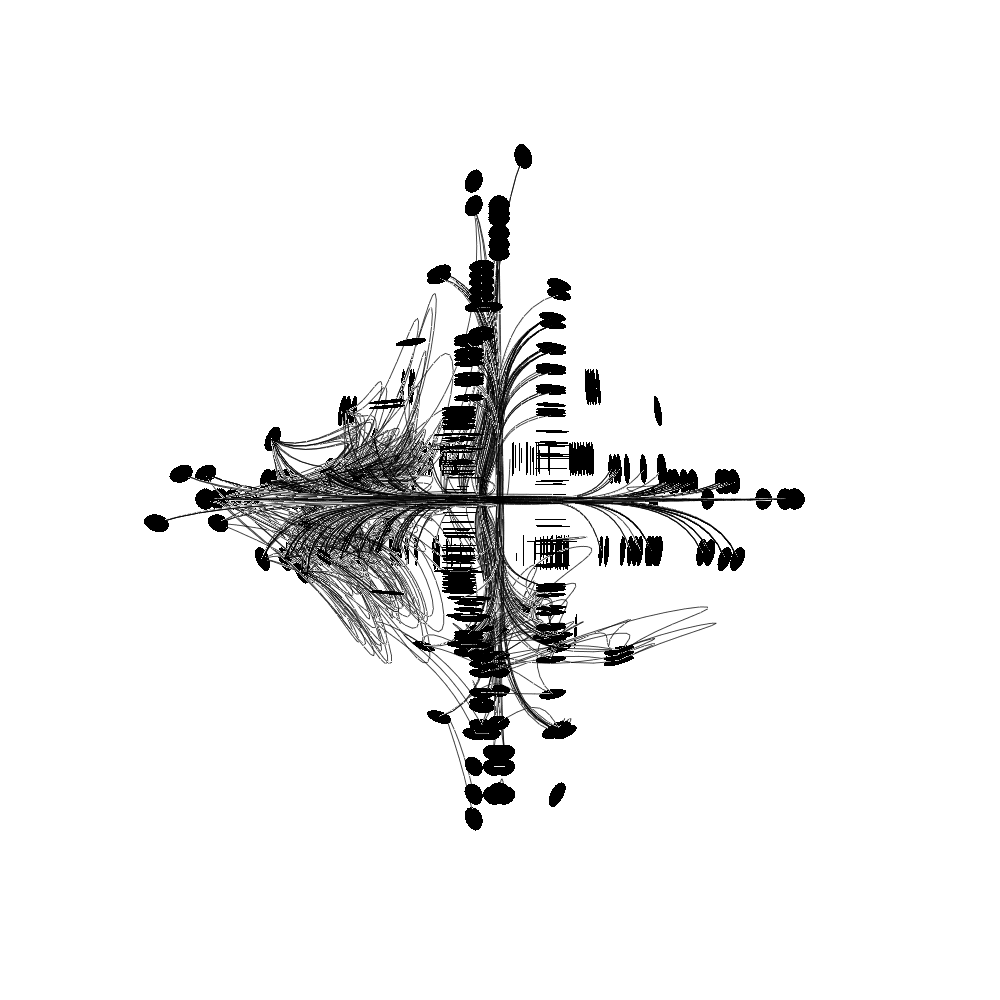}
   \caption{Rat1 in Sol}
  \end{subfigure}
  \begin{subfigure}[b]{0.3\textwidth}
   \centering
   \includegraphics[width=\textwidth]{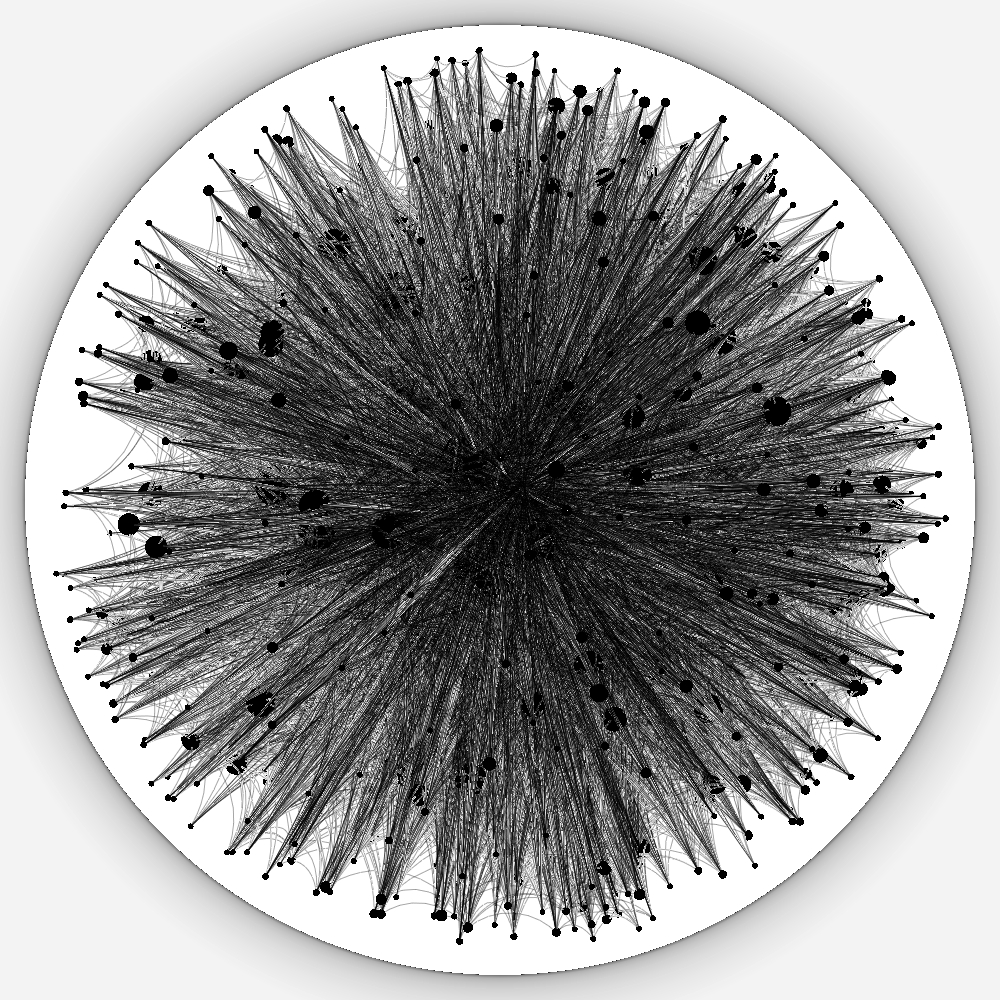}
   \caption{ZebraFinch2 in $\bbH^3$}
  \end{subfigure}
  \begin{subfigure}[b]{0.3\textwidth}
   \centering
   \includegraphics[width=\textwidth]{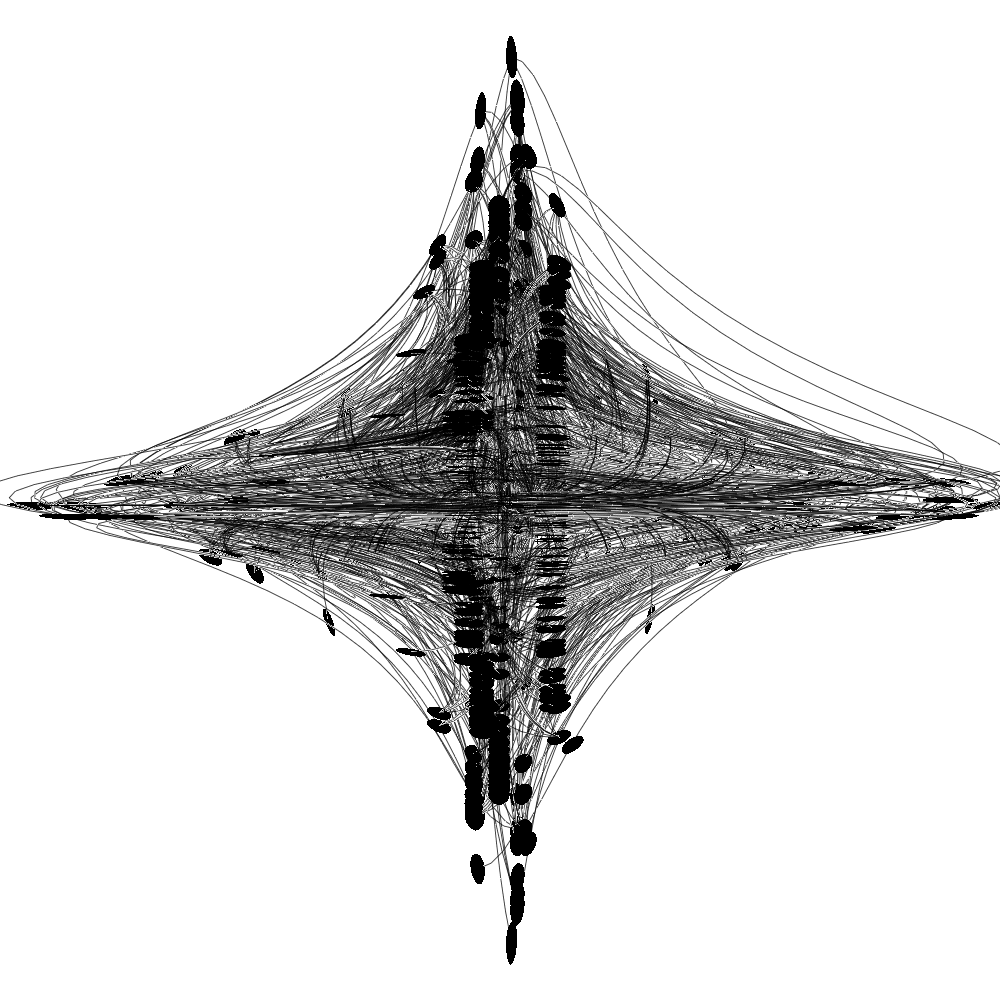}
   \caption{ZebraFinch2 in Sol}
  \end{subfigure}
\caption{A 2D projection of our embeddings. See \cite{brainvideo} for 3D visualizations. \label{fig:emb}}
\end{figure*}

\subsection{Geometric embeddings}

In low-dimensional topology, three-dimensional geometry is incredibly challenging; mainly, the Poincar\'e conjecture was the most challenging in three dimensions.
On the other hand, our interest in two-dimensional and three-dimensional geometries is based on their visualization possibilities \cite{rtviz,segmarching} and potential application to geometric embeddings.

Figure \ref{fig:tess} shows that 
hyperbolic geometry has a tree-like, hierarchical structure. This tree-likeness has found application in the visualization and modeling of hierarchical structures \cite{lampingrao,munzner},
and then in the modeling of complex networks.
The hyperbolic random graph model (HRG) \cite{bogu_internet}
is parameterized by parameters $N$, $R$, $T$, $\alpha$. Each node $i\in\{1,\ldots,n\}$ is assigned a point $m(i)$ in the hyperbolic disk of radius $R$; the parameter $\alpha$ controls the 
distribution. Then, every pair of points $a,b\in\{1,\ldots,n\}$ is connected with probability $1/(1+\exp((d-R)/T))$, where $d$ is the hyperbolic distance between $a$ and $b$.
A real-world network $(V,E)$ can be also embedded into the hyperbolic plane $\bbH^2$ by mapping its nodes $V$ to $\bbH^2$ \cite{bogu_internet};
an embedding is better if the probability of forming the actual observed connections (according to the HRG model) is higher. Moreover, graphs generated according to this model have properties typical to scale-free networks, such as high clustering coefficient and power-law degree distribution \cite{papa}.

More recently, embeddings into higher-dimensional hyperbolic spaces were studied in the network \cite{dmercator,whkl} and the machine learning community (product geometries in
\cite{productembedding}). To our knowledge, twisted product or Solv geometry have yet to be studied in this context.
We are especially interested in the intriguing suggestion of William Thurston from 1997 that the brain's architecture might be based on Solv geometry
\cite{schwartzmemory}.
Intuitively, the Solv geometry is based on two hierarchies (the hyperbolic plane $y=$const and the
hyperbolic plane $x=$const), which are opposed to each other due to the opposite sign used with $z$ in the distance formula. This gives us hope that we can use Solv geometry to represent in three dimensions hierarchies that cannot be represented using other two- or three-dimensional geometries exhibiting simpler hierarchical structure ($\bbH^2$, $\bbH^3$, $\bbH^2\times\bbR$).
A similar effect of two opposing hierarchies could also be obtained in $\bbH^2\times\bbH^2$. However, that is a four-dimensional geometry and, thus, less suitable for visualization.
An promising property of Nil geometry is that it is a three-dimensional geometry where the volume of a ball of radius $R$ has $\Theta(R^4)$ growth,
which suggests better embedding possibilities than $\bbE^3$,
but worse than the exponentially-expanding geometries.

\section{Our contribution}

We need a new embedding algorithm since the previous algorithms may be particularly tailored to the specific geometry \cite{tobias},
or assume that $d_\bbG$ is easy to compute, which is not true for Solv.
We aim to find good quality embeddings of a connectome $(V,E)$ into some geometry $\bbG$, that is, a map $m:V \rightarrow \bbG$. As in the hyperbolic random graph (HRG) model, we assume that our embedding has two parameters: $R$ and $T$.
The probability that an edge exists between $i$ and $j$ is $p_1(d) = 1/(1+\exp((d-R)/T))$, where $d$ is the distance between $m(i)$ and $m(j)$.
We use MLE method to find the embedding, that is, we aim to maximize the likelihood $\prod_{1\leq i<j\leq N} p(i,j)$, where $p(i,j)=p_1(d_{\bbG}(m(i),m(j)))$ in case if the edge between $i$ and $j$ exists,
and $p(i,j)=1-p_1(d_{\bbG}(m(i),m(j)))$ otherwise. Equivalently, we maximalize the loglikelihood $\sum_{1\leq i<j\leq N} \log p(i,j)$.

\subsection{Our embedding algorithm}

As in \cite{dhrgex}, our algorithm is based on a uniform grid in geometry $\bbG$. Natural grids exist in all Thurston geometries of interest \cite{rtviz}. In the HRG model, the network is
mapped to a disk of radius $R$; here, we map the network to the set $D$ of all grid points in $\bbG$, which are in the distance at most $d_R$ from some fixed origin. We choose $d_R$ to fix the number
of points inside $D$; in most experiments, we pick $M=20000$ points (actually, there may be slightly more points due to ties).

We compute the distance $d_\bbG$ for every pair of points in $D$, thus obtaining a $|D|\times |D|$ array that can be used to quickly find the distance between pairs of points.
In the case of Solv, it turns out that the method to compute the Solv distances from \cite{rtviz}, while
applicable to visualization, does not apply to computing this table of distances due to long ranges. Therefore, for longer distances, we approximate by $d(a,b)$ as the smallest possible
$d(a,a_1)+d(a_1,a_2)+\ldots+d(a_k,b)$, where intermediate points are also in $D$, and each pair of consecutive points is within the range of the method from \cite{rtviz}. Dijkstra's algorithm is used to find the path ($a_i$).

Now, we use the Simulated Annealing method to find the embedding. This method assumes $R$ and $T$ and starts with an arbitrary embedding $m:V \rightarrow D$. Then, we perform the following for $i=1, \ldots, N_S$:

\begin{itemize}
\item Introduce a small change $m'$ to the current embedding $m$,
\item Compute $L$, the loglikelihood of $m$, and $L'$, the loglikelihood of $m'$.
\item If $L'>L$, always replace $m$ with $m'$. Otherwise, replace $m$ with $m'$ with probability $\exp((L'-L)/exp(T))$, where the parameter $T$ (known as \emph{temperature}) depends on the iteration index.
\end{itemize}

In Simulated Annealing, we start with a very high temperature (to accept all changes and thus explore the full space of possible embeddings without getting stuck on local maxima). Then we proceed to lower and lower temperatures
(not accepting changes that yield much worse embeddings but still experimenting with crossing lower valleys), eventually accepting only the changes that improve the embedding. In our experiments, $T$ decreases linearly from
10 to -15. We consider local changes of two possible forms: move $m'(i)$ for a random $i$ to a random point in $D$, and move $m'(i)$ for a random $i$ to a random point in $D$ that is close (neighbor) to $m(i)$. These changes allow computing $L'$ (based on the earlier $L$) in time $O(|V|)$.

To obtain values of $R$ and $T$, we start with some initial values of $R$ and $T$. Occasionally, during the simulated annealing procedure, we find the values of $R$ and $T$ that best fit the current embedding,
and we use the new values for the remaining iterations. Since finding the correct values takes time, we do it relatively rarely (every $|V|$ iterations with successful moves) and
only once the simulated annealing procedure rejects most changes. In our experiments, we repeat this setup 30 times; in the following iterations, we start with the values of $R$
and $T$ that were obtained in the best embedding found so far. The time complexity of an iteration is $O(N_S\cdot |V|)$.

Our implementation uses the tessellations implemented in RogueViz \cite{rogueviz2023} and is based on the existing implementation of Simulated Annealing for finding hyperbolic visualizations \cite{hrviz}.

\subsection{Visualization}
In Figure \ref{fig:emb}, we present selected embeddings for the human cortex, rat nervous system, and zebra finch 
basal-ganglia connectomes. The embeddings exhibit different shapes: e.g., rat nervous system connectomes
are star-like, with a group of neurons in the center of the embedding connected to other neurons,
with few other connections.
Different geometries highlight these differences. E.g., such star-like networks embed
well into $\bbH^3$ or trees.
Our visualization engine lets the viewer rotate the embedding and examine the spatial relationships 
in detail. See the videos in the supplementary material.

\section{Experiments}
\label{sec:dtes}

For our experiments, we use the same set of publicly available connectomes as \cite{brainserrano}\footnote{URL: \url{https://github.com/networkgeometry/navigable_brain_maps_data/}}
(not all connectomes used there are publicly available). See Table \ref{tab:connectomes}.

\begin{table}[h!]
\centering
\begin{tabular}{|l|l|l|r|r|l|} \hline
name        & node & zone                &$|V|$& $|E|$ & source \\  \hline
CElegans    & cell & nervous system      & 279 & 2290   & \cite{connectome_celegans} \\ 
Cat1        & area & cortex              &  65 & 730   & \cite{connectome_cat1a}\\
Cat2        & area & cortex and thalamus &  95 & 1170  & \cite{connectome_cat23}\\
Cat3        & area & cortex              &  52 & 515   & \cite{connectome_cat23}\\
Drosophila1 & cell & optic medulla       & 350 & 2886  & \cite{connectome_drosophila} \\ 
Drosophila2 & cell & optic medulla       &1770 & 8904  & \cite{connectome_drosophila} \\ 
Macaque1    & area & cortex              &  94 & 1515  & \cite{connectome_macaque1} \\
Macaque2    & area & cortex              &  71 & 438   & \cite{connectome_macaque2} \\
Macaque3    & area & cortex              & 242 & 3054  & \cite{connectome_macaque3} \\
Macaque4    & area & cortex              &  29 & 322   & \cite{connectome_macaque4} \\
Mouse2      & cell & retina              & 916 & 77584 & \cite{connectome_mouse23} \\ 
Mouse3      & cell & retina              &1076 & 90810 & \cite{connectome_mouse23} \\ 
Human1      & area & cortex              & 493 & 7773  & \cite{connectome_human12} \\
Human2      & area & cortex              & 496 & 8037  & \cite{connectome_human12} \\
Human6      & area & whole brain         & 116 & 1164  & \cite{connectome_human678} \\ 
Human7      & area & whole brain         & 110 & 965   & \cite{connectome_human678} \\ 
Human8      & area & whole brain         & 246 & 11060 & \cite{connectome_human678} \\
Rat1        & area & nervous system      & 503 & 23029 & \cite{connectome_rat} \\ 
Rat2        & area & nervous system      & 502 & 24655 & \cite{connectome_rat} \\ 
Rat3        & area & nervous system      & 493 & 25978 & \cite{connectome_rat} \\ 
ZebraFinch2 & cell & basal-ganglia (Area X)&610& 15342 & \cite{connectome_zebrafinch} \\ \hline 
\end{tabular}
\caption{Connectomes in our experiments. From \cite{brainserrano} (some labels and sizes fixed to match actual data).\label{tab:connectomes}}
\end{table}

We run 30 iterations of SA to find the best $R$ and $T$, with $N_S = 10000 \cdot |V|$. We evaluate the quality of embeddings using the following five measures (all ranging from from 0 -- worst to 1 -- perfect).
\begin{itemize}
\item[SC] Greedy routing success rate. SC is the probability that, for random pair of vertices $(x,y) \in V^2$, the greedy routing algorithm starting at $x$
eventually successfully reaches the target $y$. This routing algorithm moves in the first step from $x$ to $x_1$, the neighbor of $x$ the closest to $y$ (that is, $d_\bbG(m(x_1),m(y))$
is the smallest). If $x_1\neq y$, we continue to $x_2$, the neighbor of $x_1$ the closest to $y$, and so on.
\item[IST] Greedy routing stretch. Stretch is the expected ratio of the route length found in the greedy routing procedure to the shortest route length, conditional
that greedy routing was successful. IST is the reciprocal of stretch.
\item[IMR] For an edge $(x,y) \in E$, rank$(x,y)$ is one plus the number of
vertices that are closer to $x$ than $y$ but not connected with an edge. MeanRank is the expected value of $(x,y)$ over all edges. We use IMR=1/MeanRank.
\item[MAP] For an edge $(x,y) \in E$, $P(x,y)$ is the ratio of vertices in distance of at most
$d_\bbG(m(x),m(y))$ to $x$ which are connected with $x$. $AP(x)$ is the average of $P(x,y)$ for all $y$ connected with $x$, and MAP is the average of $AP(X)$ over all $X$.
\item[NLL] Last but not least, log-likelihood (LL), which is directly maximized in our algorithm, as well as in many other embedding algorithms \cite{tobias,mercatorembedding}.
For a given connectome $(V,E)$, the best theoretically possible log-likelihood is obtained when an edge between $x$ and $y$ occurs if and only if the distance $d_\bbG(m(x),m(y))$
is below some threshold value, and thus, edges can be predicted with full certainty based on the distance (log-likelihood = 0) and the worst possible is obtained when the distance
gives no information on edges, and thus, the probability of each edge is predicted as $|E|/{|V| \choose 2}$ (log-likelihood = H). Normalized log-likelihood, NLL, is defined as 
1-LL/H.
\end{itemize}

\emph{Greedy routing} measures are standards in the network science community (e.g., \cite{bogu_internet}) and MeanRank/mAP measures in the machine learning community (e.g., \cite{nickel}). The computations of SC, STR, MR, and MAP care on the order of nodes $y \in V$ by distance from $x \in V$. However, since we are using a discrete set $D$, it is possible that
$d_\bbG(m(x),m(y))=d_\bbG(m(x),m(z))$ for $y \neq z$. Thus, we assume that the tied nodes are ordered randomly in the case of a tie.



We work with the 15 tessellations listed in Table~\ref{tab:geoms}.
Most of our tessellations are hyperbolic. Subdivided($d$) means that each cube of the honeycomb has been subdivided into $d \times d \times d$ subcubes,
and the point $D$ consists of the vertices and centers of these subcubes, approximating the set of centers of cells of the Euclidean bitruncated cubic honeycomb.
In the case of Nil and Solv, we do not get actual cubes, so this construction is approximate.

\begin{table*}[h!]
  \centering
  \resizebox{\columnwidth}{!}{%
\begin{tabular}{|l|l|l|l|l|l|l|l|l|l|}
  \hline
name                           &  dim &  geometry   &  closed &  nodes &  diameter   & description of the set $D$ \\ \hline 
$\bbH^2$                       & 2    &  hyperbolic &  F      & 20007  & 304         & bitruncated $\{7,3\}$ (Figure \ref{fig:tess}a) \\ 
$\bbH^2\&$                     & 2    &  hyperbolic &  T      & 17980  & 157         & closed hyperbolic manifold \\ 
tree                           & 2    &  tree       &  F      & 20002  & 396         & $\{3,\infty\}$ (Figure \ref{fig:tess}b) \\ 
$\bbE^3$                       & 3    &  euclid     &  F      & 20107  & 1070        & bitruncated cubic honeycomb \\ 
$\bbE^3\&$                     & 3    &  euclid     &  T      & 19683  & 450         & torus subdivided into $27 \times 27 \times 27$ cells \\ 
$\bbH^3$                       & 3    &  hyperbolic &  F      & 21365  & 201         & $\{4,3,5\}$ hyperbolic honeycomb \\ 
$\bbH^3*$                      & 3    &  hyperbolic &  F      & 20039  & 146         & $\{4,3,5\}$ subdivided(2) \\ 
$\bbH^3\&$                     & 3    &  hyperbolic &  T      & 9620   & 102         & subdivided(2) closed hyperbolic manifold \\ 
Nil                            & 3    &  nil        &  F      & 20009  & 1000        & integer coordinates \\ 
Nil*                           & 3    &  nil        &  F      & 20208  & 290         & integer coordinates, subdivided(2) \\ 
Twist                          & 3    &  twist      &  F      & 20138  & 152         & twisted $\{5,4\} \times \bbZ$ \\ 
$\bbH^2 \times \bbR$           & 3    &  product    &  F      & 20049  &  29         & bitruncated $\{7,3\} \times \bbZ$ \\ 
Solv                           & 3    &  solv       &  F      & 20017  & 246         & analog of Figure \ref{fig:tess}c\\ 
Solv*                          & 3    &  solv       &  F      & 20000  & 143         & analog of Figure \ref{fig:tess}c, subdivided(2) \\ 
$\bbS^3$                       & 3    &  sphere     &  T      & 21384  & 628         & 8-cell, each cell subdivided(11) \\ 
\hline
\end{tabular}%
}
\caption{Details on tessellations used in our study; * denotes finer grids. \label{tab:geoms}}
\end{table*}



\section{Comparison at maximum performances}

\begin{figure}[h!]
  \centering
  \begin{subfigure}[b]{0.93\textwidth}
   \centering
   \includegraphics[width=\textwidth]{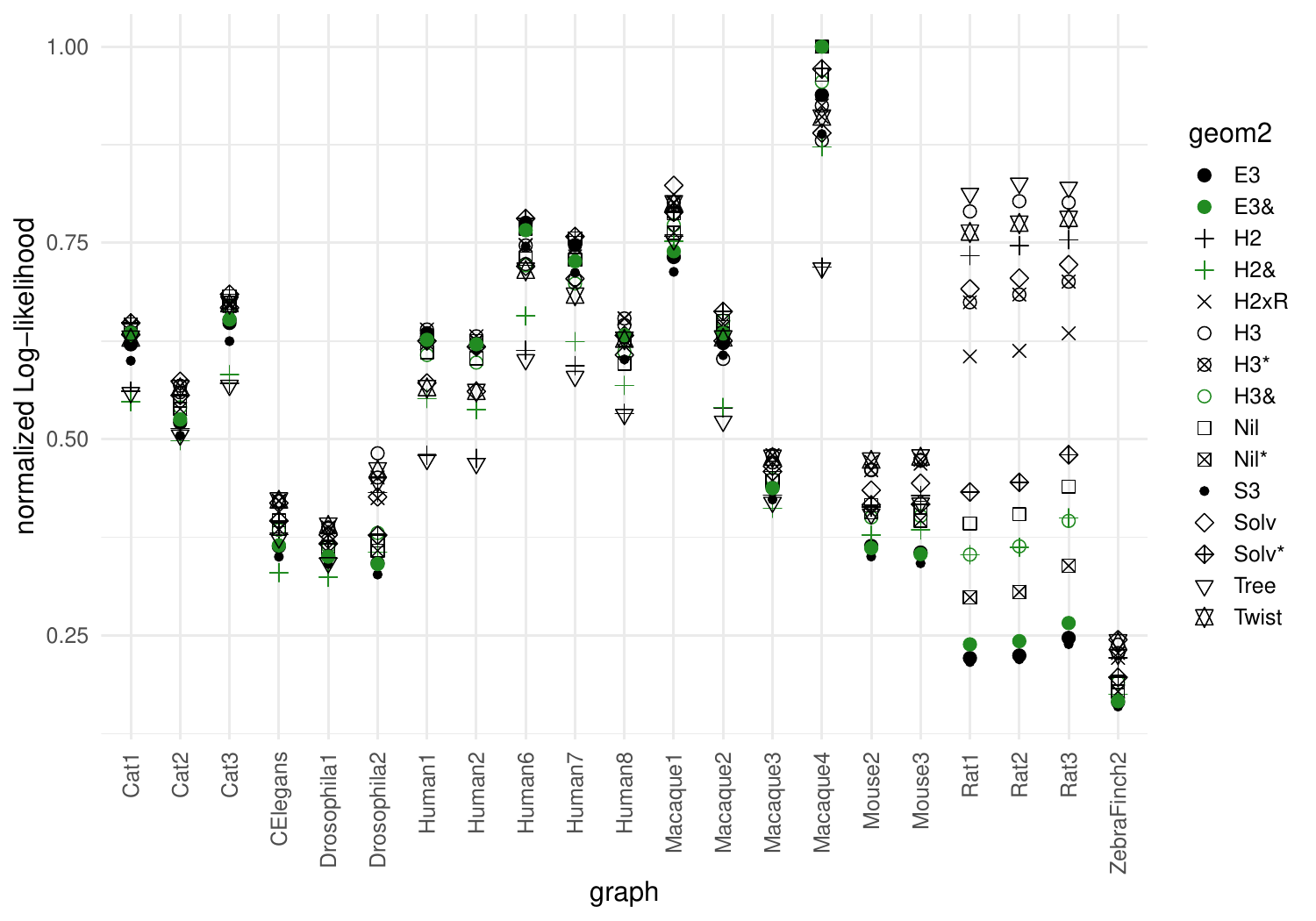}
   \caption{Normalized log-likelihood}
  \end{subfigure}\\
  \begin{subfigure}[b]{\textwidth}
   \centering
   \includegraphics[width=0.93\textwidth]{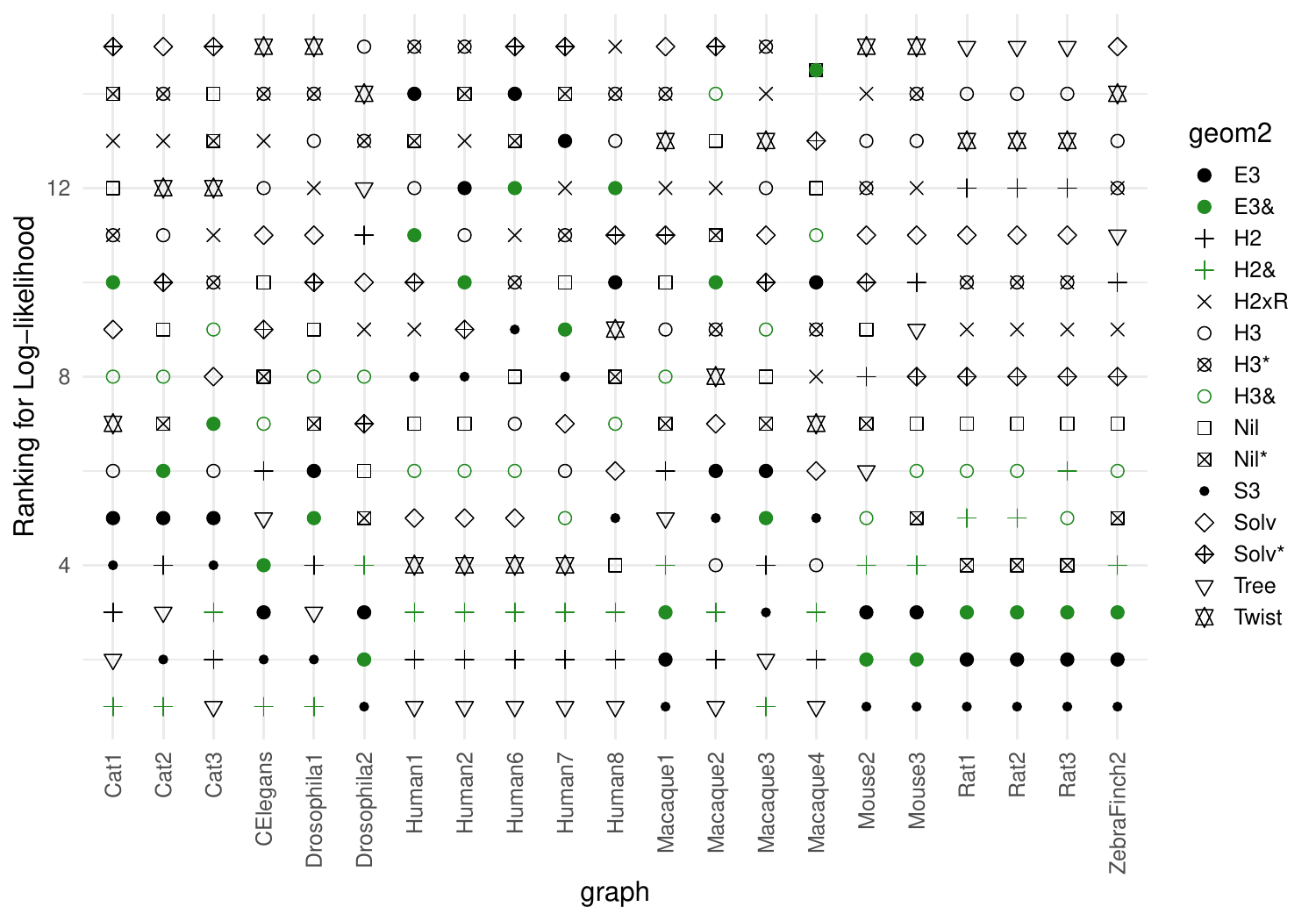}
   \caption{Normalized log-likelihood -- ranks}
  \end{subfigure}
\caption{Our best embeddings -- log-likelihood. Top = best embedding obtained, bottom = worst embedding obtained, * = fine grid. \label{fig:bestlogliks1}}
\end{figure}

\begin{figure}[h!]
  \centering
  \begin{subfigure}[b]{0.93\textwidth}
   \centering
   \includegraphics[width=\textwidth]{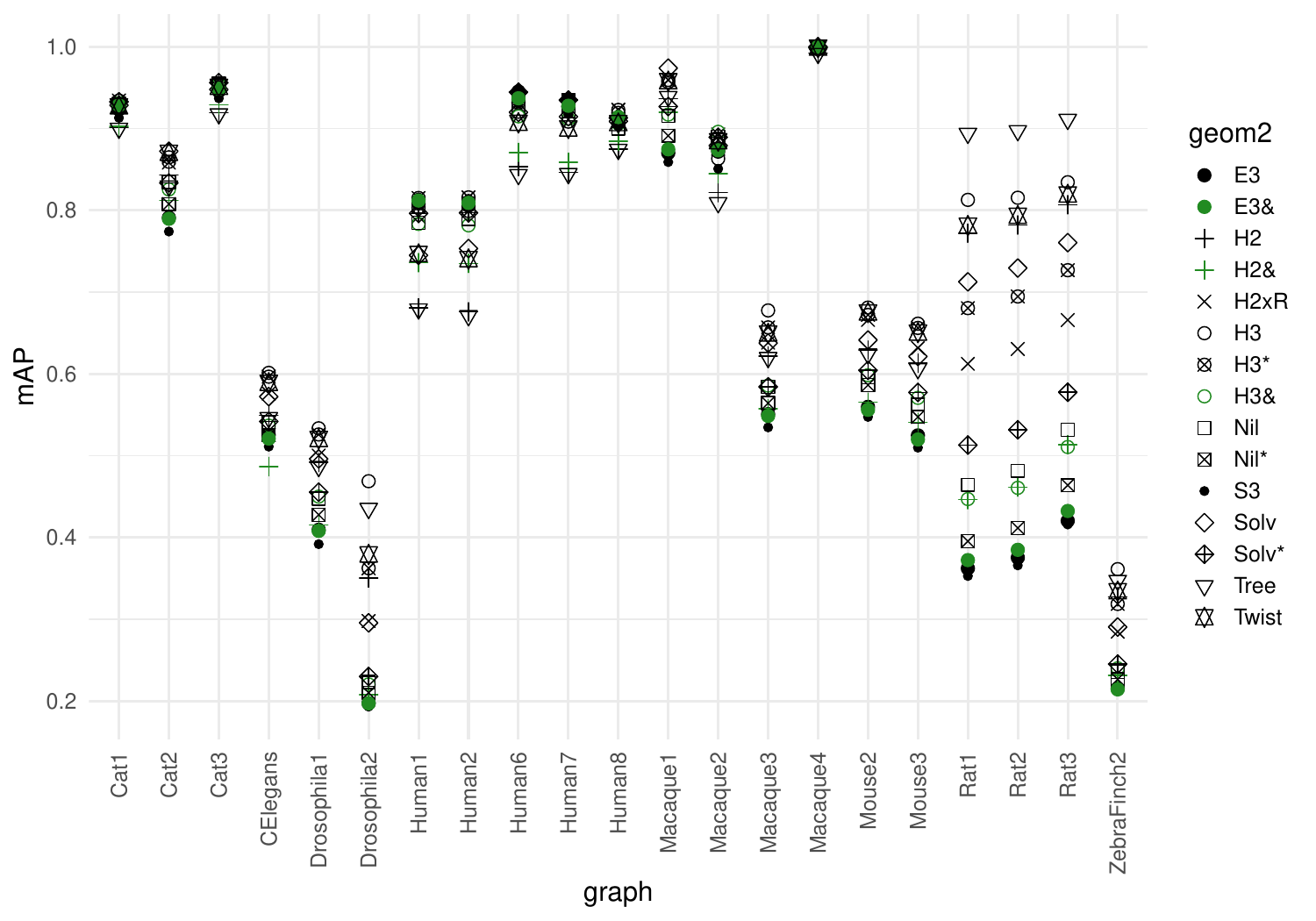}
   \caption{mAP}
  \end{subfigure}\\
  \begin{subfigure}[b]{\textwidth}
   \centering
   \includegraphics[width=0.93\textwidth]{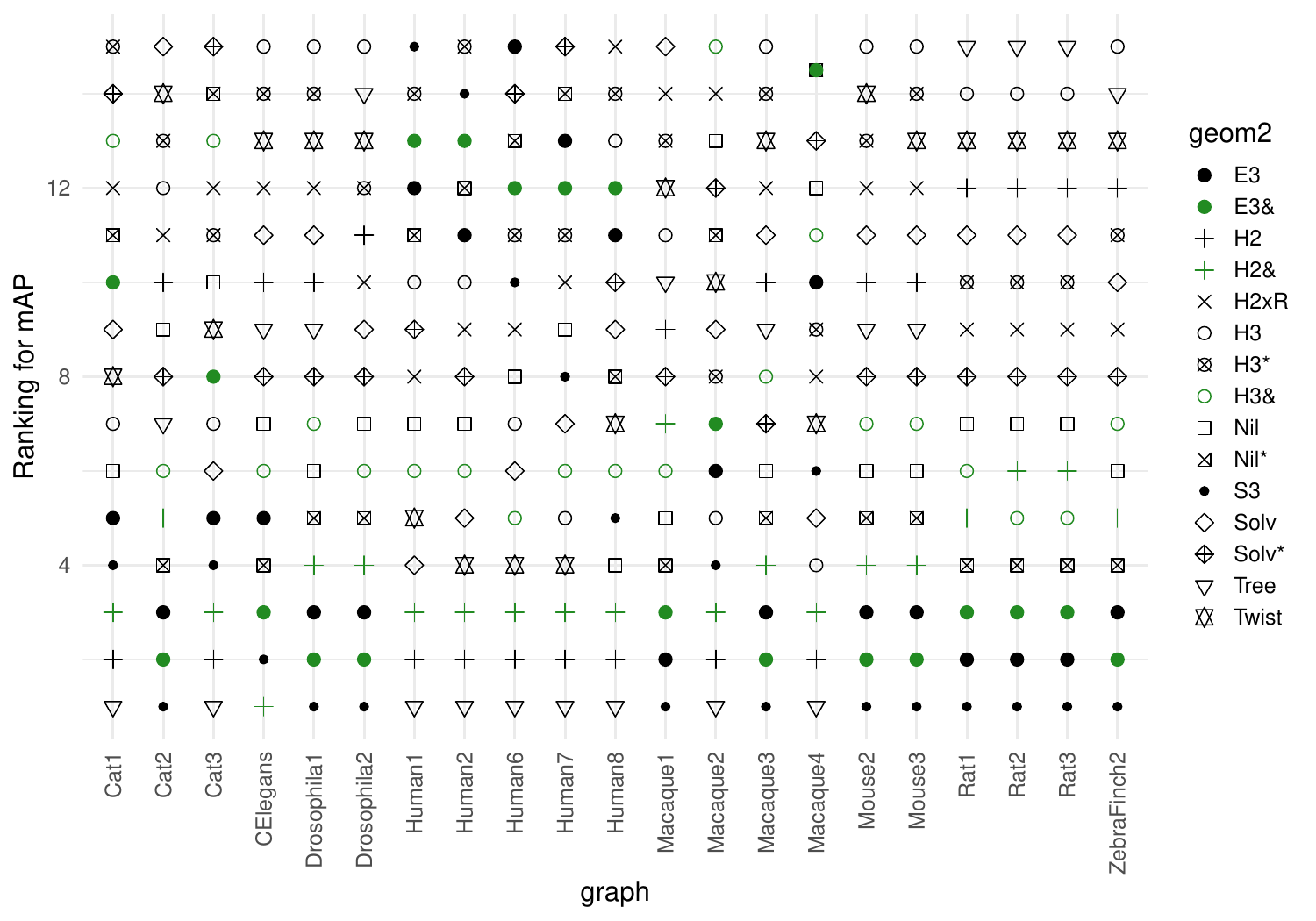}
   \caption{mAP -- ranks}
  \end{subfigure}
\caption{Our best embeddings -- mAP. Top = best embedding obtained, bottom = worst embedding obtained, * = fine grid. \label{fig:bestlogliks2}}
\end{figure}

\begin{figure}[h!]
  \centering   
 \begin{subfigure}[b]{0.93\textwidth}
   \centering
   \includegraphics[width=\textwidth]{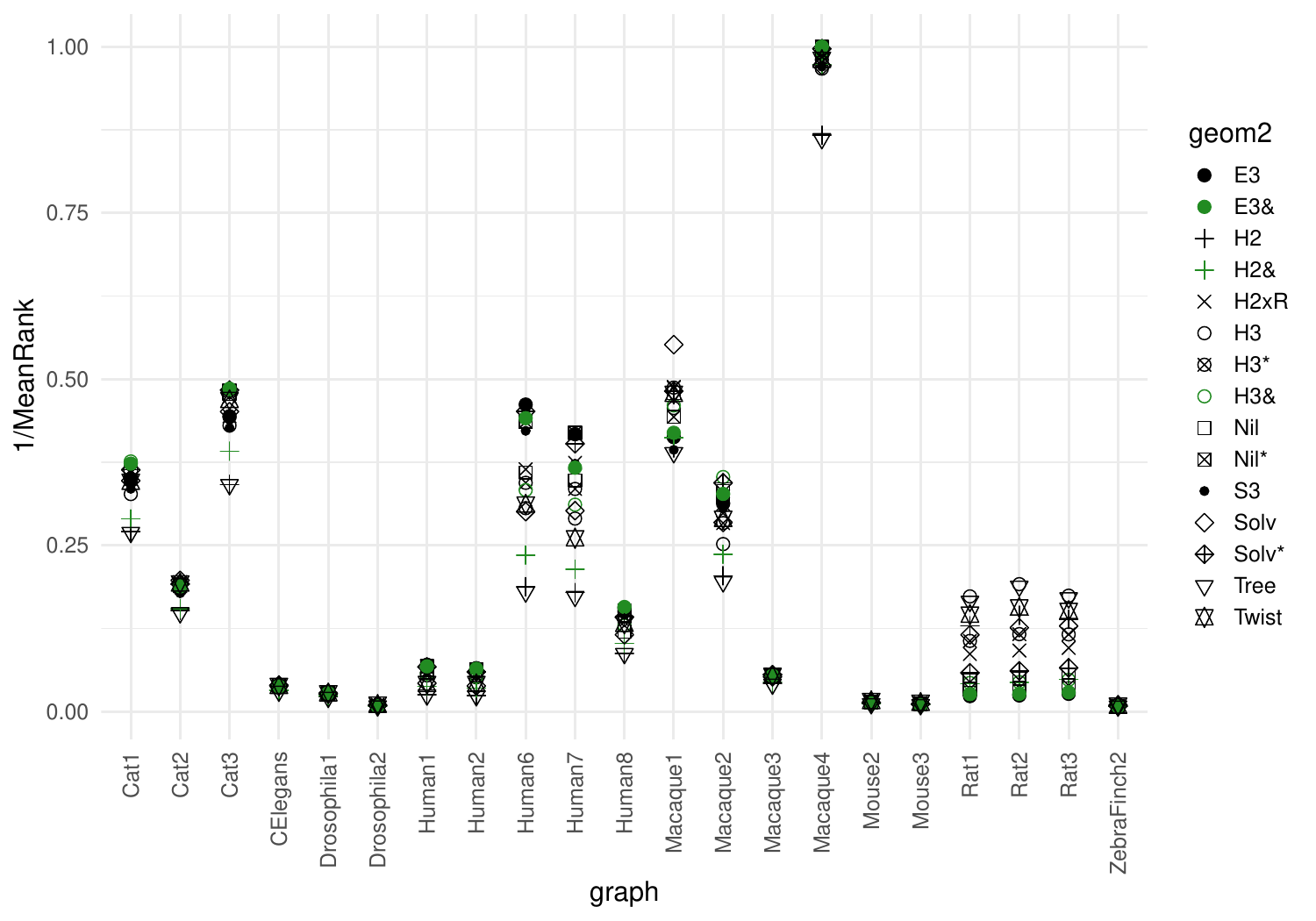}
   \caption{1/MeanRank}
  \end{subfigure}
  \begin{subfigure}[b]{0.93\textwidth}
   \centering
   \includegraphics[width=\textwidth]{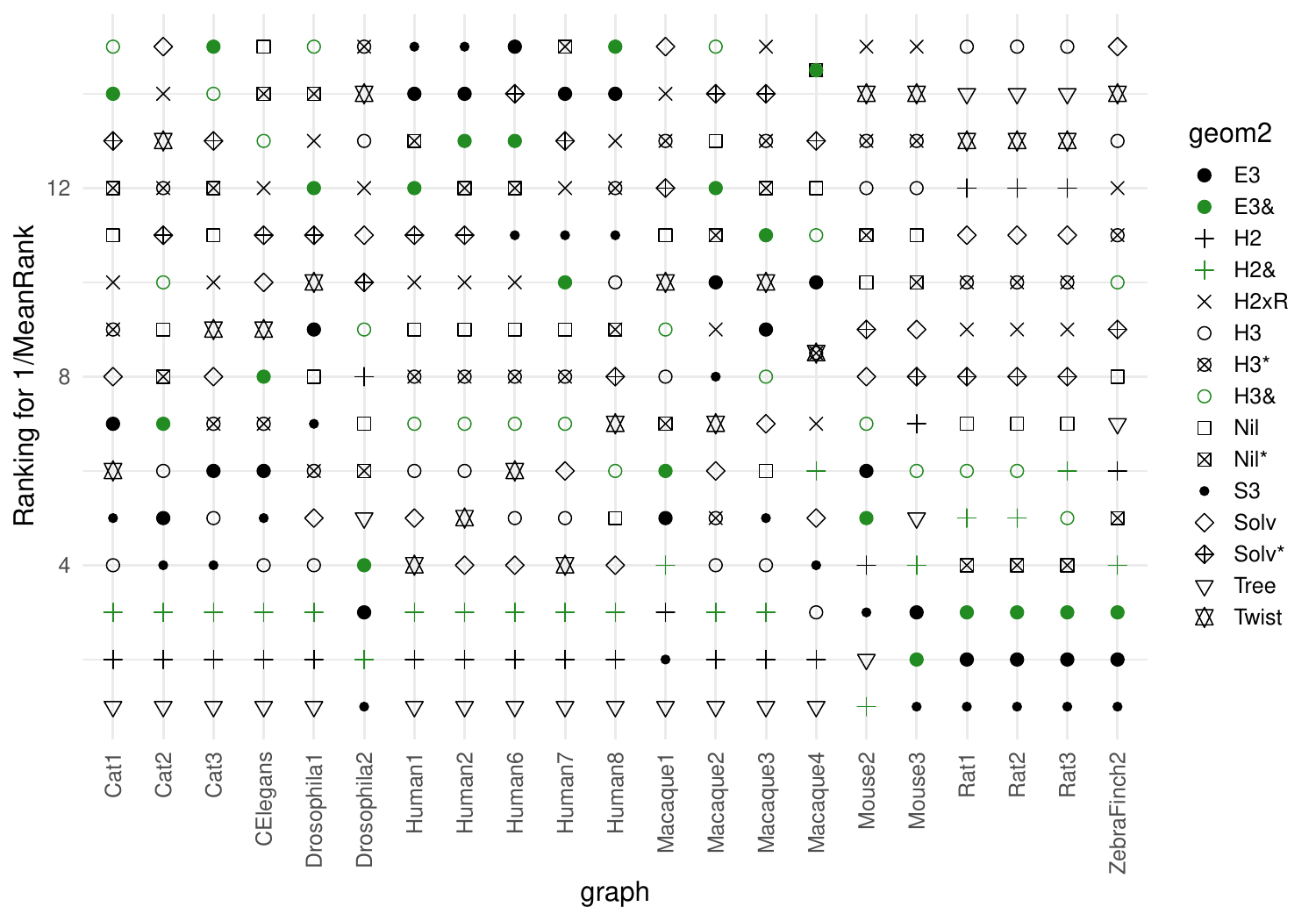}
   \caption{1/MeanRank -- ranks}
  \end{subfigure}
 \caption{Our best embeddings -- MeanRank. Top = best embedding obtained, bottom = worst embedding obtained, * = fine grid. \label{fig:bestlogliks5}}
\end{figure}

\begin{figure}[h!]
  \centering  
  \begin{subfigure}[b]{0.93\textwidth}
   \centering
  \includegraphics[width=\textwidth]{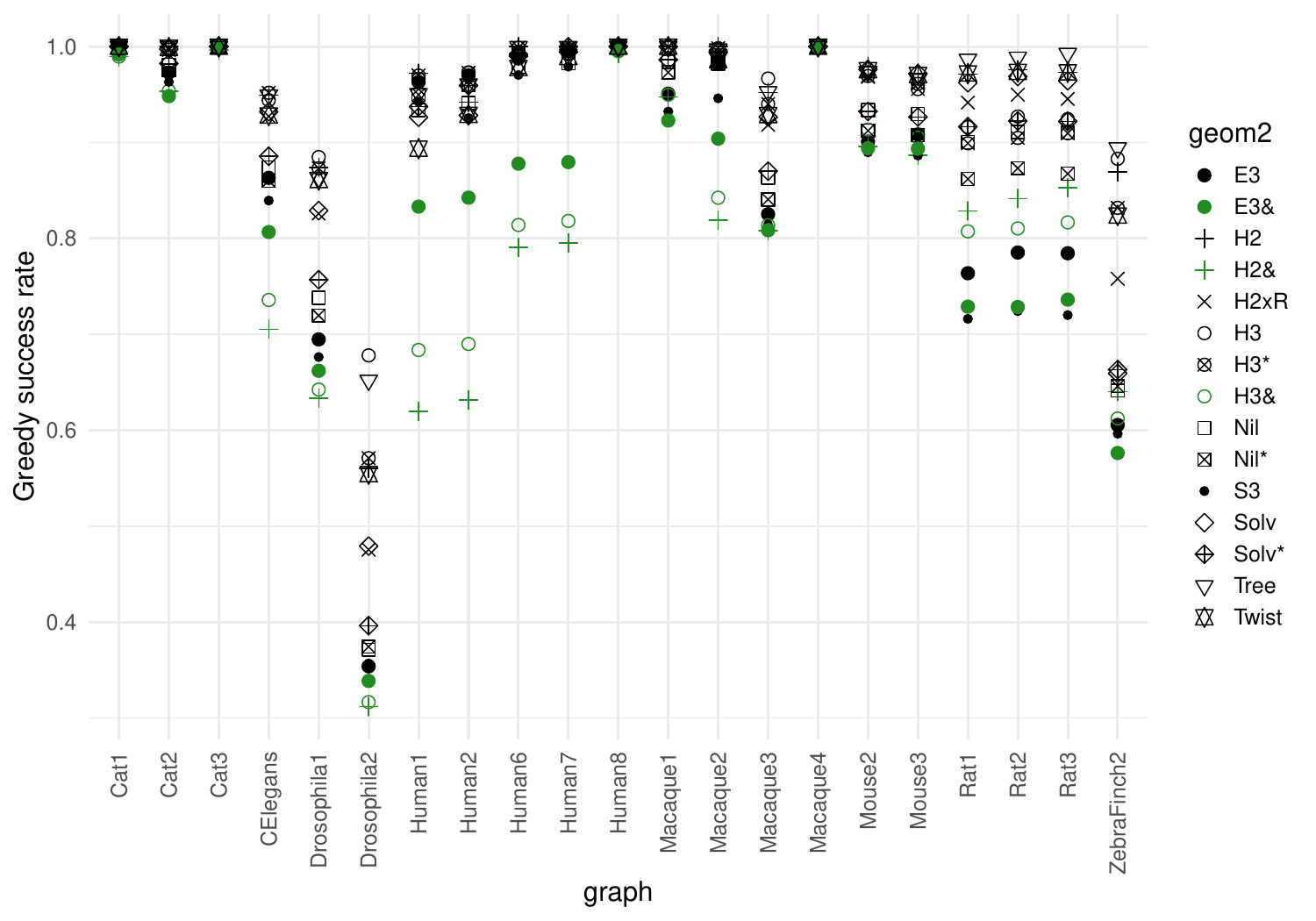}
    \caption{Greedy success rate}
  \end{subfigure}\\
  \begin{subfigure}[b]{0.93\textwidth}
   \centering
   \includegraphics[width=\textwidth]{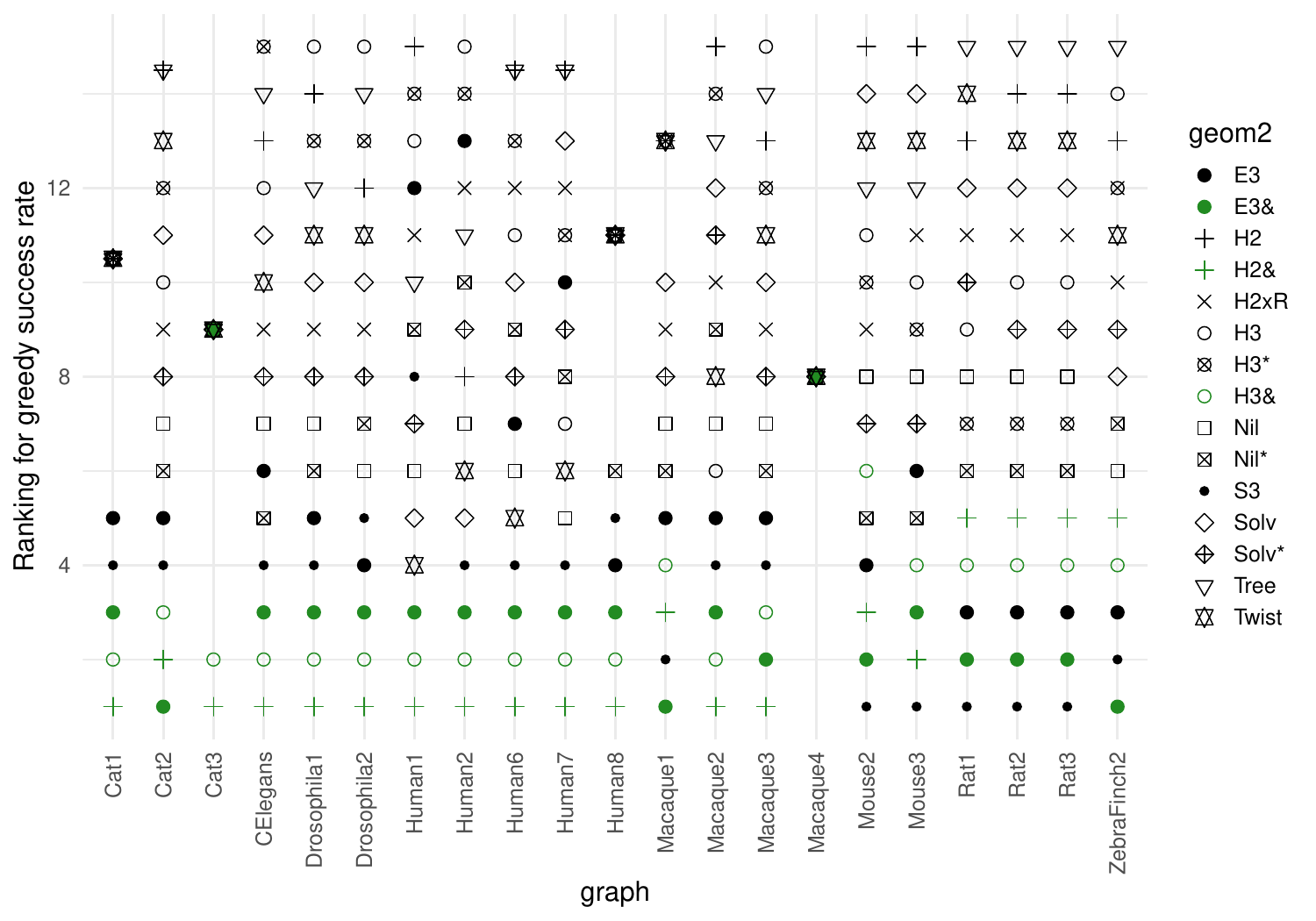}
   \caption{Greedy success rate -- ranks}
  \end{subfigure}
\caption{Our best embeddings -- greedy success rate. Top = best embedding obtained, bottom = worst embedding obtained, * = fine grid. \label{fig:bestlogliks4}}
\end{figure}

\begin{figure}[h!]
  \centering  
  \begin{subfigure}[b]{0.93\textwidth}
   \centering
  \includegraphics[width=\textwidth]{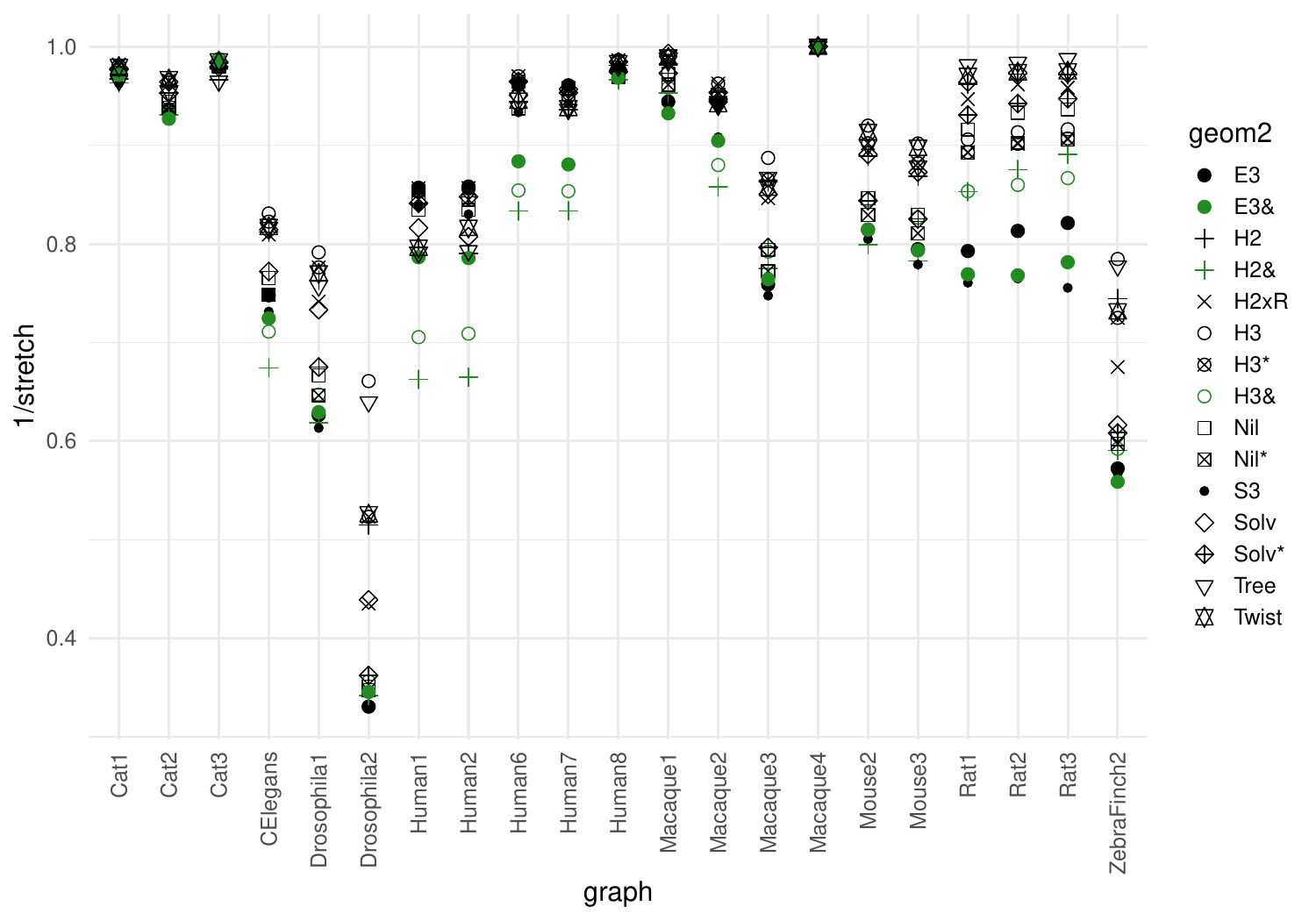}
    \caption{1/stretch}
  \end{subfigure}\\
  \begin{subfigure}[b]{0.93\textwidth}
   \centering
   \includegraphics[width=\textwidth]{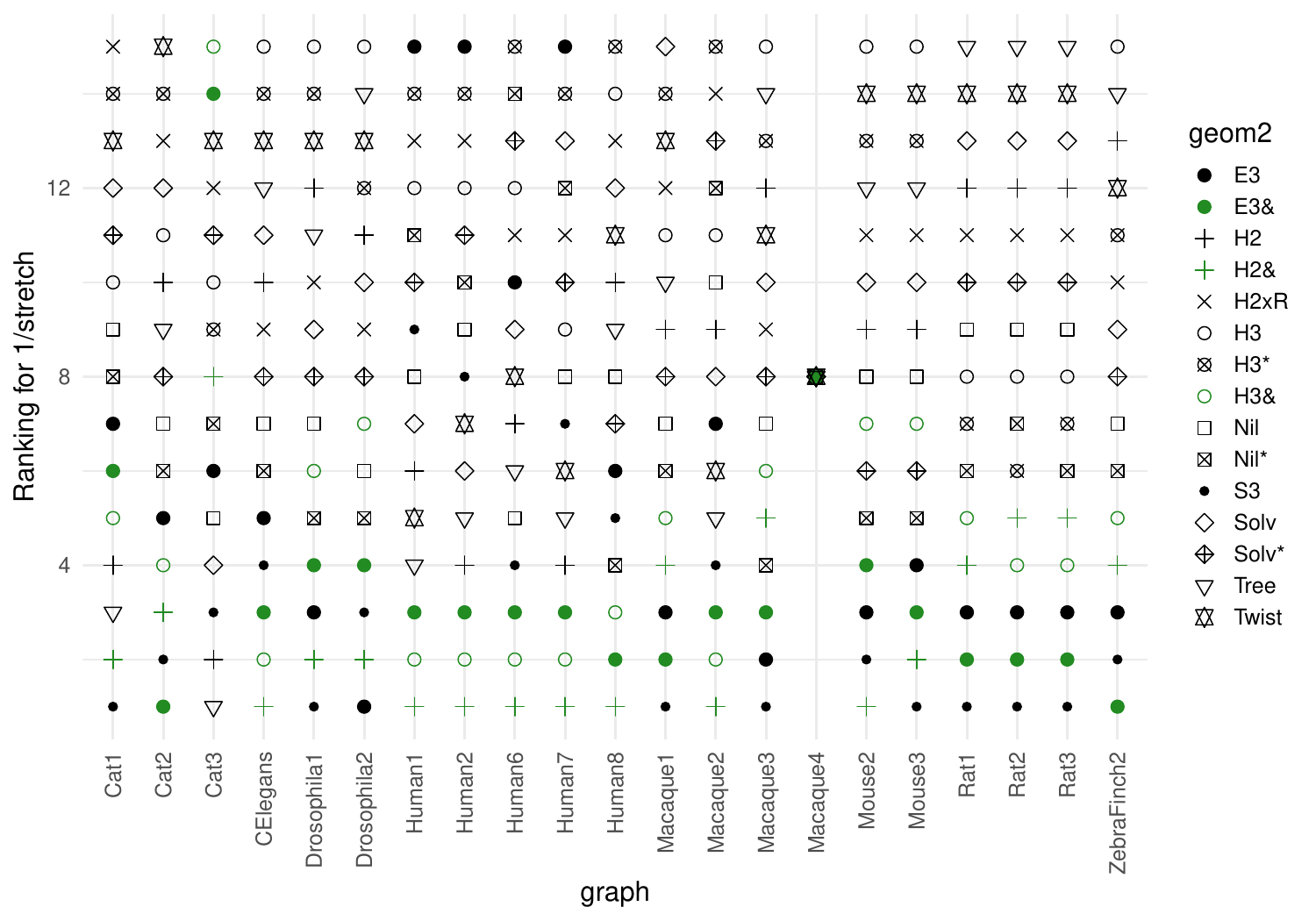}
   \caption{1/stretch -- ranks}
  \end{subfigure}
\caption{Our best embeddings -- stretch. Top = best embedding obtained, bottom = worst embedding obtained, * = fine grid. \label{fig:bestlogliks3}}
\end{figure}

We start with a naive comparison among the tessellations based on the best results obtained for each tessellation for each connectome.
Figures~\ref{fig:bestlogliks1}, \ref{fig:bestlogliks2}, \ref{fig:bestlogliks3}, \ref{fig:bestlogliks4}, and \ref{fig:bestlogliks5}  
visualize the rankings of the tessellations. 1s (the top) are the highest results, and 0s (the bottom) are the lowest for a given connectome.




\begin{table}[h!]
  \centering
\begin{tabular}{|l|r|r|r|r|r|}
  \hline 
connectome & NLL       & MAP    & IMR     & SC    & IST \\ \hline
Cat1         & 5.47   & 1.29   & 10.28  & 0.40  & 0.65 \\
Cat2         & 4.84   & 3.75   & 8.94   & 1.94  & 1.63 \\
Cat3         & 6.22   & 1.35   & 11.04  & 0.09  & 0.66 \\
CElegans     & 7.46   & 6.05   & 8.38   & 8.89  & 6.30 \\
Drosophila1  & 5.46   & 10.15  & 8.34   & 12.19 & 9.47 \\
Drosophila2  & 12.52  & 32.87  & 11.48  & 27.32 & 25.87 \\
Human1       & 9.13   & 5.95   & 29.08  & 11.94 & 7.06 \\
Human2       & 9.19   & 6.20   & 28.38  & 11.62 & 7.00 \\
Human6       & 7.69   & 3.52   & 26.79  & 7.29  & 4.53 \\
Human7       & 8.13   & 3.45   & 25.58  & 7.23  & 4.34 \\
Human8       & 6.38   & 1.72   & 17.92  & 0.23  & 0.74 \\
Macaque1     & 3.95   & 3.93   & 10.21  & 2.87  & 2.21 \\
Macaque2     & 7.22   & 3.02   & 16.74  & 6.11  & 3.30 \\
Macaque3     & 4.99   & 7.52   & 9.05   & 6.88  & 5.84 \\
Macaque4     & 9.44   & 0.27   & 4.51   & 0.00  & 0.00 \\
Mouse2       & 9.68   & 7.54   & 10.86  & 3.78  & 4.94 \\
Mouse3       & 10.85  & 8.84   & 10.98  & 3.58  & 5.14 \\
Rat1         & 44.60  & 32.51  & 66.25  & 10.25 & 8.18 \\
Rat2         & 44.32  & 31.33  & 68.97  & 10.02 & 8.13 \\
Rat3         & 40.76  & 27.42  & 62.36  & 9.85  & 7.96 \\
ZebraFinch2  & 14.83  & 19.70  & 7.06   & 16.29 & 12.50 \\ \hline
  \end{tabular}
\caption{Coefficients of variations (CV, in \%) for the maximum performance of the geometries per connectome \label{tab:cv}}
\end{table}

\begin{table}[h!]
  \centering
  \resizebox{\columnwidth}{!}{%
\begin{tabular}{|l|c|c|c|c|c|c|c|c|c|c|c|c|c|c|r|}
  \hline 
 & \multicolumn{5}{|c}{MIN} & \multicolumn{5}{|c}{MED}  & \multicolumn{5}{|c|}{MAX} \\ \hline 
 geometry              & NLL & MAP & IMR & SC & IST & NLL & MAP  & IMR  & SC  & IST   & NLL    & MAP   & IMR   & SC   & IST \\ \hline
 $\bbH^2$              & 2   & 2  & 2  & 2 & 8  & 4    & 10  & 2  & 13   & 9   & 12    & 12    & 12   & 15   & 13 \\
 $\bbH^2\&$            & 1   & 1  & 1  & 1 & 1  & 3    & 4   & 3  & 1    & 2   & 6     & 7     & 6    & 8    & 8  \\
 tree                  & 1   & 1  & 1  & 1 & 8  & 3    & 9   & 1  & 13   & 10  & 15    & 15    & 14   & 15   & 15 \\
 $\bbE^3$              & 2   & 2  & 2  & 1 & 3  & 5    & 3   & 6  & 5    & 5   & 14    & 15    & 15   & 13   & 15 \\
 $\bbE^3\&$            & 2   & 2  & 2  & 1 & 1  & 5    & 3   & 10 & 3    & 3   & 14.5  & 14.5  & 15   & 9    & 14 \\
 $\bbH^3$              & 4   & 4  & 3  & 8 & 6  & 12   & 13  & 6  & 11   & 12  & 15    & 15    & 15   & 15   & 15 \\
 $\bbH^3*$             & 9   & 8  & 5  & 6 & 7  & 12   & 13  & 10 & 12   & 14  & 15    & 15    & 15   & 15   & 15 \\
 $\bbH^3\&$            & 5   & 5  & 5  & 2 & 2  & 7    & 6   & 8  & 2    & 5   & 14    & 15    & 15   & 8    & 15 \\
 Nil                   & 4   & 4  & 5  & 5 & 5  & 8    & 7   & 9  & 7    & 8   & 14    & 13    & 15   & 11   & 10 \\
 Nil*                  & 4   & 4  & 4  & 4 & 5  & 7    & 5   & 11 & 6    & 6   & 14.5  & 14.5  & 15   & 10.5 & 14 \\
 Twist                 & 4   & 4  & 4  & 5 & 4  & 13   & 13  & 10 & 11   & 13  & 15    & 14    & 14   & 14   & 15 \\
 $\bbH^2 \times \bbR$  & 8   & 8  & 7  & 8 & 8  & 12   & 11  & 12 & 10   & 11  & 15    & 15    & 15   & 12   & 15 \\
 Solv                  & 5   & 4  & 4  & 4 & 5  & 11   & 10  & 8  & 10.5 & 10  & 15    & 15    & 15   & 14   & 15 \\
 Solv*                 & 7   & 7  & 8  & 6 & 7  & 10   & 8   & 11 & 8    & 8   & 15    & 15    & 14   & 11   & 13 \\
 $\bbS^3$              & 1   & 1  & 1  & 1 & 1  & 2    & 1   & 4  & 4    & 2   & 9     & 15    & 15   & 9    & 9  \\ \hline
  \end{tabular}}
\caption{Descriptive statistics (minimum, median, maximum) for ranks obtained by geometries (at the maximum performance) \label{tab:bestranks}}
\end{table}

\begin{table}[h!]
  \centering
\begin{tabular}{|l|r|r|r|r|r|r|r|r|r|r|}
  \hline 
 & \multicolumn{5}{|c}{Top 5 ranks} & \multicolumn{5}{|c|}{Bottom 5 ranks} \\ \hline 
geometry              & NLL     & MAP    & IMR    & SC     & IST    & NLL    & MAP    & IMR   & SC    & IST \\ \hline
$\bbH^2$              & 19.05   & 23.81  & 14.29  & 80.95  & 33.33  & 57.14  & 42.86  & 71.43 & 0.00  & 19.05 \\
$\bbH^2\&$            & 0.00    & 0.00   & 0.00   & 0.00   & 0.00   & 95.24  & 85.71  & 90.48 & 95.24 & 90.48 \\
tree                  & 23.81   & 23.81  & 14.29  & 80.95  & 47.62  & 66.67  & 42.86  & 80.95 & 0.00  & 28.57 \\
$\bbE^3$              & 19.05   & 23.81  & 23.81  & 9.52   & 14.29  & 57.14  & 66.67  & 38.10 & 61.90 & 57.14 \\
$\bbE^3\&$            & 19.05   & 28.57  & 47.62  & 0.00   & 4.76   & 52.38  & 57.14  & 33.33 & 90.48 & 85.71 \\
$\bbH^3$              & 66.67   & 61.90  & 33.33  & 52.38  & 66.67  & 9.52   & 14.29  & 42.86 & 0.00  & 0.00  \\
$\bbH^3*$             & 66.67   & 76.19  & 38.10  & 61.90  & 76.19  & 0.00   & 0.00   & 4.76  & 0.00  & 0.00  \\
$\bbH^3\&$            & 9.52    & 19.05  & 28.57  & 0.00   & 4.76   & 14.29  & 14.29  & 4.76  & 90.48 & 66.67 \\
Nil                   & 19.05   & 9.52   & 33.33  & 4.76   & 0.00   & 4.76   & 9.52   & 4.76  & 4.76  & 9.52  \\
Nil*                  & 38.10   & 38.10  & 57.14  & 0.00   & 19.05  & 28.57  & 57.14  & 19.05 & 14.29 & 28.57 \\
Twist                 & 61.90   & 57.14  & 38.10  & 57.14  & 71.43  & 19.05  & 19.05  & 14.29 & 9.52  & 4.76 \\
$\bbH^2 \times \bbR$  & 66.67   & 52.38  & 52.38  & 42.86  & 71.43  & 0.00   & 0.00   & 0.00  & 0.00  & 0.00 \\
Solv                  & 52.38   & 47.62  & 33.33  & 47.62  & 42.86  & 14.29  & 14.29  & 28.57 & 9.52  & 4.76 \\
Solv*                 & 38.10   & 28.57  & 61.90  & 9.52   & 23.81  & 0.00   & 0.00   & 0.00  & 0.00  & 0.00 \\
$\bbS^3$              & 0.00    & 9.52  & 23.81  & 0.00   & 0.00   & 80.95   & 76.19  & 66.67  & 85.71 & 80.95 \\ \hline
  \end{tabular}
\caption{Percentages how many times a given geometry occurred within top or bottom five ranks (at the maximum performance) \label{tab:bestperc}}
\end{table}

According to Figures \ref{fig:bestlogliks1}-\ref{fig:bestlogliks5} and Tables \ref{tab:bestranks} and \ref{tab:bestperc}, we notice that
the assessment of the performance of the geometry may vary concerning the quality measure; there are also differences across species.
In general, trees perform poorly in measures other than greedy success rate, and no matter the measure, they are always the best
choice for Rat's connectomes (nervous system). Results for Rat's and Drosophila2's connectomes are also characterized by the relatively high variation among species (Table \ref{tab:cv}).
For other species, the best performances are similar with respect to a quality measure: the differences in best performance
among geometries measured with MAP, greedy rate success, and stretch are slight (in most cases, values of CVs are under 10\%); especially for Cat's connectomes, they tend to be negligible (values of CVs even under 1\%). 

The results suggest that $\bbH^2\&$ and $\bbS^3$ seem to be inefficient choices: the first one never enters the top five ranks; both often occur within the bottom five ranks, at their best performance being even the worst choices no matter the quality measure. In contrast, $\bbH^3$ and $\bbH^2 \times \bbR$ perform very well -- they rarely occur within the bottom five ranks. Twist and Solv or Solv$*$ never happen to be the worst choices; they all perform relatively well. Interestingly, the usage of finer grids may not increase the chance of obtaining the best performance, no matter the quality measure: while for $\bbH^3*$ vs. $\bbH^3$ and Solv* vs. Solv, we notice that it reduces the chance of occurring within the bottom five ranks, the best performances of non-fine grids still outperform them when it comes to the occurrences within the five top ranks. On the contrary, a finer grid for Nil significantly increases the percentage of occurrences among the five best ranks. When it comes to Euclidean geometry, the results are inconsistent. The best performances of $\bbE^3$ and $\bbE^3\&$ often occur among the bottom five ranks of the geometries. However, there are cases in which those geometries perform excellently, e.g., for Human connectomes.

\subsection{Distribution-based comparison}

Comparison of the maximum performance from the previous section gives us intuition about the optimistic scenarios and the limits for our embeddings.
However, due to the nature of Simulated Annealing, the maximum values we obtained are still realizations of random variables; that is why a closer inspection,
including information about the distributions of the simulation results, is needed. To this end, we will compare geometries using voting rules. In particular,
we will be interested in finding Condorcet winners and losers. As Condorcet winner may not exist in the presence of ties, we will refer to its simple modification:
Copeland rule \cite{copeland}.

Geometry A wins against geometry B if the probability that (for a given quality measure) a randomly chosen simulation result obtained by A is greater
than a randomly chosen simulation result obtained by B exceeds 0.5. If that probability is equal to 0.5, we have a tie between A and B;
otherwise, A loses against B. To compute the score for a given geometry, we add 1 for every winning scenario, 0 for every tie, and -1 for every losing scenario. The geometries with the highest and lowest scores become Copeland winners and losers, respectively (we allow for multiple candidates in both cases).

\begin{table}[h!]
  \begin{tabular}{|l|c|c|c|c|c|c|c|c|c|c|r|} \hline
              & \multicolumn{5}{|c}{Copeland winners} & \multicolumn{5}{|c|}{Copeland losers} \\ \hline 
 connectome    & NLL         & MAP         & IMR                   & SC           & IST                   & NLL           & MAP         & IMR         & SC             & IST       \\ \hline
 Cat1          & Solv*       & $\bbH^3*$   & Solv*                 & $\bbH^3*$    & Solv*                 & $\bbH^2\&$    & tree        & tree        & $\bbH^2\&$     & tree      \\
 Cat2          & $\bbH^3*$   & $\bbH^3*$   & $\bbH^2 \times \bbR$  & Twist        & $\bbH^2 \times \bbR$  & $\bbH^2\&$    & $\bbS^3$    & tree        & $\bbH^3\&$     & tree  \\
 Cat3          & Solv*       & Solv*       & $\bbH^3\&$            & Nil*         & $\bbH^3\&$            & $\bbH^2\&$    & tree        & tree        & $\bbH^2\&$     & tree  \\
 CElegans      & $\bbH^3*$   & $\bbH^3$    & Nil                   & $\bbH^3*$    & Nil                   & $\bbH^2\&$    & $\bbH^2\&$  & tree        & $\bbH^2\&$     & tree  \\
 Drosophila1   & Twist       & $\bbH^3$    & $\bbH^3\&$            & $\bbH^3$     & $\bbH^3\&$            & $\bbH^2\&$    & $\bbS^3$    & tree        & $\bbH^2\&$     & tree \\
 Drosophila2   & $\bbH^3$    & $\bbH^3$    & $\bbH^3*$             & $\bbH^3$     & $\bbH^3*$             & $\bbS^3$      & $\bbS^3$    & $\bbS^3$    & $\bbH^3\&$     & $\bbS^3$  \\
 Human1        & $\bbE^3$    & $\bbS^3$    & $\bbS^3$              & $\bbH^3*$    & $\bbS^3$              & tree          & tree        & tree        & $\bbH^2\&$     & tree  \\
 Human2        & $\bbE^3$    & $\bbS^3$    & $\bbS^3$              & $\bbH^3*$    & $\bbS^3$              & tree          & tree        & tree        & $\bbH^2\&$     & tree  \\
 Human6        & $\bbE^3$    & $\bbE^3$    & $\bbE^3$              & $\bbH^3*$    & $\bbE^3$              & tree          & tree        & tree        & $\bbH^2\&$     & tree  \\
 Human7        & $\bbE^3$    & $\bbE^3$    & $\bbE^3$              & Solv         & $\bbE^3$              & tree          & tree        & tree        & $\bbH^2\&$     & tree  \\
 Human8        & $\bbH^3*$   & $\bbH^3*$   & $\bbE^3$              & $\bbH^2$     & $\bbE^3$              & tree          & tree        & tree        & $\bbH^2\&$     & tree  \\
 Macaque1      & Solv        & Solv        & Solv                  & $\bbH^3*$    & Solv                  & $\bbS^3$      & $\bbS^3$    & tree        & $\bbE^3\&$     & tree  \\
 Macaque2      & Nil         & Nil         & Nil*                  & $\bbH^2$     & Nil*                  & tree          & tree        & tree        & $\bbH^2\&$     & tree  \\
 Macaque3      & $\bbH^3*$   & $\bbH^3*$   & $\bbH^2 \times \bbR$  & $\bbH^2$     & $\bbH^2 \times \bbR$  & $\bbH^2\&$    & $\bbS^3$    & tree        & $\bbH^2\&$     & tree  \\
 Macaque4      & $\bbE^3\&$  & $\bbE^3\&$  & $\bbE^3\&$            & Twist        & $\bbE^3\&$            & tree          & tree        & tree        & $\bbE^3$       & tree  \\
 Mouse2        & Twist       & $\bbH^3$    & $\bbH^2 \times \bbR$  & $\bbH^2$     & $\bbH^2 \times \bbR$  & $\bbS^3$      & $\bbS^3$    & $\bbH^2\&$  & $\bbS^3$       & $\bbH^2\&$  \\
 Mouse3        & Twist       & $\bbH^3$    & $\bbH^2 \times \bbR$  & $\bbH^2$     & $\bbH^2 \times \bbR$  & $\bbS^3$      & $\bbS^3$    & $\bbS^3$    & $\bbH^2\&$     & $\bbS^3$  \\
 Rat1          & tree        & tree        & $\bbH^3$              & tree         & $\bbH^3$              & $\bbS^3$      & $\bbS^3$    & $\bbS^3$    & $\bbS^3$       & $\bbS^3$  \\
 Rat2          & tree        & tree        & $\bbH^3$              & tree         & $\bbH^3$              & $\bbS^3$      & $\bbS^3$    & $\bbS^3$    & $\bbE^3\&$     & $\bbS^3$  \\
 Rat3          & tree        & tree        & $\bbH^3$              & tree         & $\bbH^3$              & $\bbS^3$      & $\bbS^3$    & $\bbS^3$    & $\bbS^3$       & $\bbS^3$  \\
 ZebraFinch2   & Solv        & $\bbH^3$    & Solv                  & $\bbH^3$     & Solv                  & $\bbS^3$      & $\bbS^3$    & $\bbS^3$    & Solv           & $\bbS^3$  \\ \hline
\end{tabular}
\caption{Voting rules: Copeland winners and losers. \label{fig:voting}}
\end{table}

The winners based on the Copeland method beat most of the other candidates in pairwise contests. They should be the best options for embeddings.
Based on the data in Table~\ref{fig:voting}, we cannot name one universal winner. While it seems that $\bbH^3$ is a sound choice, we also notice that Solv and Twist are worthy of attention. Interestingly, for Human connectomes, $E^3$ outperforms other geometries.

\begin{figure}[h!]
  \centering
   \includegraphics[width=\textwidth]{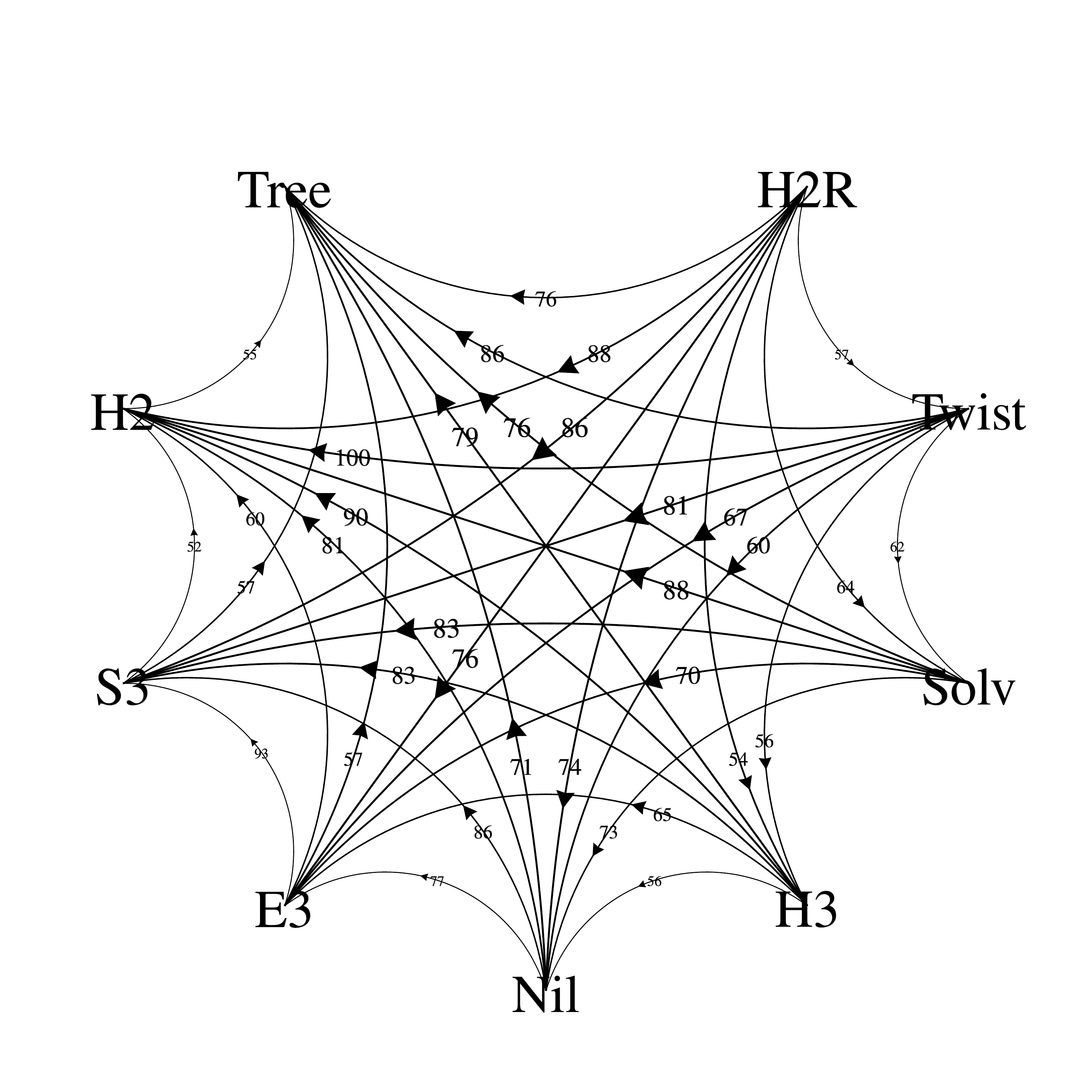}
   \caption{Normalized log-likelihood \label{fig:graphvoting1}}
\end{figure}

\begin{figure}[h!]
   \centering
   \includegraphics[width=\textwidth]{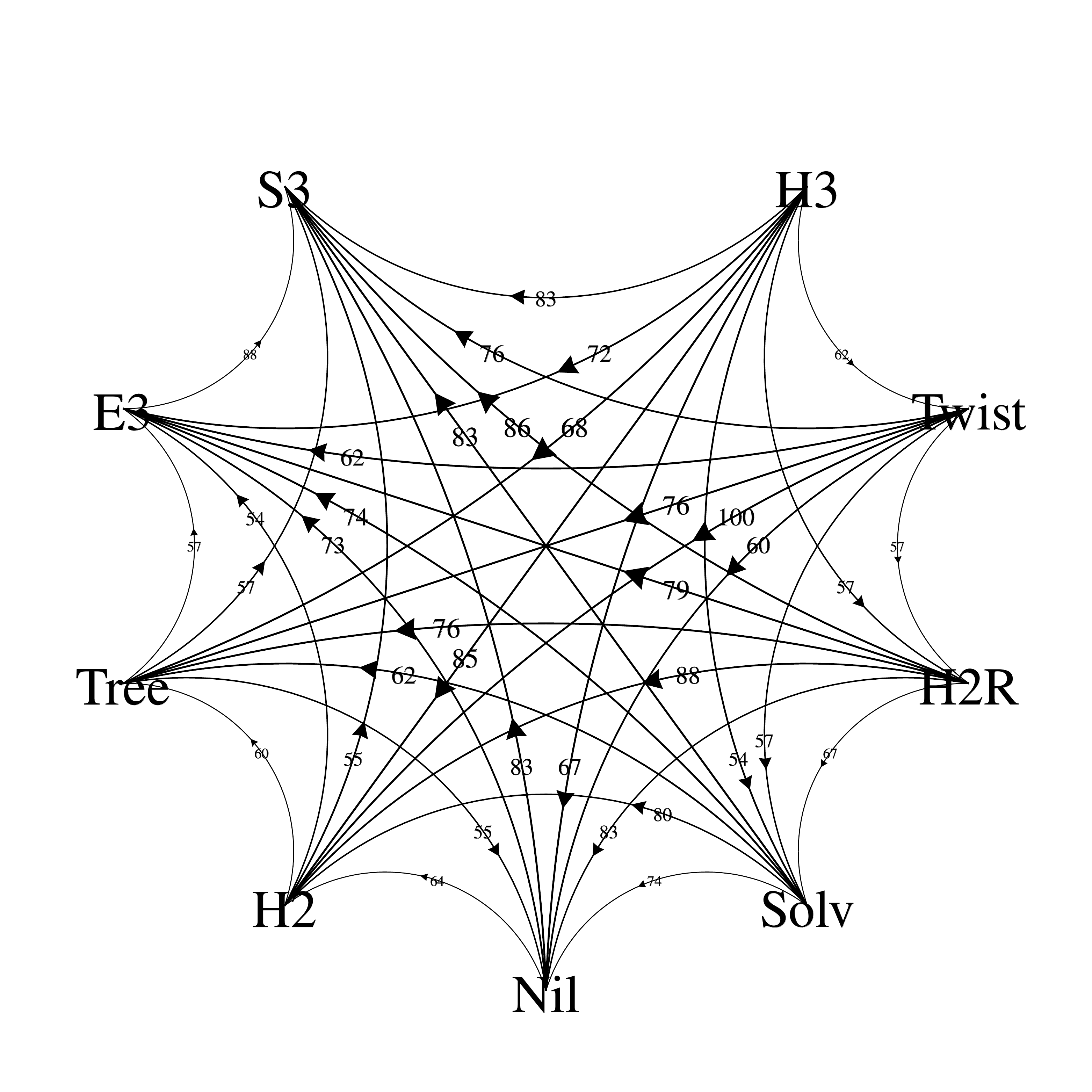}
   \caption{MAP \label{fig:graphvoting2}}
   \end{figure}

\begin{figure}[h!]
   \centering
   \includegraphics[width=\textwidth]{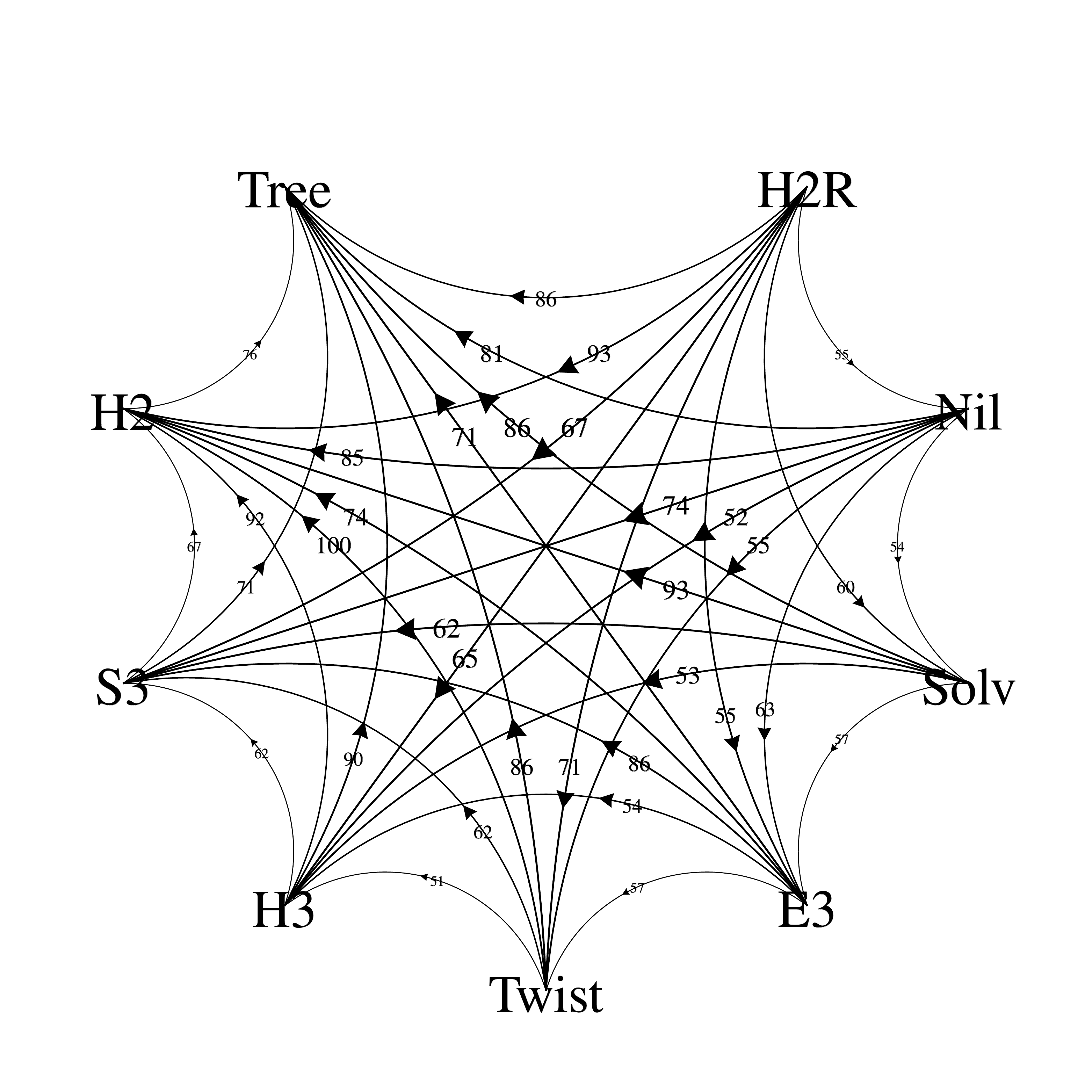}
    \caption{IMR \label{fig:graphvoting3}}
  \end{figure}

\begin{figure}[h!]
   \centering
   \includegraphics[width=\textwidth]{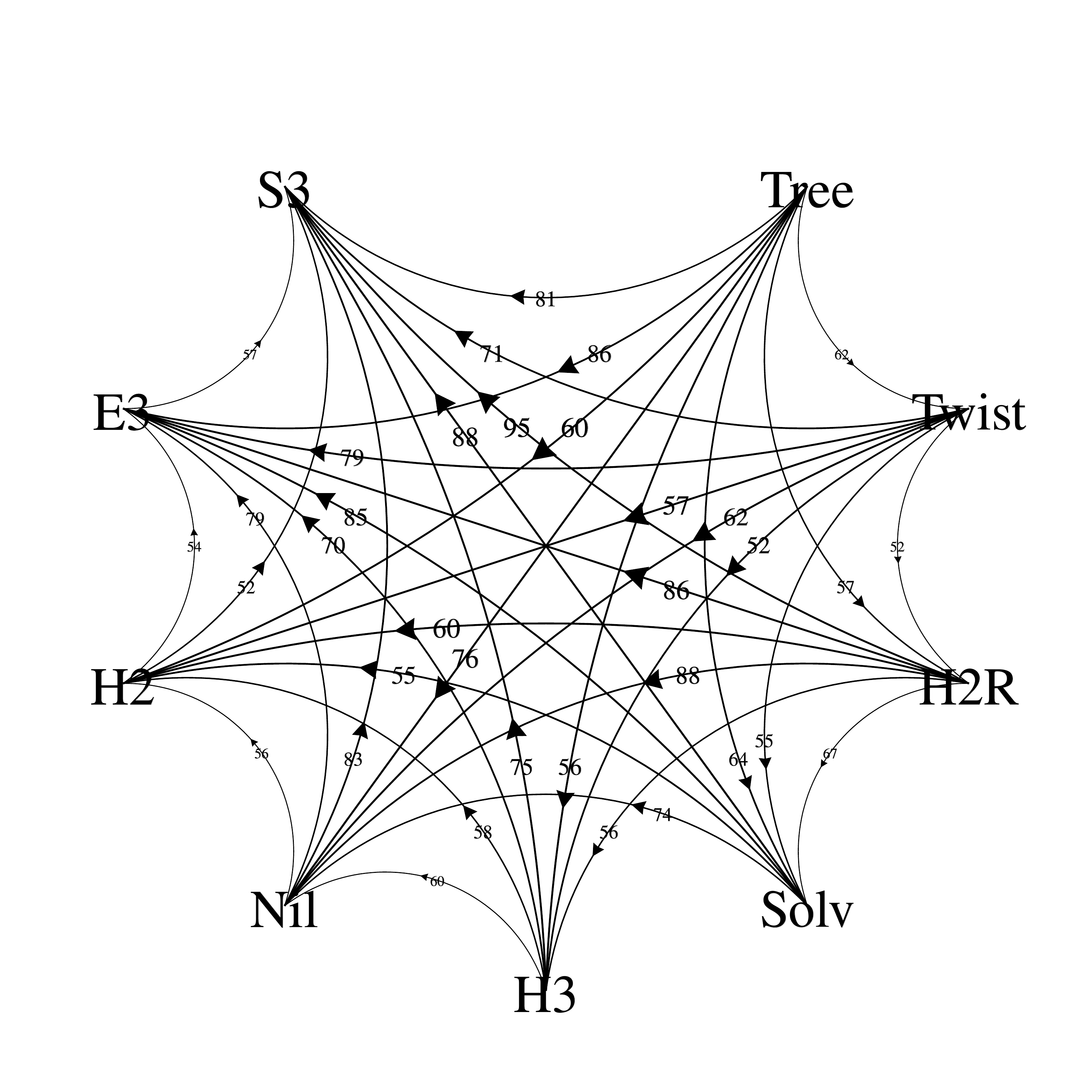}
   \caption{SC \label{fig:graphvoting4}}
\end{figure}

  \begin{figure}[h!]
   \centering
   \includegraphics[width=\textwidth]{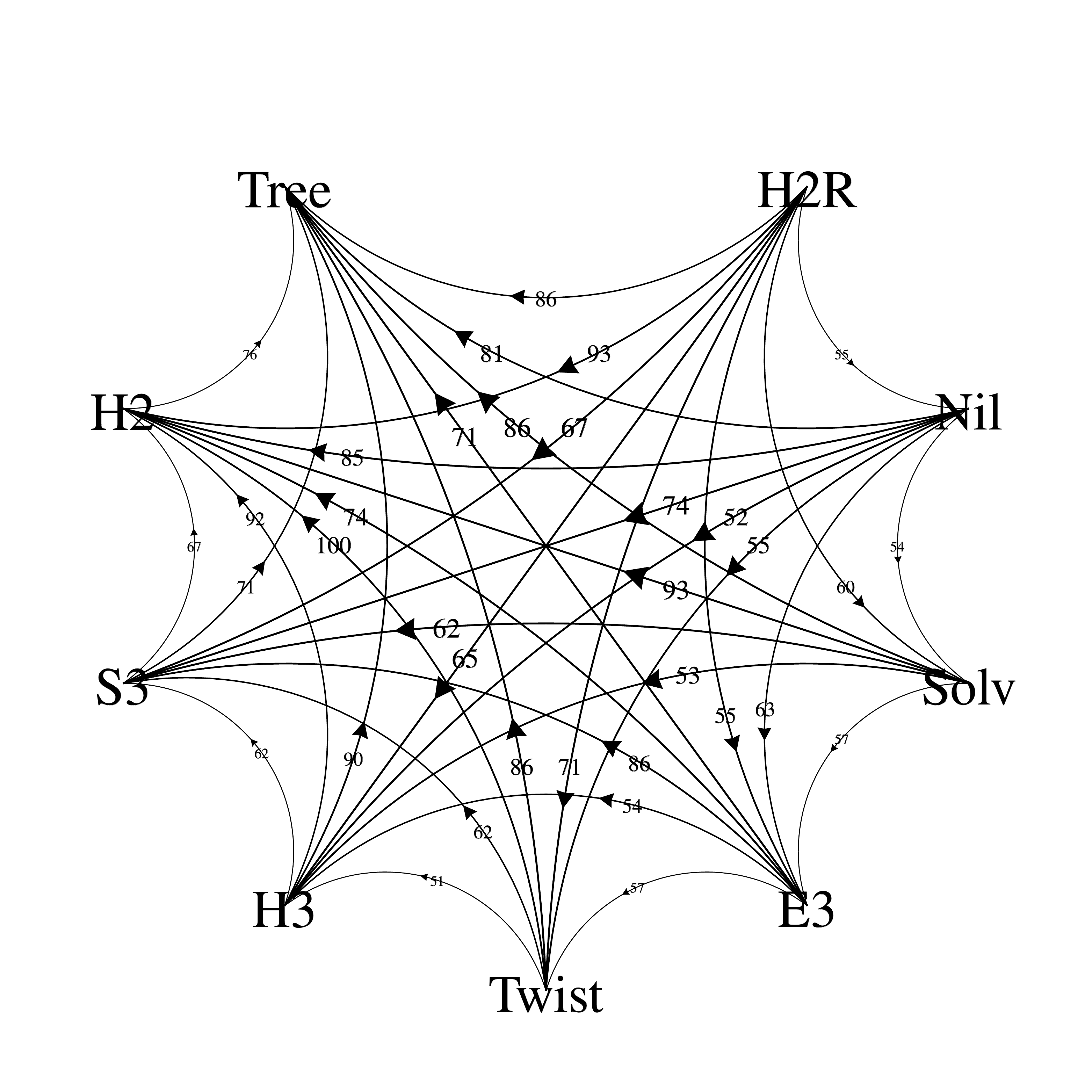}
  \caption{ISTR  \label{fig:graphvoting5}}
\end{figure}

In Figures \ref{fig:graphvoting1}-\ref{fig:graphvoting5}, we provide weighted directed networks constructed upon the voting rules to allow for generalizations. The weights correspond to the percent of connectomes for which the source geometry in the edge beats the target geometry. 
Embeddings to Twist have a 100\% success rate over embeddings in $\bbH^2$ (for quality measures different than greedy routing success).

\subsection{Zone-function-based comparison}
We have already shown that, contrary to previous results from the literature, we cannot name one universal best geometry to model any connectome.
A new interesting question arises if there are relationships between the function of the connectome (based on its zone) and the suitability of the geometries (using the rankings from the distribution-based comparisons).
To find out, we analyze the values of intraclass correlation coefficients (ICC), a widespread tool in the assessment of consistency among multiple raters \cite{icc_raters} when the rating scale is ordinal to continuous. The literature suggests that the values of ICC below 0.50 indicate poor agreement, between 0.50 and 0.90 suggest moderate to good agreement, and above 0.90: an excellent one \cite{icc_interpret}. 
To obtain an aggregate ranking for the whole zone, for each measure and each geometry
we computed the median of its minus rank (the minus gives us information 50\% of ranks achieved by the given geometry are at least this high in modelling given zone, the lower values, the better).
Medians were chosen due to their immunity to outliers.


\begin{table}[h!]
  \resizebox{\columnwidth}{!}{%
  \begin{tabular}{|l|c|c|c|r|} \hline
measure & cortex & nervous & other & inter-zone \\ \hline

NLL    & 0.918  & 0.944  & 0.834  & 0.659 \\
       & (0.836;0.968) & (0.877;0.979) & (0.663;0.936) & (0.164;0.878) \\
       & p = 7.61e-16 & p = 6.98e-13 & p = 6.31e-08  & p = 0.00964 \\ \hline
MAP    & 0.8  & 0.977  & 0.8  & 0.693  \\
       & (0.597;0.922) & (0.949;0.991) & (0.593;0.923) & (0.25;0.89) \\
       & p = 1e-06  & p = 9.83e-20  & p = 1.59e-06  & p = 0.00505 \\ \hline
IMR    & 0.918  & 0.944  & 0.834  & 0.659  \\
       & (0.836;0.968) & (0.877;0.979) & (0.663;0.936) & (0.164;0.878) \\
       & p = 7.61e-16 & p = 6.98e-13  & p = 6.31e-08  & p = 0.00964 \\ \hline
IST    & 0.918  & 0.944  & 0.834  & 0.659 \\
       & (0.836;0.968) & (0.877;0.979) & (0.663;0.936) & (0.164;0.878) \\
       & p = 7.61e-16  & p = 6.98e-13 & p = 6.31e-08 & p = 0.00964 \\ \hline
SC     & 0.891  & 0.97  & 0.943  & 0.931  \\
       & (0.782;0.958) & (0.935;0.989) &  (0.884;0.978) & (0.834;0.975) \\
       & p = 1.98e-12 & p = 9.57e-18 & p = 9.06e-20  & p = 5.83e-09 \\ \hline
\end{tabular}}
\caption{The agreements between the rankings obtained for intra-zone and inter-zone rankings of geometries (values of two-way mixed effects, absolute agreement, multiple xraters ICCs). 95\% confidence intervals in brackets. \label{tab:icc}}
\end{table}

\def\hprod{\bbH^2 {\hskip -1mm} \times {\hskip -1mm} \bbR}
According to data in Table \ref{tab:icc}, the intra-zone comparisons suggest good to excellent agreement no matter the quality measure. 
Rankings for connectomes from ``other'' zone (usually specific cells, e.g., from retina or optic medula) show relatively lower agreement.
On the contrary, the inter-zone comparison suggests poor to moderate agreement between the rankings. While the p-values in significance tests
for ICCs in inter-zone comparisons suggest significance at any reasonable significance level (even after Bonferroni corrections), the results
for intra-zone comparison after Bonferroni corrections appear insignificant (so any similarities might be random), apart from the result for
the SC measure. Those results are promising for us. They suggest that the choice of the suitable geometry may depend on the function of the connectome.
For example, our results suggest that the trees are best choice in modelling nervous systems (no matter the quality measure), for cortex $\bbH^3*$, $\hprod$, 
Nil* or Solv* would be a suitable choice, and Twist may be beneficial for modelling specific cells.

While a natural question arises about the anatomical implications of different best fits for geometries, as well as why different connectomes might have different best geometries,
answering this question in a statistically robust manner would require a more detailed study on a bigger number of sample connectomes.

\section{Robustness checks and threats to validity} 

Ideally, there exists optimal embedding of $(V,E)$ into the whole geometry $\bbG$, where $m_{\mbox{opt}}:V\rightarrow \bbG$, and some values of $R$ and $T$ are used.
Unfortunately, the embedding $m$ found by Simulated Annealing might be worse than $m_{\mbox{opt}}$ due to the following issues:

\begin{itemize}
\item The radius $d_R$ is too small, making $m_{\mbox{opt}}$ simply not fit,
\item The grid used is too coarse, hence the necessity of making $m(i)$ the grid point to closest to $m_{\mbox{opt}}(i)$, and thus reducing the log-likelihood,
\item The number of iterations of Simulated Annealing, $N_S$, is too small -- while Simulated Annealing is theoretically guaranteed to find the optimal embedding for given $R$ and $T$ with high probability as $N_S$ tends to infinity, in practice, we are constrained by computation time limits,
\item The values of the parameters $R$ and $T$ have not been chosen correctly.
\end{itemize}

In this section, we will explain how we combated those issues. We will also check if they affected our results.

\paragraph{Possibly insufficient size of grids.}
For comparability, we aimed to keep the number of neurons as close to 20,000 as possible. However, one could argue if this is enough. To combat the first two issues, in some geometries, we consider coarser and finer grids: coarser grids are better at handling the first issue, and finer grids are better at handling the second issue -- 
in both cases, we expect that increasing $d_R$ and grid density beyond some threshold yields diminishing returns. That is why, based on the results from the previous sections, we have added the so-called \emph{big} versions -- coarser but larger grids ($M=100000$) -- for selected, promising manifolds ($\bbH^3$, $\bbH^3*$, $\bbH^2 \times \bbR$, Solv, and Twist). We will denote them with **. See Table \ref{tab:geoms2} for the details.

\begin{table}[h!]
  \centering
  \resizebox{\columnwidth}{!}{%
\begin{tabular}{|l|l|l|l|l|l|l|l|l|l|}
  \hline
name                           &  dim &  geometry   &  closed &  nodes &  diameter   & description of the set $D$ \\ \hline 
$\bbH^3**$                     & 3    &  hyperbolic &  F      &  100427 &       233  & $\{4,3,5\}$ hyperbolic honeycomb \\ 
$(\bbH^3*)**$                  & 3    &  hyperbolic &  F      &  100641 &       179  & $\{4,3,5\}$ subdivided(2) \\ 
Twist**                        & 3    &  twist      &  F      &  101230 &       184  & twisted $\{5,4\} \times \bbZ$ \\ 
$\bbH^2 \times \bbR**$         & 3    &  product    &  F      &  100030 &       282  & bitruncated $\{7,3\} \times \bbZ$ \\ 
Solv                           & 3    &  solv       &  F      &  100041 &       310  & as in \cite{rtviz} \\ 
\hline
\end{tabular}%
}

\caption{Details on tessellations used in our study (\emph{big} versions); * denotes finer grids. \label{tab:geoms2}}
\end{table}

\begin{figure}[h!]
  \centering
  \begin{subfigure}[b]{0.4\textwidth}
   \centering
   \includegraphics[width=\textwidth]{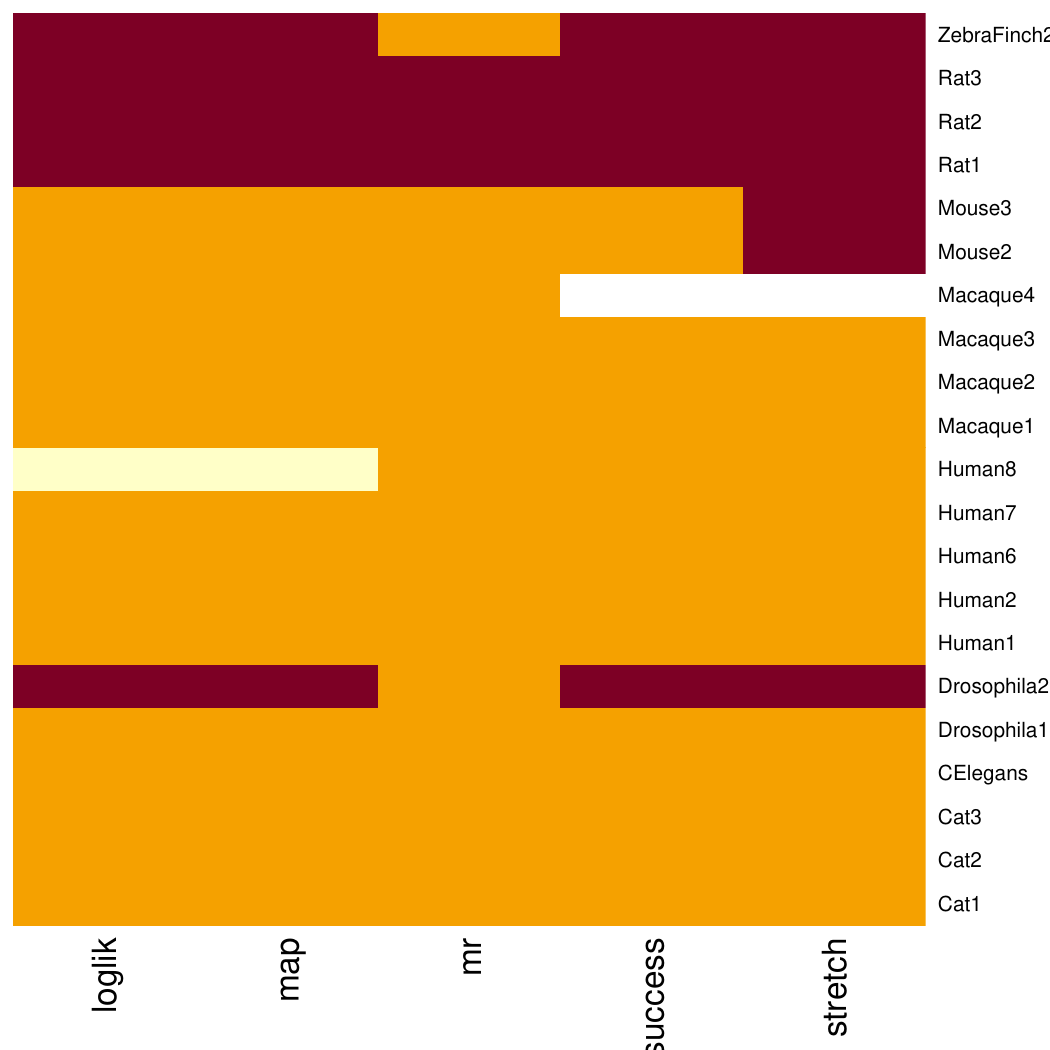}
   \caption{$\bbH^3$}
  \end{subfigure}
  \hfill
  \begin{subfigure}[b]{0.4\textwidth}
   \centering
   \includegraphics[width=\textwidth]{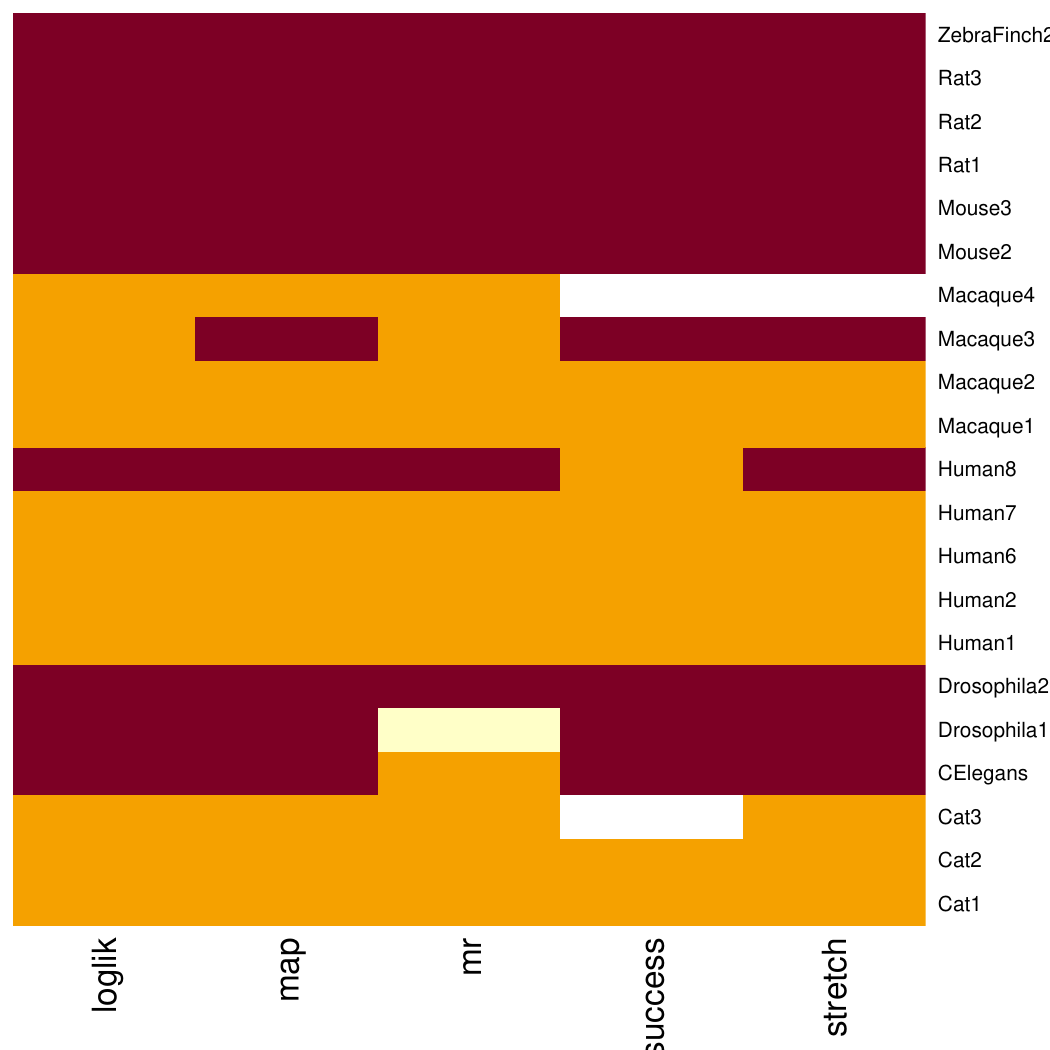}
   \caption{$\bbH^3*$}
  \end{subfigure}
\\
  \begin{subfigure}[b]{0.4\textwidth}
   \centering
   \includegraphics[width=\textwidth]{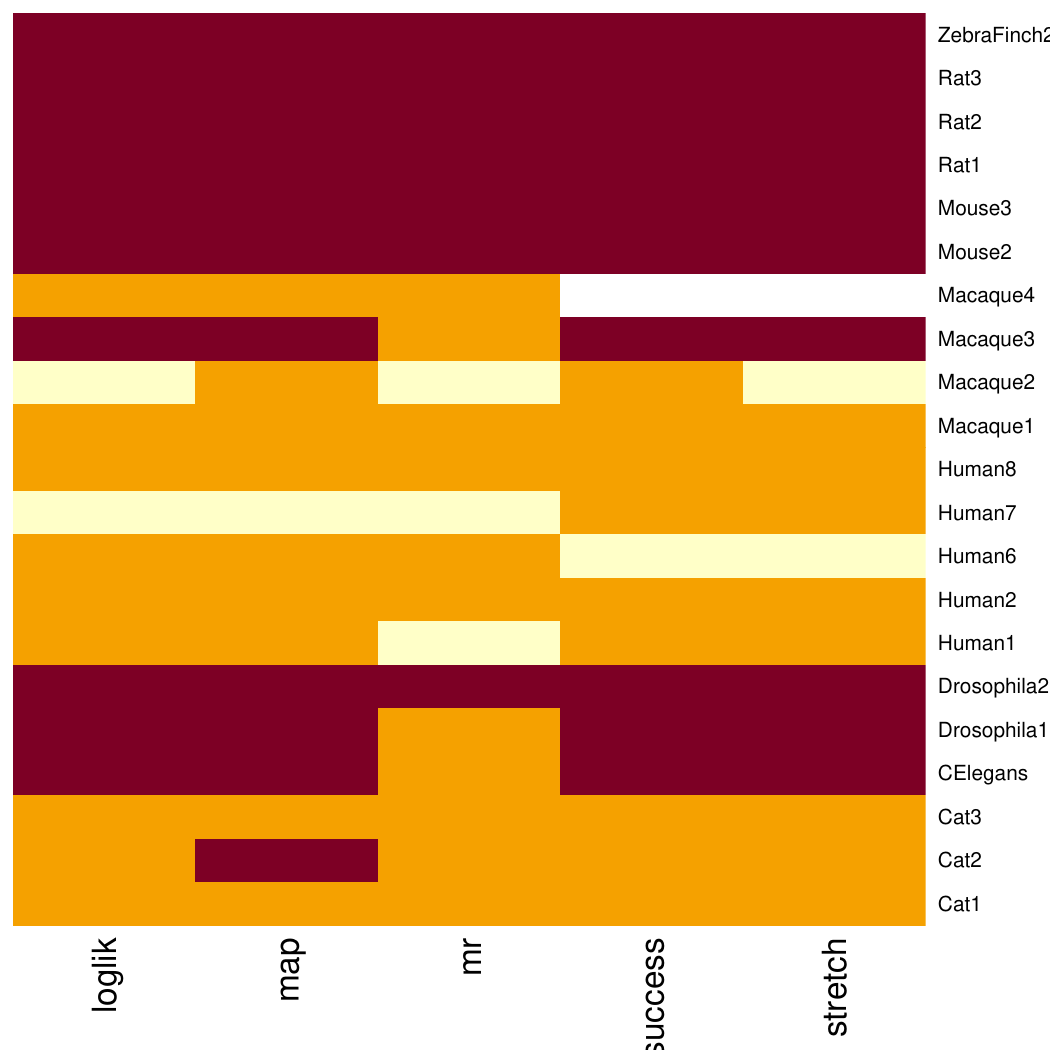}
   \caption{$\bbH^2 \times \bbR$}
  \end{subfigure}
  \hfill
  \begin{subfigure}[b]{0.4\textwidth}
   \centering
   \includegraphics[width=\textwidth]{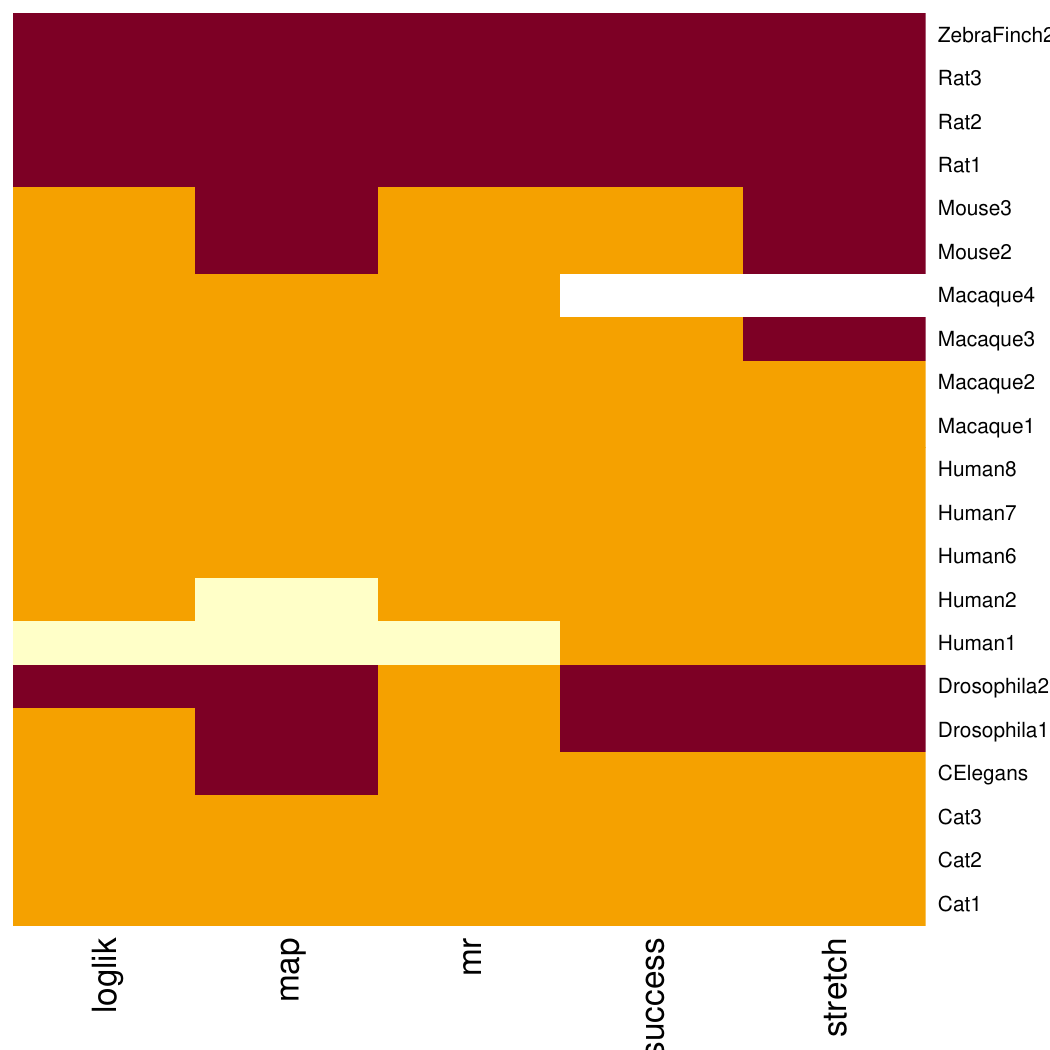}
   \caption{Twisted}
  \end{subfigure}
  \\
  \begin{subfigure}[b]{0.4\textwidth}
   \centering
   \includegraphics[width=\textwidth]{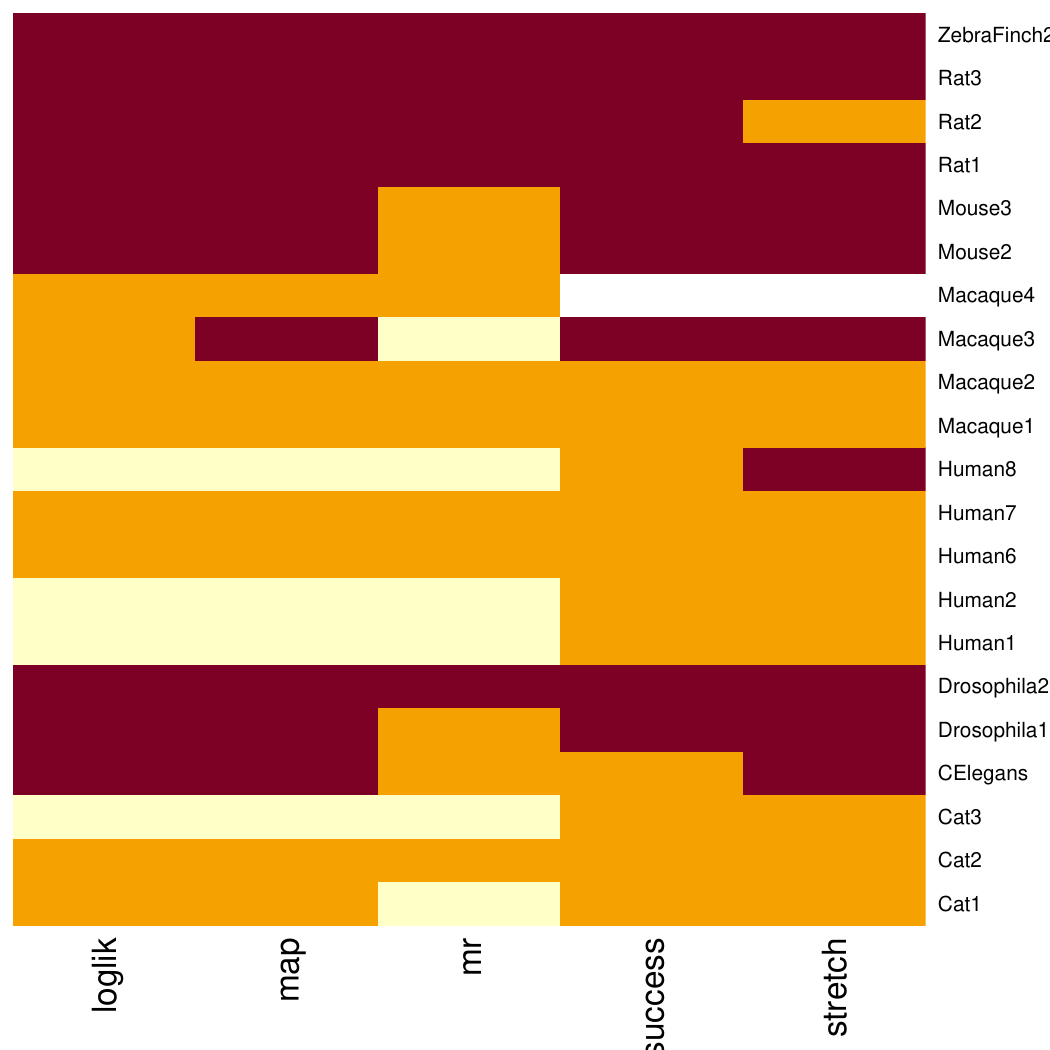}
   \caption{Solv}
  \end{subfigure}
\caption{Comparison of the goodness of fit between regular and \emph{big} versions of manifolds. Red suggests that the \emph{big} version yields better results and the difference is significant; orange suggests lack of significant difference, and yellow suggests significantly worse results for \emph{big} version. \label{fig:bigvsshort}}
\end{figure}

We started by checking for significant differences in favor of \emph{big} versions of manifolds; to this end, we performed Wilcoxon tests with Bonferroni correction for multiple comparisons. Figure \ref{fig:bigvsshort} depicts the procedure results. According to our results, in most cases, the differences are insignificant, which suggests that the size of the manifold is not a severe threat to validity. Usage of \emph{big} versions usually results in better embeddings for Rat connectomes, which might correlate with a different function of those connectomes compared to others in the sample (they describe nervous systems). Rarely, \emph{big} versions yield worse embeddings than the standard ones -- usually for Human connectomes; however, no pattern-enabling explanation is noticeable here.

Next, we checked if the size of the manifolds affects rankings. To this end, we computed weighted Cohen's kappas \cite{kappa}. In kappas, 0 represents the amount of agreement expected from random chance, and 1 signifies a perfect agreement between the raters. Initially, kappas take into account only the agreements of the raters. The weighted kappas allow disagreements to be weighted differently, which is more suitable for us -- we are more interested in the relative placement of the pairs of the geometries in the ranking than in the actual places. If there are slight differences in ranks by two raters, e.g., by one, the ranks should remain similar to us as embeddings yielding comparable quality results should still be close to each other. Although there are no universal guidelines for interpreting of those coefficients, the literature suggests that the values over 0.61 indicate moderate to substantial agreement between raters and values exceeding 0.81 -- strong to almost perfect agreement \cite{kappa_int}. 

\begin{table}[h!]
  \resizebox{\columnwidth}{!}{%
  \begin{tabular}{|l|c|c|c|c|c|c|c|c|c|r|} \hline
    Pair of rankings & \multicolumn{5}{|c}{Max performance} & \multicolumn{5}{|c|}{Copeland} \\ 
    & NLL & MAP & IMR & SC & IST & NLL & MAP & IMR & SC & IST \\ \hline
Standard$_{SA:10,000}$ vs Big$_{SA:10,000}$ & 0.80       & 0.75        & 0.84       & 0.57       &   0.61        & 0.86        & 0.81        & 0.91        & 0.71        &  0.78 \\
& (0.72;0.88) & (0.66;0.84) & (0.77;0.90) & (0.40;0.73) & (0.47;0.74) & (0.79;0.93) & (0.74;0.88) & (0.87;0.95) & (0.61;0.82) & (0.69;0.88)
\\ \hline
Standard$_{SA:100,000}$ vs Big$_{SA:100,000}$ & 0.78       & 0.75        & 0.85        & 0.52        & 0.65        & 0.85        & 0.83        & 0.84        & 0.75      & 0.82 \\
                                   & (0.69;0.88) & (0.65;0.85) & (0.79;0.91) & (0.34;0.70) & (0.53;0.77) & (0.77;0.92) & (0.76;0.89) & (0.77;0.91) & (0.66;0.84) & (0.73;0.90) \\ \hline
Standard$_{SA:10,000}$ vs Standard$_{SA:100,000}$ & 0.95 & 0.93 &        0.93        & 0.86        & 0.87        & 0.84        & 0.82        & 0.83        & 0.77        & 0.83 \\
& (0.92;0.97) & (0.90;0.96) & (0.90;0.96) & (0.80;0.92) & (0.81;0.92) & (0.77;0.90) & (0.74;0.89) & (0.76;0.91) & (0.66;0.88) & (0.75;0.91) \\ \hline
Big$_{SA:10,000}$ vs Big$_{SA:100,000}$ & 0.76   & 0.70       &  0.78       & 0.73       & 0.60  &        0.79        & 0.75        & 0.86        & 0.69        & 0.76 \\
& (0.66;0.86) & (0.59;0.82) & (0.69;0.87) & (0.59;0.86) & (0.44;0.75) & (0.70;0.88) & (0.66;0.84) & (0.80;0.91) & (0.55;0.83) & (0.66;0.87) \\ \hline
Standard$_{SA:10,000}$ vs Big$_{SA:100,000}$ & 0.74 & 0.74             & 0.79        & 0.51        & 0.68 & 0.76              & 0.76        & 0.82       & 0.58         & 0.79 \\
                                    & (0.63;0.85) & (0.64;0.83) & (0.70;0.89) & (0.35;0.68) & (0.56;0.8) & (0.67;0.85) & (0.67;0.85) & (0.75;0.90) & (0.44;0.72) & (0.69;0.89) \\ \hline
\end{tabular}}
\caption{The agreements between the rankings obtained for different simulation setups (values of Cohen's kappa). Standard includes: $\bbH^3$, $\bbH^3*$, $\bbH^2 \times \bbR$, Solv, and Twist. 95\% confidence intervals in brackets. \label{tab:kappas}}
\end{table}

According to data in Table \ref{tab:kappas}, rankings obtained from \emph{big} versions of manifolds in the standard setup of Simulated Annealing ($N_s = 10,000$ iterations) are at least in substantial agreement with rankings based on standard versions. The high agreement in rankings based on voting rules is unsurprising. It aligns with the results depicted in Figure \ref{fig:bigvsshort} -- we include more information from the distributions, so the results should be more robust than those based on max performance (outliers). However, we recommend cautiously treating the results for greedy routing success and stretch.  

\paragraph{Possibly insufficient number of iterations.}

\begin{figure}[h!]
  \centering
  \begin{subfigure}[b]{0.30\textwidth}
   \centering
   \includegraphics[width=\textwidth]{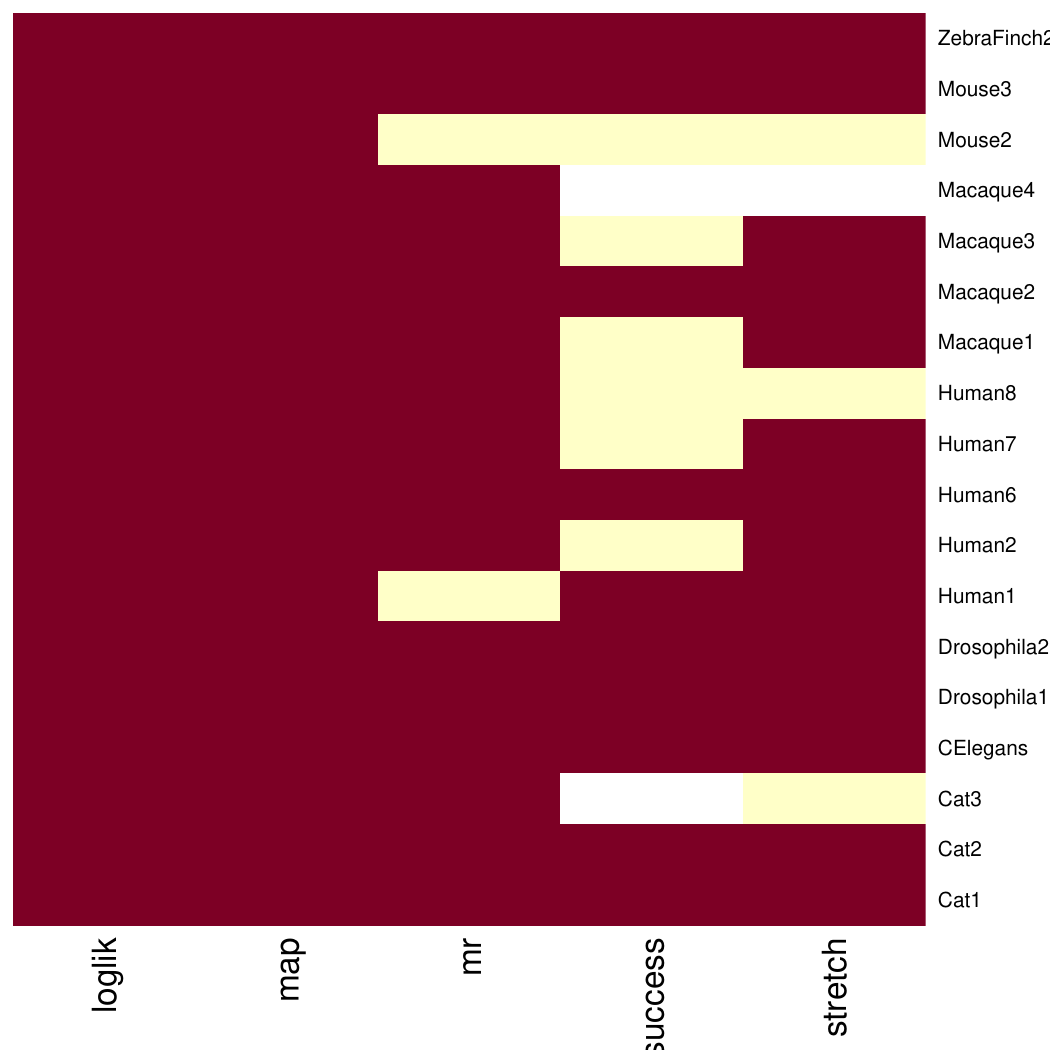}
   \caption{$\bbH^3$}
  \end{subfigure}
  \hfill
  \begin{subfigure}[b]{0.30\textwidth}
   \centering
   \includegraphics[width=\textwidth]{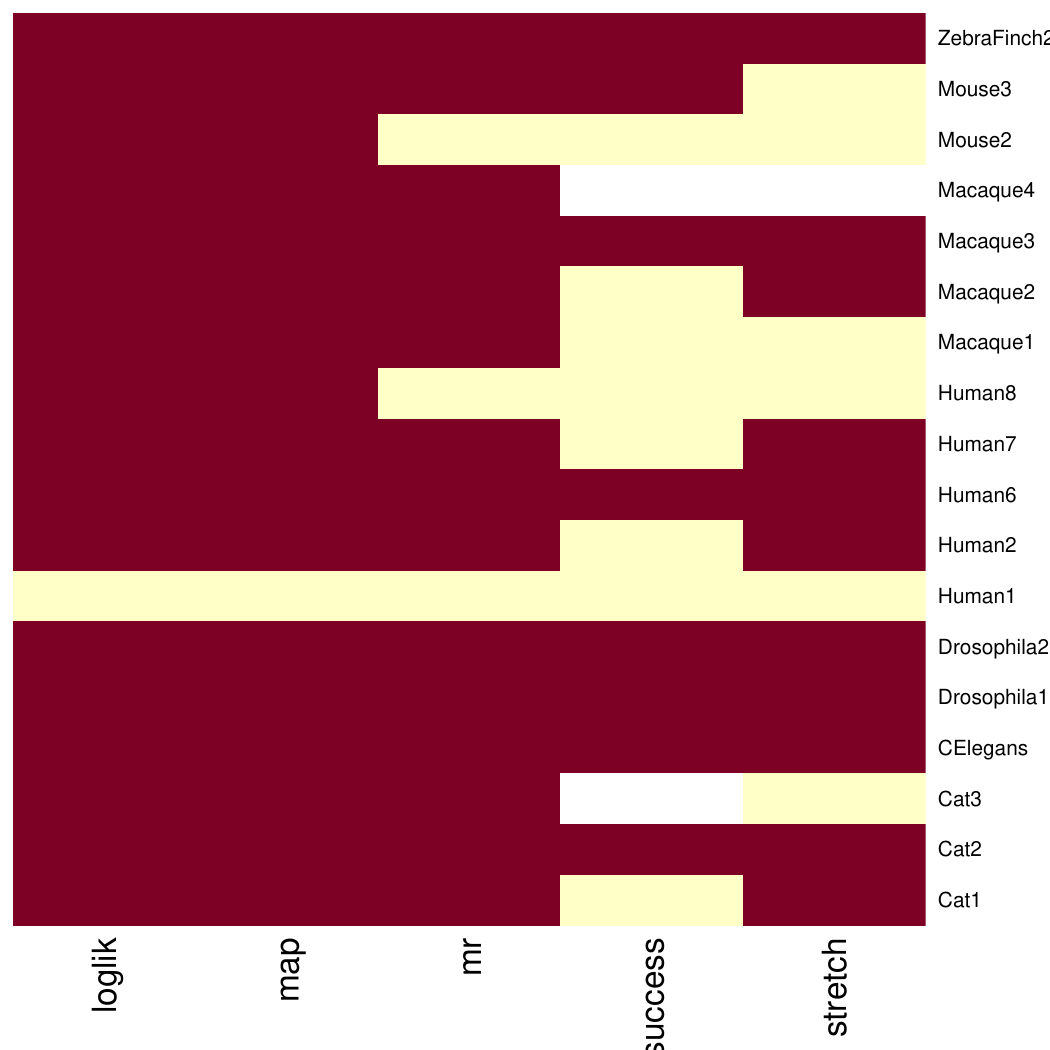}
   \caption{$\bbH^3*$}
  \end{subfigure}
    \hfill
  \begin{subfigure}[b]{0.30\textwidth}
   \centering
   \includegraphics[width=\textwidth]{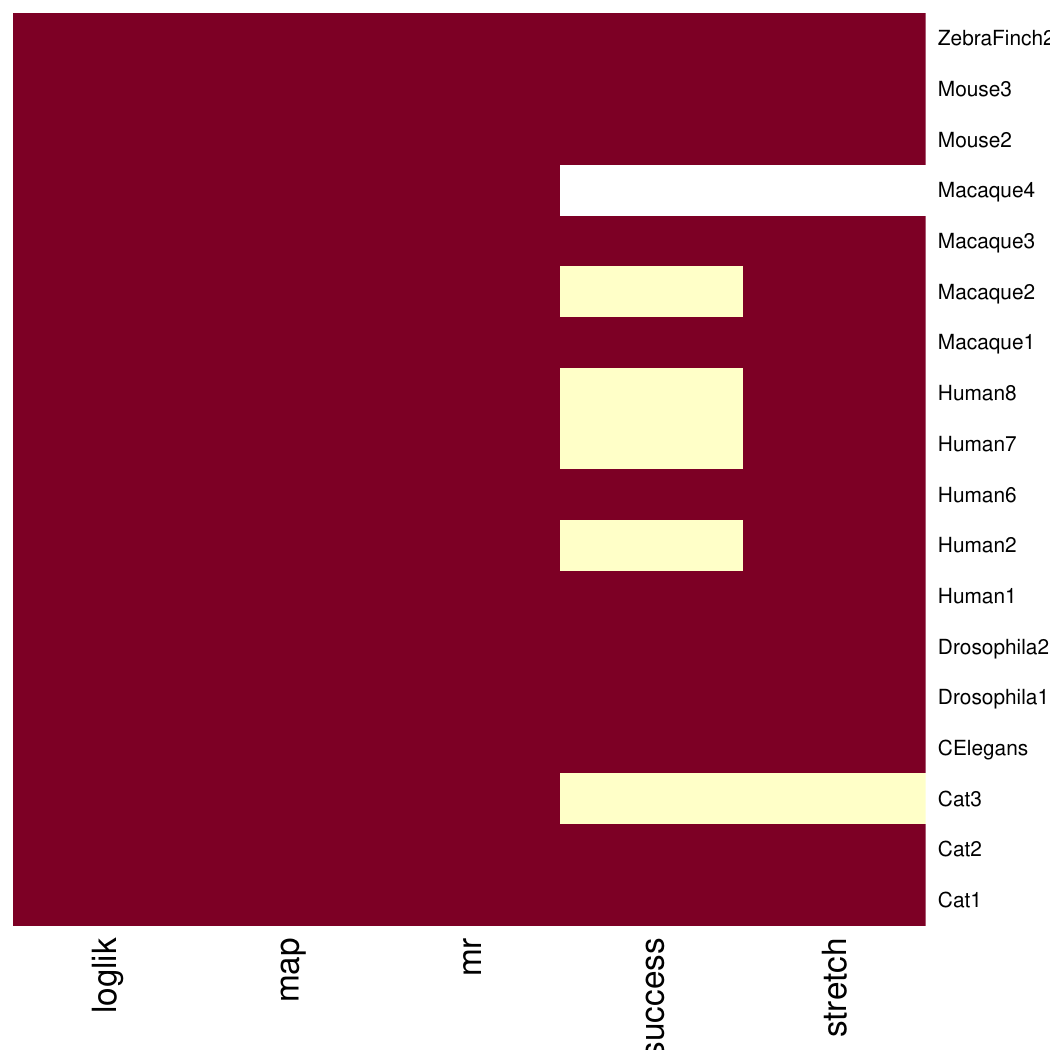}
   \caption{$\bbH^3**$}
  \end{subfigure}
  \\
    \begin{subfigure}[b]{0.30\textwidth}
   \centering
   \includegraphics[width=\textwidth]{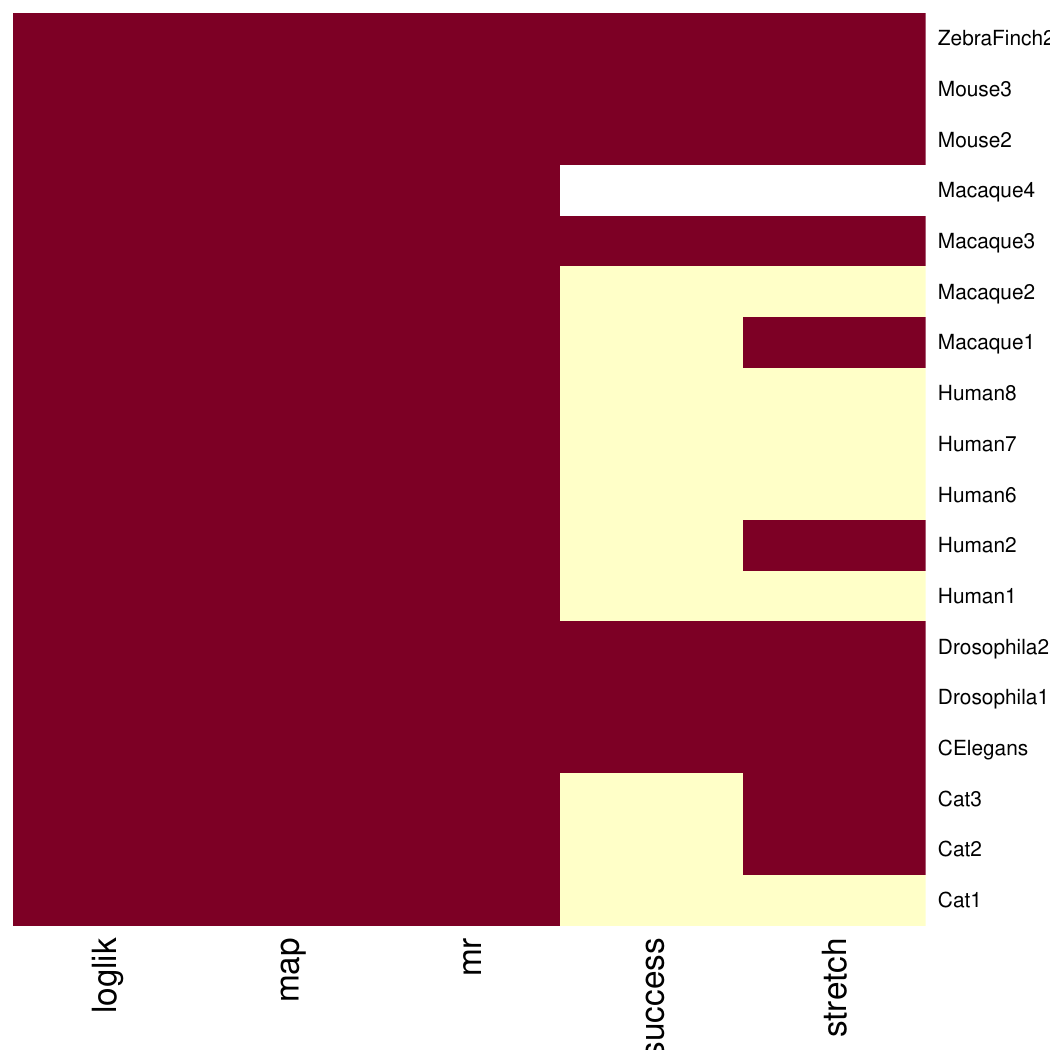}
   \caption{Solv}
  \end{subfigure}
  \hfill
  \begin{subfigure}[b]{0.30\textwidth}
   \centering
   \includegraphics[width=\textwidth]{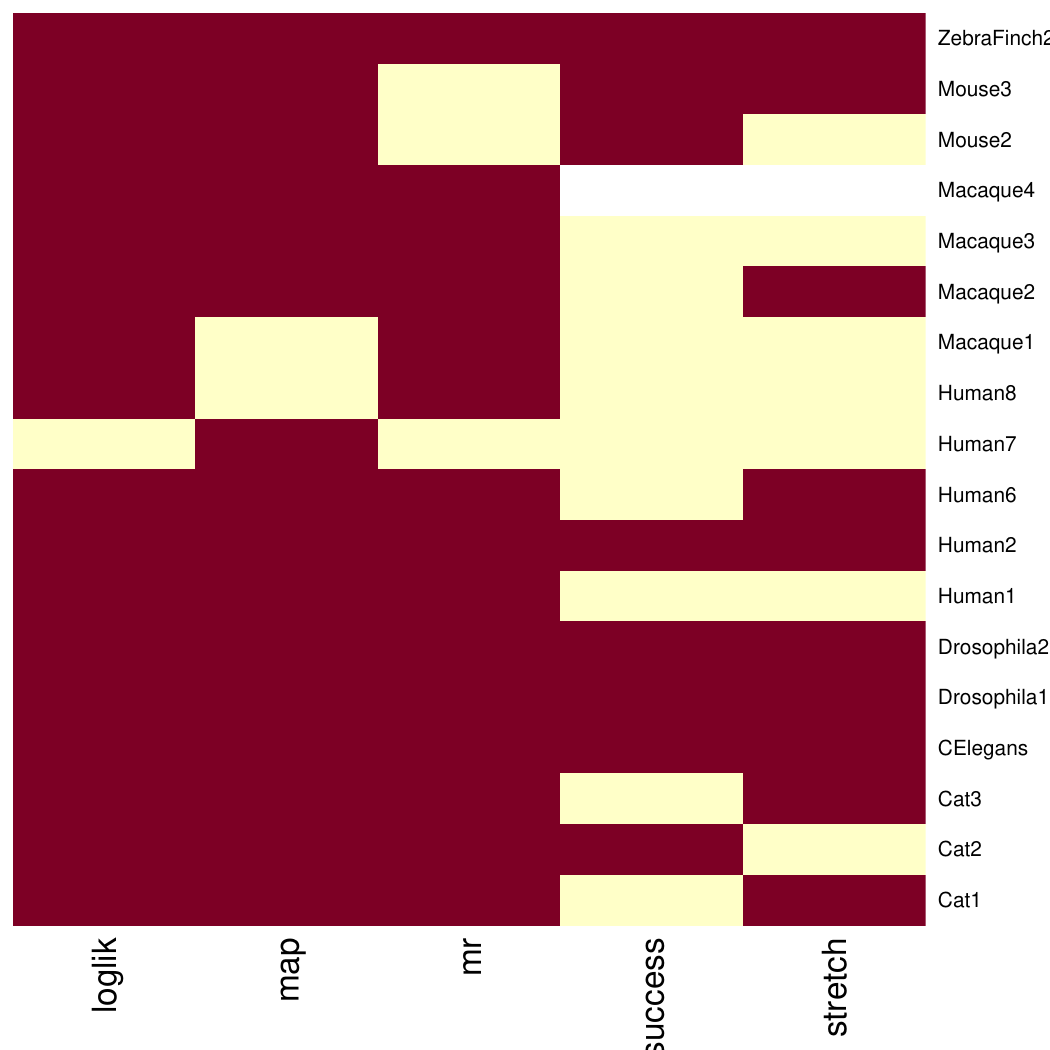}
   \caption{Solv*}
  \end{subfigure}
    \hfill
  \begin{subfigure}[b]{0.30\textwidth}
   \centering
   \includegraphics[width=\textwidth]{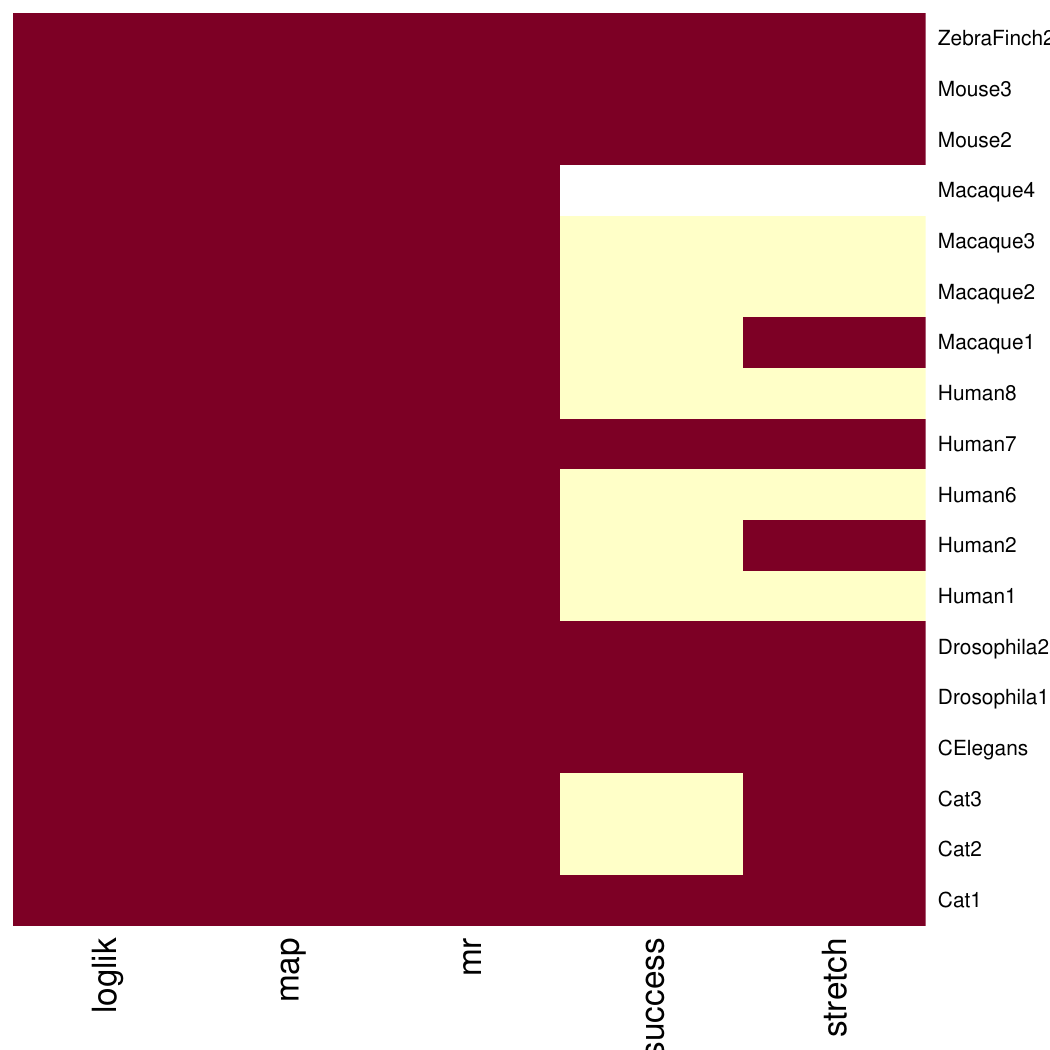}
   \caption{Solv**}
  \end{subfigure}
  \\
    \begin{subfigure}[b]{0.30\textwidth}
   \centering
   \includegraphics[width=\textwidth]{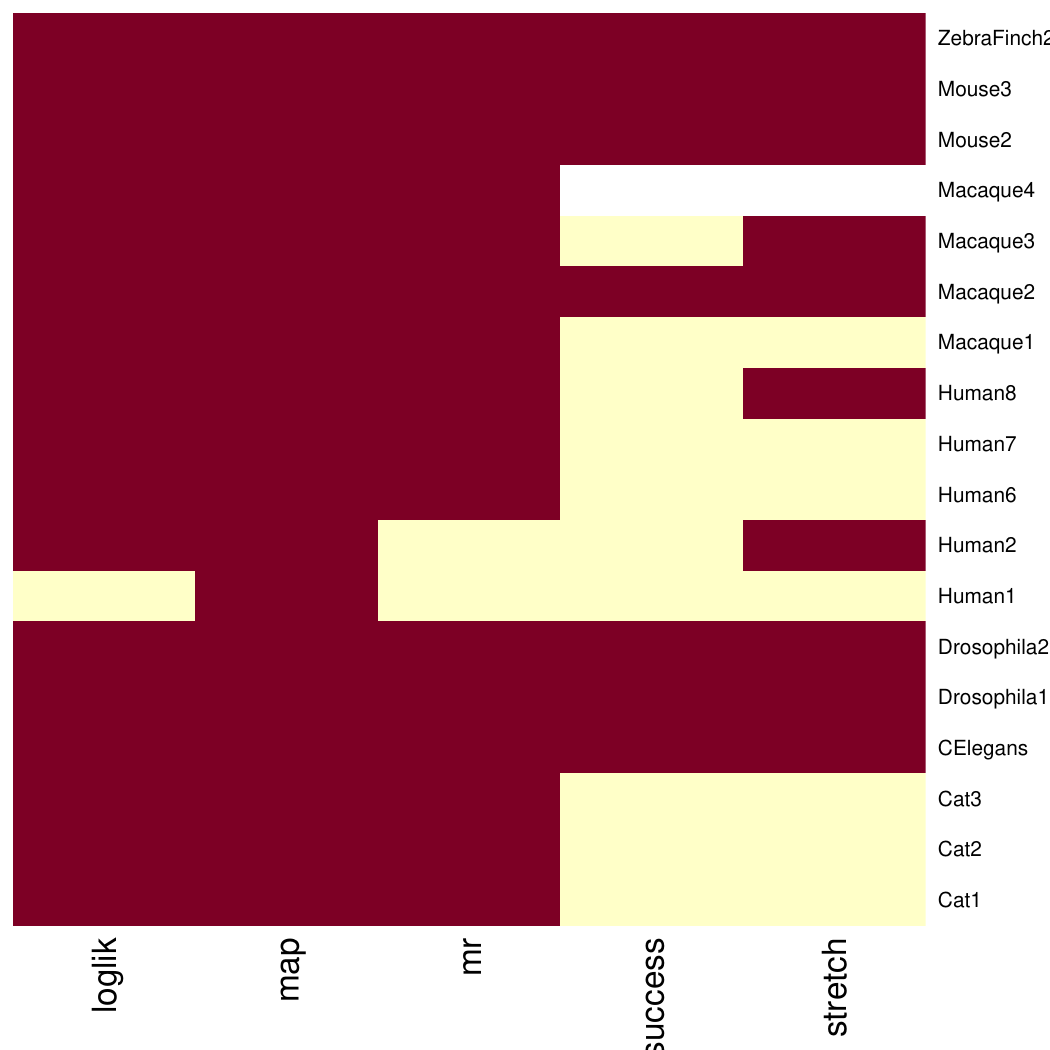}
   \caption{$\bbH^2 \times \bbR **$}
  \end{subfigure}
  \hfill
  \begin{subfigure}[b]{0.30\textwidth}
   \centering
   \includegraphics[width=\textwidth]{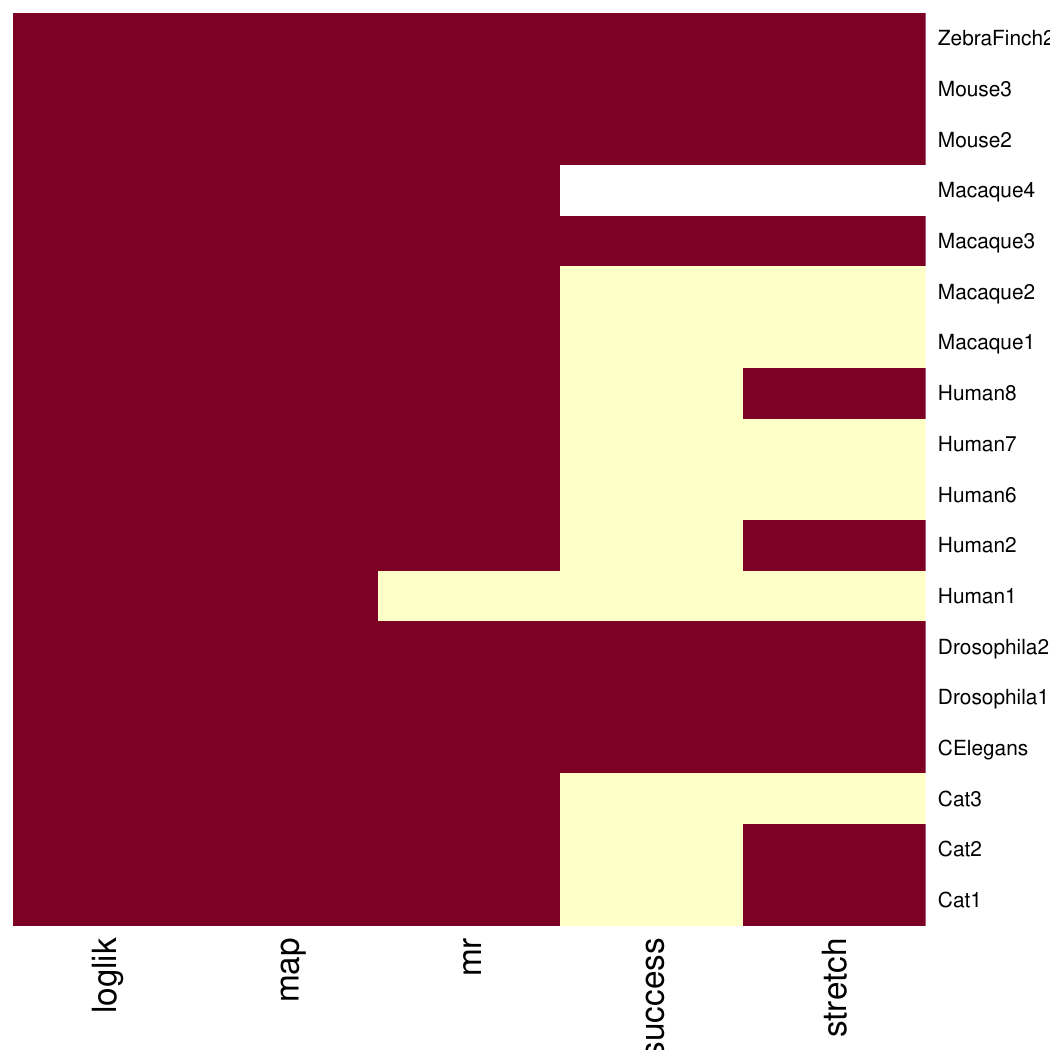}
   \caption{Twist}
  \end{subfigure}
    \hfill
  \begin{subfigure}[b]{0.30\textwidth}
   \centering
   \includegraphics[width=\textwidth]{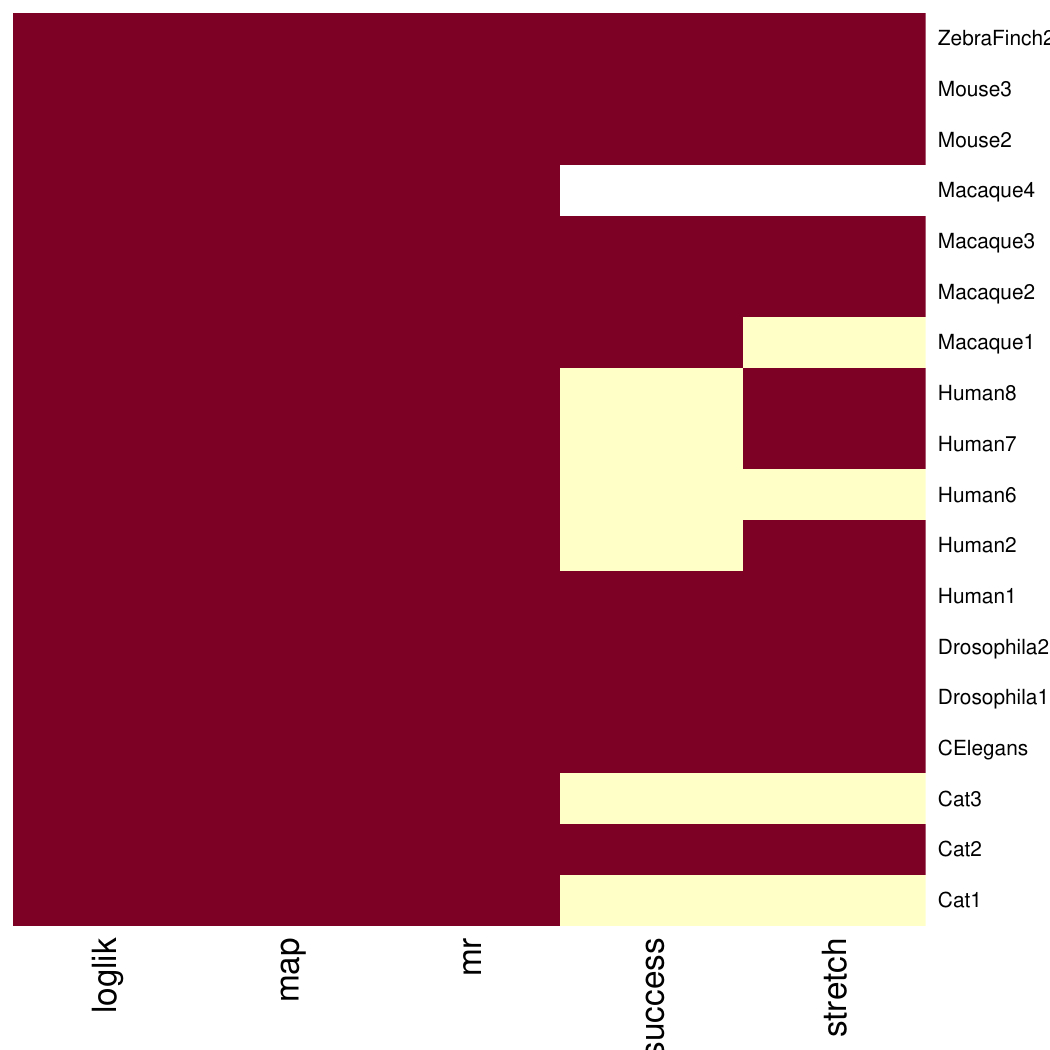}
   \caption{Twist**}
  \end{subfigure}
  \\
    \centering
  \begin{subfigure}[b]{0.31\textwidth}
   \centering
   \includegraphics[width=\textwidth]{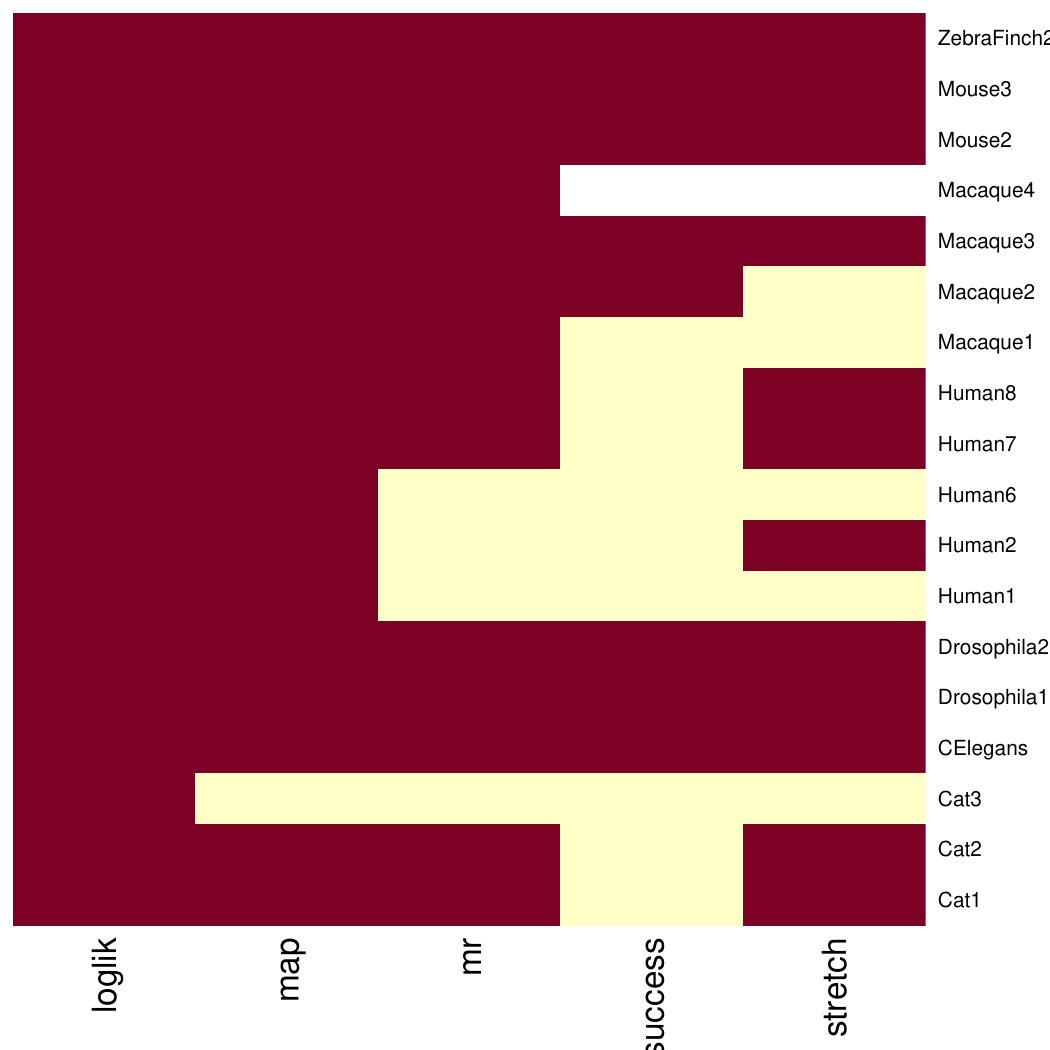}
   \caption{$\bbH^2$}
  \end{subfigure}
  \hfill
  \begin{subfigure}[b]{0.31\textwidth}
   \centering
   \includegraphics[width=\textwidth]{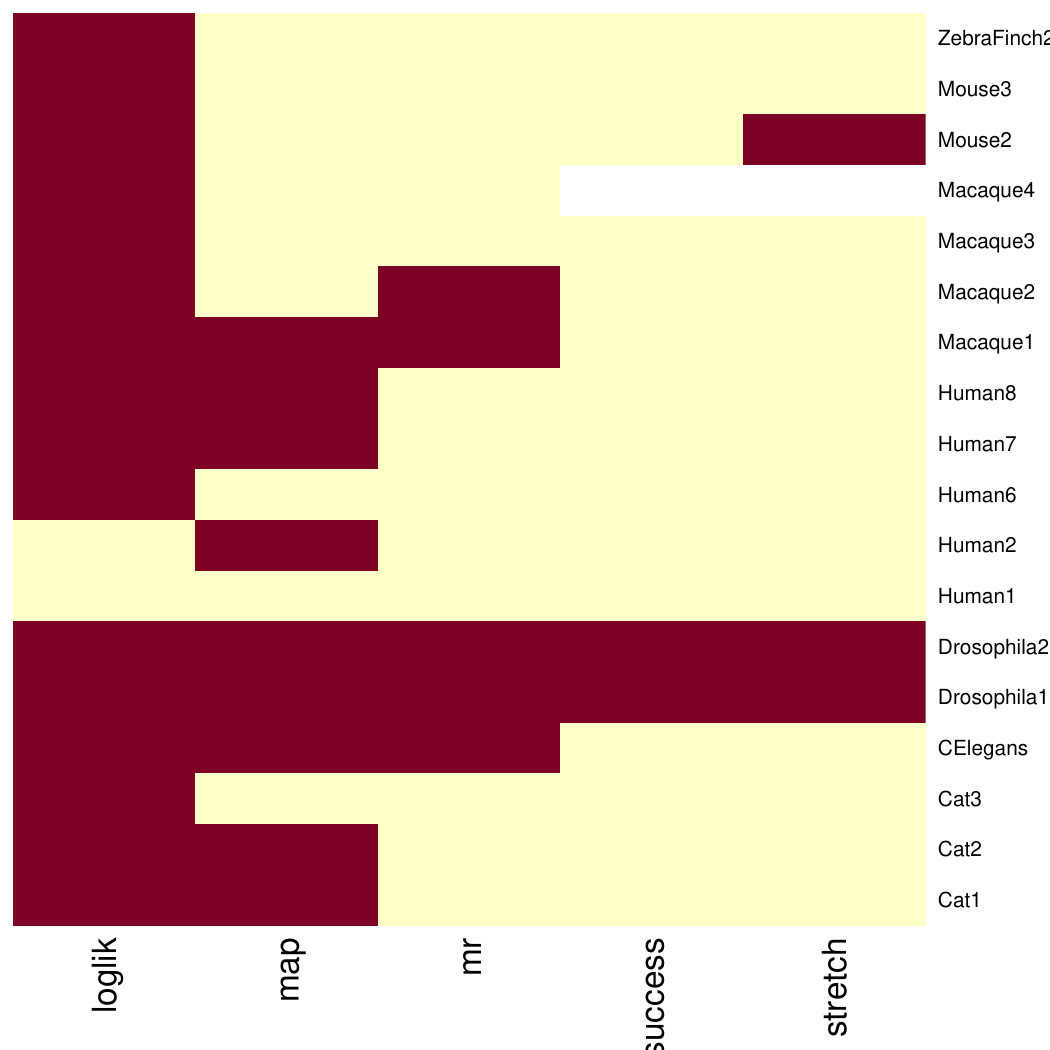}
   \caption{$\bbE^3$}
  \end{subfigure}
    \hfill
\caption{Comparison of the goodness of fit between results of Simulated Annealing with 10.000 vs. 100.000 steps per iteration. Red suggests that the longer version yields better results and the difference is significant; orange suggests lack of significant difference, and yellow suggests significantly worse results for longer version. \label{fig:longlongsa}}
\end{figure}

As Simulated Annealing is a probabilistic technique for approximating the global optimum of a given function, one could argue that, e.g., increasing the number of iterations could improve our results (the third issue). The main paper describes the results obtained with Simulated Annealing with $N_s = 10,000 \cdot |V|$ iterations per simulation iteration. We also checked if our results differ if we perform Simulated Annealing with $N_s = 100,000 \cdot |V|$ iterations per simulation iteration instead. As expected, for log-likelihood, MAP, and MR, we cannot reject the hypotheses that the results obtained with larger numbers of iterations are usually better. However, surprisingly, for greedy success rate and stretch, the results worsen with the increase in the number of iterations (Figure \ref{fig:longlongsa} depicts the results of Wilcoxon tests with Bonferroni correction for multiple comparisons).

We checked if the number of iterations for Simulated Annealing $N_s$ affected our results regarding rankings with Cohen's kappas. Based on the data in Table \ref{tab:kappas}, we notice that pairs of rankings are in at least substantial agreement. The results regarding standard grids (presented in the main part of the paper) are robust to $N_s$ -- the values of kappas for optimistic scenarios are over 0.85. For rankings based on voting rules, they usually exceed 0.80. Although the agreements of rankings, if we change $N_s$ for \emph{big} grids, are still satisfying (most values of kappas over 0.75), the results from comparison of rankings based on standard grids with shorter time for Simulated Annealing against the \emph{big} grids with longer time for Simulated Annealing suggests that the \emph{big} versions of grids might be affected by $N_s$. Again, we notice that greedy routing success rate and stretch are less immune to the setup of Simulating Annealing, so we suggest caution while generalizing the results obtained for them.

\paragraph{Alternative methods of obtaining $R$ and $T$}
The fourth issue is challenging. As explained in Section \ref{sec:dtes}, the values of $R$ and $T$ have been obtained by dynamically adjusting them during 
the simulated annealing process (A). We have also experimented with other methods: $R$ is changed, but $T$ remains fixed (B),
and both $R$ and $T$ remain fixed. We run 30 iterations using method (A), then 30 iterations using method (B), then 30
iterations using method (C). The fixed values of $R$ and $T$ are based on the best result (by log-likelihood) obtained
in the earlier iterations.

\begin{figure}[h!]
  \centering
  \begin{subfigure}[b]{0.9\textwidth}
   \centering
   \includegraphics[width=\textwidth]{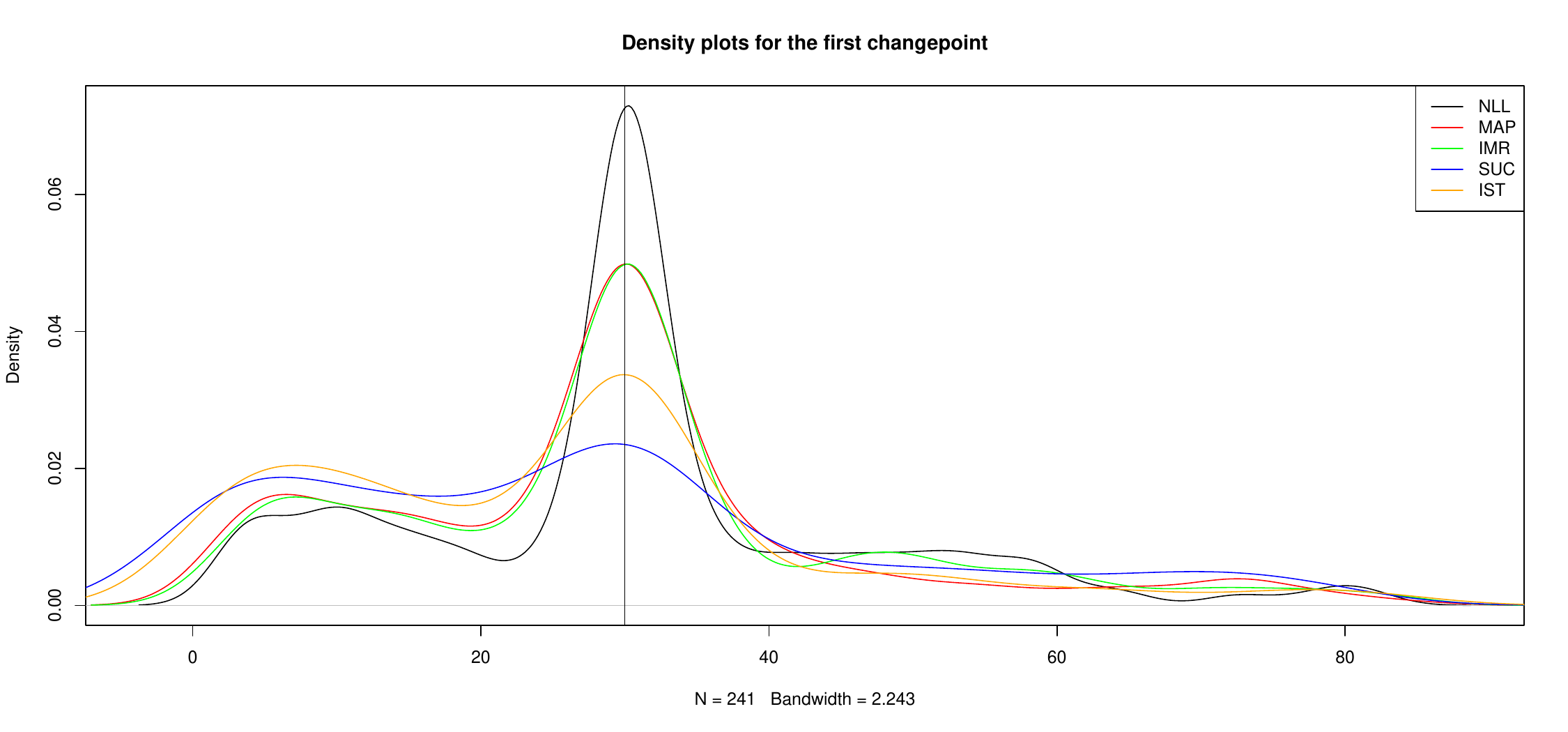}
   \caption{First changepoint}
  \end{subfigure}
  \hfill
  \begin{subfigure}[b]{0.9\textwidth}
   \centering
   \includegraphics[width=\textwidth]{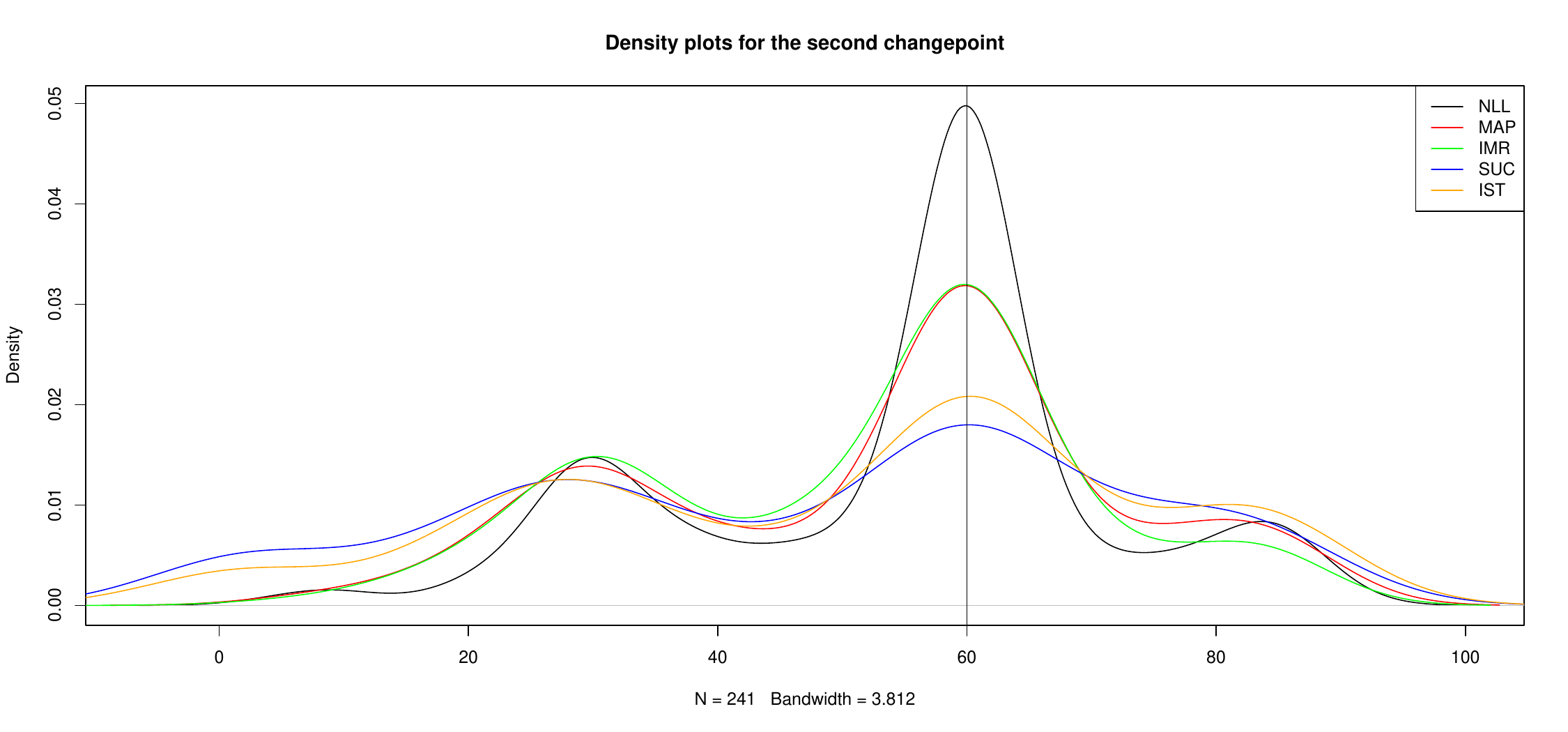}
   \caption{Second changepoint}
  \end{subfigure}
  \caption{Density plots for changepoints in time series of measures (indicators for level shifts) \label{fig:changepoints}}
\end{figure}

If the methods change the results, we should notice level shifts in the time series of the quality measures' values -- level shifts appear as a parallel movement of the trend line. That is why we started by identifying possible locations of the level shifts in our results. Most of the time series (determined by a pair animal and geometry) has two level shifts -- around the 30th and 60th iterations that correspond to the starting points of new methods (Figure \ref{fig:changepoints}).

We use OLS regressions to understand the impact of the change in method on the values of the quality measures. We control for the characteristics of connectomes: number of nodes in the connectome $n$, number of edges in the connectome $m$, its \emph{density}, \emph{assortativity} and \emph{clustering} coefficients, and the \emph{zone} of the connectome; we also take into account the number of available \emph{cells} in the grid and its \emph{geometry}.

\begin{table}[h!]
\def\zero{0.00} 

\resizebox{\columnwidth}{!}{
\begin{tabular}{|l|rr|rr|rr|rr|rr|} \hline
                              & \multicolumn{2}{c|}{$100\cdot$NLL}                 & \multicolumn{2}{c|}{$100\cdot$mAP}                 & \multicolumn{2}{c|}{$100\cdot$IMR}                 & \multicolumn{2}{c|}{$100\cdot$SC}                & \multicolumn{2}{c|}{$100\cdot$ISTR} \\ \hline
                              &  Estimate &$P(>|t|)$ &  Estimate &$P(>|t|)$ &  Estimate & $P(>|t|)$&  Estimate &$P(>|t|)$&  Estimate &$P(>|t|)$ \\ \hline
(Intercept)                   & 4.737e+00 &   \zero & 2.719e+01 &   \zero &-4.323e+01 &   \zero & 8.118e+01 &   \zero & 6.930e+01 & 2.205e-01\\
n                             & 4.876e-03 &   \zero &-1.050e-02 &   \zero &-1.561e-02 &   \zero &-1.857e-02 &   \zero &-1.799e-02 & 1.011e-04\\
m                             &-6.801e-05 &1.65e-11 &-1.553e-04 &   \zero & 7.865e-05 &1.81e-10 & 1.032e-04 &   \zero & 6.741e-05 & 5.579e-06\\
density                       &-2.428e-01 &   \zero & 9.127e-02 &   \zero &-8.497e-01 &   \zero & 1.545e-01 &   \zero & 8.551e-02 & 5.370e-03\\
assort                        & 1.302e+01 &   \zero & 3.379e+01 &   \zero &-1.687e+01 &   \zero & 1.431e+01 &   \zero & 3.645e+00 & 1.991e-01\\
cluster                       & 8.659e+01 &   \zero & 8.163e+01 &   \zero & 1.428e+02 &   \zero & 1.539e+01 &   \zero & 2.998e+01 & 2.854e-01\\

nervous        &-1.061e+00 &5.37e-07 &-6.122e+00 &   \zero &-3.455e+01 &   \zero & 1.285e+00 &   \zero &-2.679e+00 & 1.169e-01\\
other          & 1.564e+00 &   \zero &-3.702e+00 &   \zero &-2.141e+00 &   \zero &-3.542e+00 &   \zero &-3.940e-01 & 9.186e-02\\

cells          & 2.347e+00 &2.94e-15 & 4.730e+00 &   \zero & 3.912e+01 &   \zero & 1.425e+00 &2.36e-15 & 4.654e-01 & 1.642e-01\\

hyperbolic & 8.233e+00 &   \zero & 9.441e+00 &   \zero &-3.735e-01 &0.109924 & 8.125e+00 &   \zero & 7.031e+00 & 1.057e-01\\
other      & 9.250e+00 &   \zero & 9.527e+00 &   \zero & 3.351e-01 &0.202089 & 6.862e+00 &   \zero & 6.473e+00 & 1.189e-01\\
product    & 8.502e+00 &   \zero & 7.357e+00 &   \zero & 7.195e-01 &0.006128 & 6.931e+00 &   \zero & 6.187e+00 & 1.188e-01\\
solv       & 7.399e+00 &   \zero & 6.787e+00 &   \zero & 8.688e-01 &0.000372 & 5.140e+00 &   \zero & 4.747e+00 & 1.104e-01\\
nodes                         & 3.950e-05 &   \zero & 4.645e-05 &   \zero & 1.631e-05 &   \zero & 2.202e-05 &   \zero & 2.445e-05 & 6.950e-07\\

B            &-7.985e-01 &5.76e-12 &-5.009e-01 &4.18e-07 &-2.787e-01 &0.049061 &-4.759e-01 &1.19e-11 &-5.433e-01 & 6.406e-02\\
C            & 5.925e-01 &3.30e-07 & 4.841e-01 &1.02e-06 & 6.684e-01 &2.41e-06 & 3.999e-01 &1.23e-08 & 3.746e-01 & 6.412e-02\\
\hline \hline
$R^2$ (adjusted $R^2$)        & \multicolumn{2}{c|}{0.8061 (0.806)}      & \multicolumn{2}{c|}{0.8948 (0.8948)}     & \multicolumn{2}{c|}{0.8426 (0.8245)}     & \multicolumn{2}{c|}{0.8194 (0.8193)} & \multicolumn{2}{c|}{0.8784 (0.8784)}    \\
p-value for F-test            & \multicolumn{2}{c|}{\zero}           & \multicolumn{2}{c|}{\zero}           & \multicolumn{2}{c|}{\zero}          & \multicolumn{2}{c|}{\zero} & \multicolumn{2}{c|}{\zero}           \\ \hline
\end{tabular}
}
\caption{OLS regression results for the determinants of the quality measures. Number of observations = 24,536.\label{fig:regregres}}
\end{table}

For all the quality measures, we notice that, on average, method B leads to lower values of the respective quality measures, and method C increases the
values of the respective quality measures in comparison to results obtained with method A, ceteris paribus (Table \ref{fig:regregres}). The differences are statistically significant. However, even if we are aware that with the increase in the number of observations, the p-values drop to zero, we work here with the multilevel categorical variables, so we are unable to comment on the size of the effect (available methods based on partial regressions and $R^2$ coefficients would allocate the impact to the constant term). The regressions have substantial explanatory power ($R^2$ coefficients at least 80\%).

To sum up, if one is interested in optimizing the quality measures, we recommend using method B. We know that our choice of method A for Simulated Annealing may affect the final results; in particular, our ``best'' evaluations may not be optimal. However, there are at least two advantages of our approach. First, allowing the algorithm to optimize does not favor any of the tessellations -- all of them have the same chances to find an optimal solution. This way, we ensure the comparability of our results. We have observed that the first iteration may yield worse results due to the poor initially guessed values of $R$ and $T$; however, in further experiments, while the initial values of $R$ and $T$ change, it only affects the results a little. Eventually, we get independently distributed data, i.e., there is no serial correlation (all p-values in the Ljung-Box test smaller than $10^{-5}$), which makes statistical analysis of the results significantly easier.

\paragraph{Alternative methods of computing distances.}
By a distance between two points $a,b\in D$, we mean the length of the shortest geodesic
between $a$ and $b$. Another option is the graph distance, where two points in $D$ are connected when they correspond to adjacent tessellation
tiles. In the case of the product geometry $\bbH^2\times\bbR$, we could compute the angular distance, which is, intuitively, how small an object at $b$ appears to an observer placed at $a$, assuming that the light travels along geodesics. 
To pick the method of measuring the distances in our experiment, we started with the preliminary list of tessellations shown in Table \ref{tab:pregeoms}.

\begin{table}[h!]
  \centering
  \resizebox{\columnwidth}{!}{%
\begin{tabular}{|l|l|l|l|l|l|l|l|l|l|}
  \hline
name                           &  dim &  geometry   &  closed &  nodes &  diameter   & description of the set $D$ \\ \hline 
$\bbH^2$ (d)                   & 2    &  hyperbolic &  F      & 27000  & 560         & bitruncated $\{7,3\}$ (Figure \ref{fig:tess}a) \\ 
$\bbH^2$ (c)                   & 2    &  hyperbolic &  F      & 27007  & 316         & bitruncated $\{7,3\}$ (Figure \ref{fig:tess}a) \\ 
tree (c)                       & 2    &  tree       &  F      & 20002  & 396         & $\{3,\infty\}$ (Figure \ref{fig:tess}b) \\ 
tree (d)                       & 2    &  tree       &  F      & 24574  & 520         & binary tree \\ 
$\bbH^3$ (c)                   & 3    &  hyperbolic &  F      & 40979  & 214         & $\{4,3,5\}$ hyperbolic honeycomb \\ 
$\bbH^3$ (d)                   & 3    &  hyperbolic &  F      & 41511  & 280         & $\{4,3,5\}$ hyperbolic honeycomb \\ 
$\bbH^2 \times \bbR$ (c)       & 3    &  product    &  F      & 20049  & 222         & bitruncated $\{7,3\})$ times $\bbZ$ \\ 
$\bbH^2 \times \bbR$ (a)       & 3    &  product    &  F      & 20022  & 5637        & bitruncated $\{7,3\})$ times $\bbZ$ \\ 
\hline
\end{tabular}%
}
\caption{Details on the preliminary tessellations used in our study.
\label{tab:pregeoms}}
\end{table}

\begin{figure}[h!]
  \centering
  \begin{subfigure}[b]{0.49\textwidth}
   \centering
   \includegraphics[width=\textwidth]{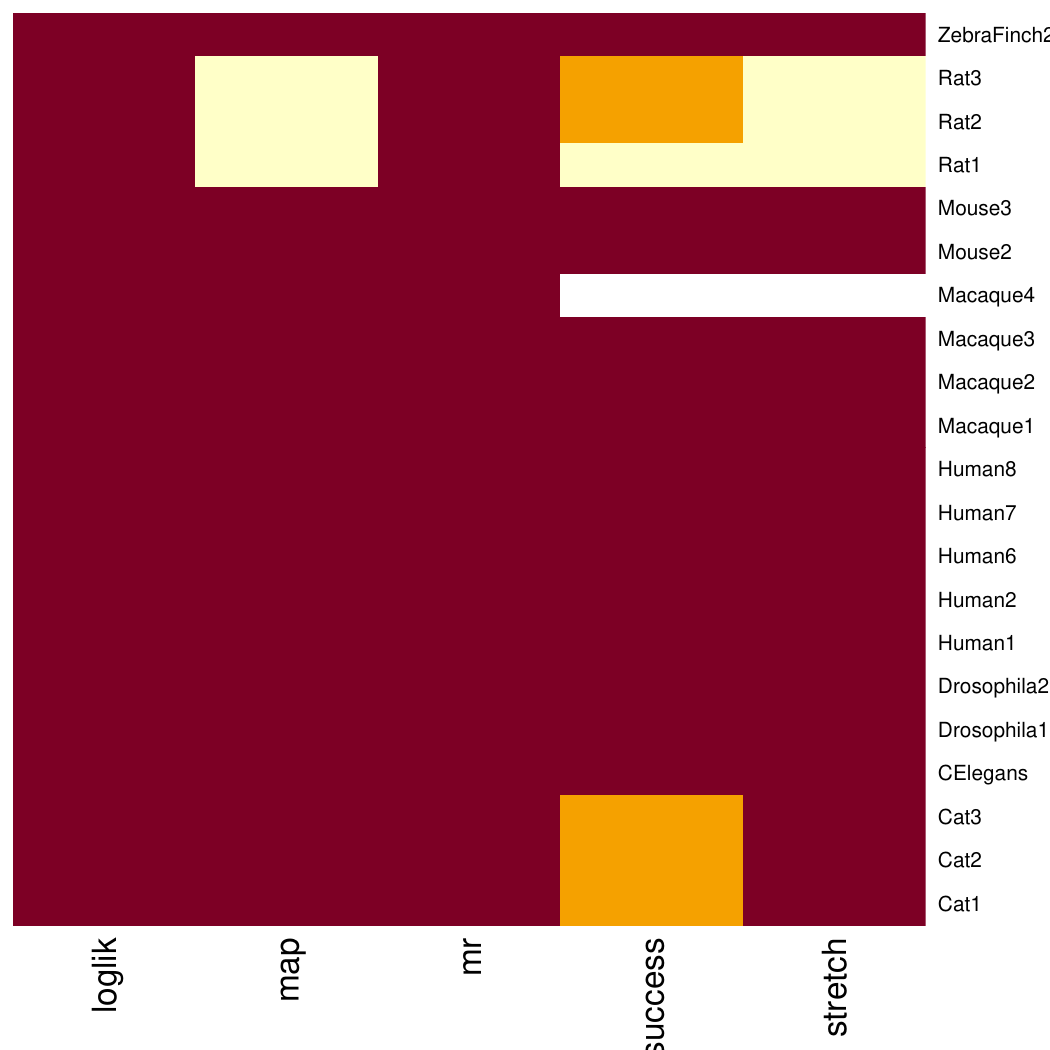}
   \caption{$\bbH^3$}
  \end{subfigure}
  \hfill
  \begin{subfigure}[b]{0.49\textwidth}
   \centering
   \includegraphics[width=\textwidth]{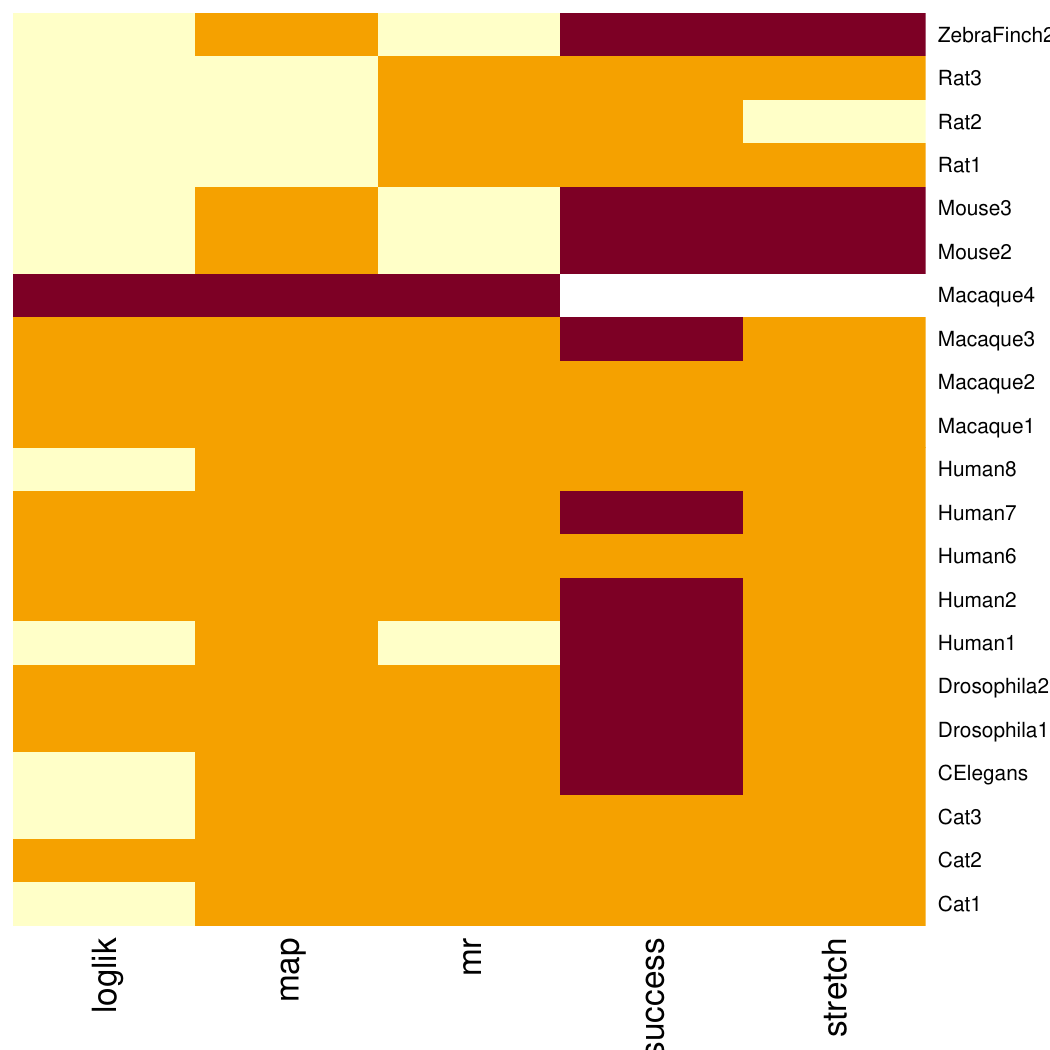}
   \caption{$\bbH^2$}
  \end{subfigure}
  \\
  \begin{subfigure}[b]{0.49\textwidth}
   \centering
   \includegraphics[width=\textwidth]{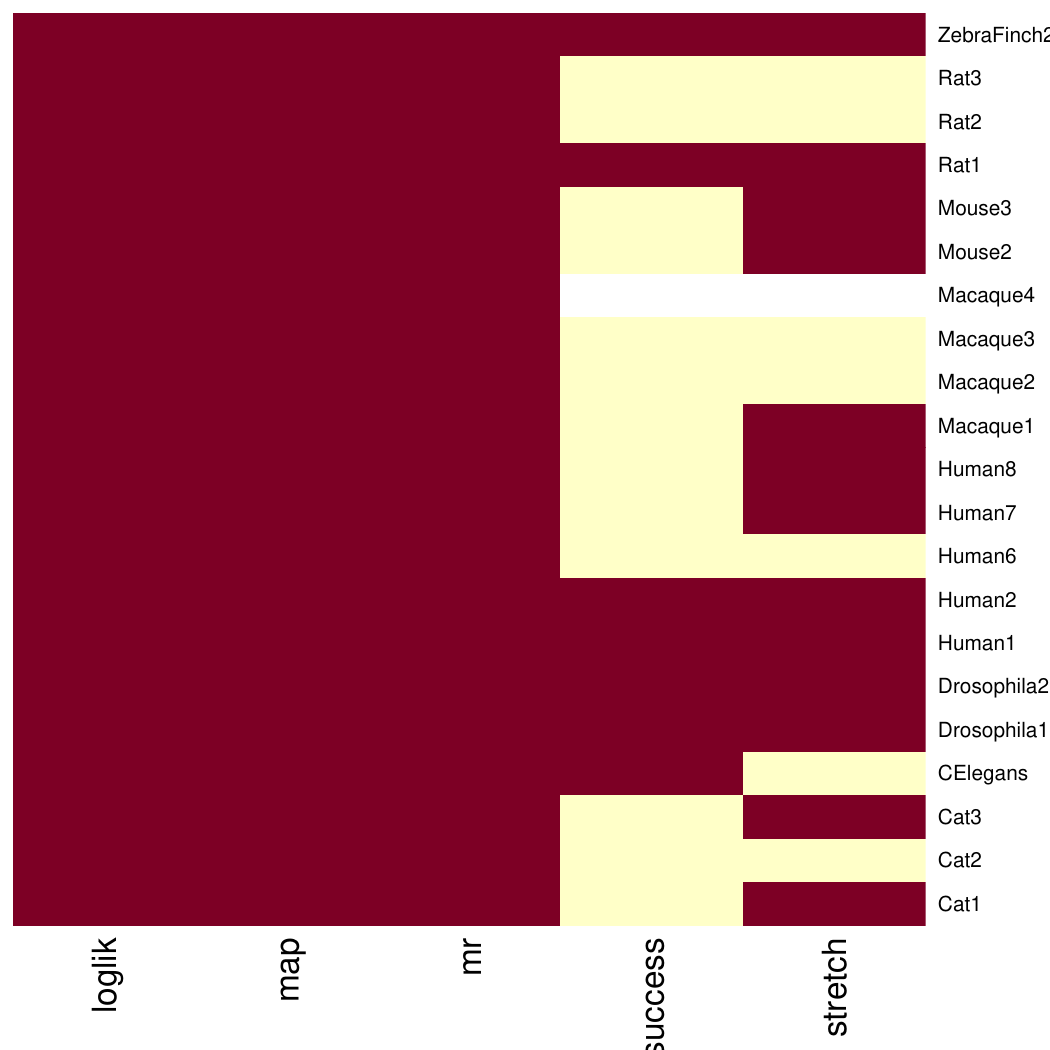}
    \caption{Tree}
  \end{subfigure}
  \hfill
  \begin{subfigure}[b]{0.49\textwidth}
   \centering
   \includegraphics[width=\textwidth]{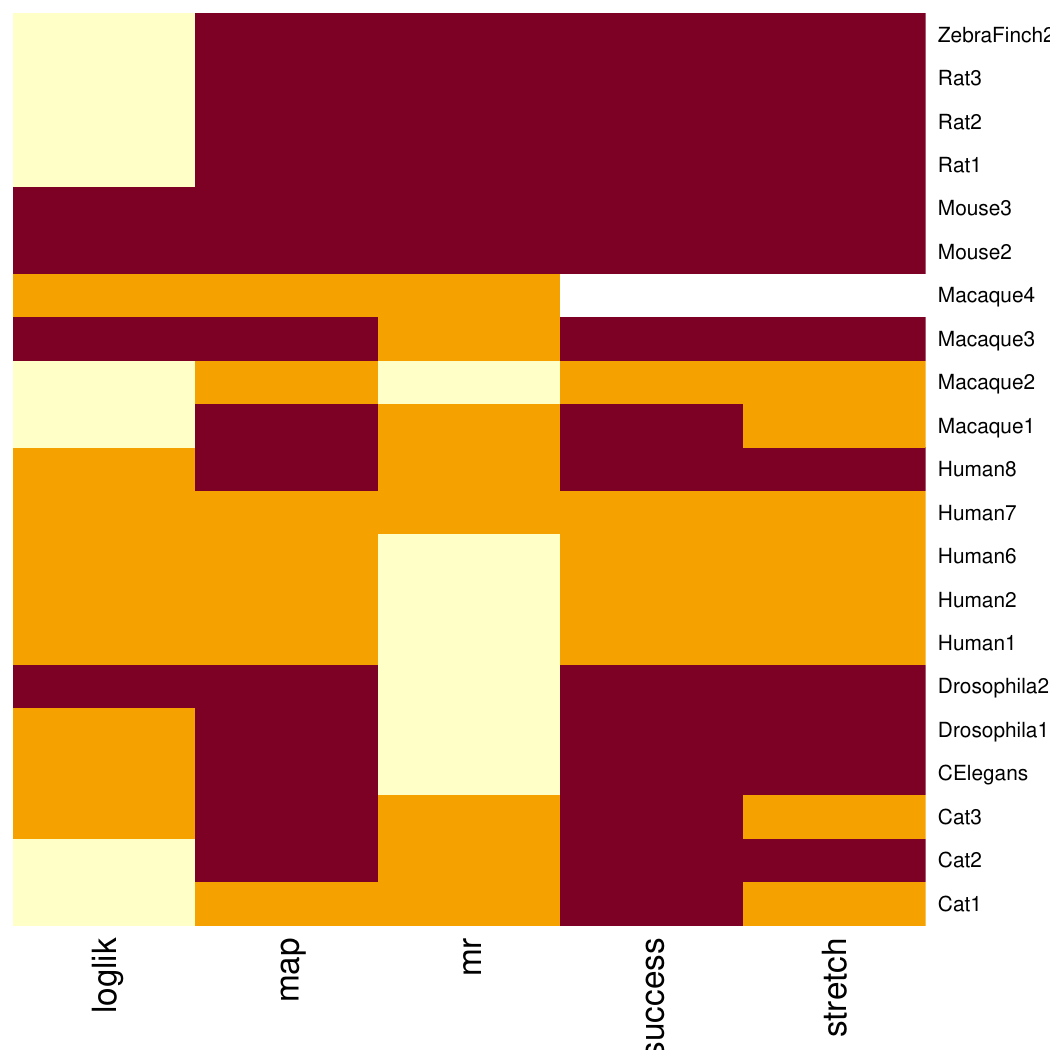}
   \caption{$\bbH^2 \times \bbR$}
  \end{subfigure}
\caption{Comparison of the goodness of fit between pairs of tessellations. Red suggests that the continuous (non-angular) version yields better results and the difference is significant; orange suggests lack of significant difference, and yellow suggests significantly worse results for continuous (non-angular) version, respectively.  \label{fig:comparisons}}
\end{figure}

In the tessellations marked with (d), distances are computed as the lengths of the shortest paths in the graph $(D,E_D)$ where two points in $D$ are connected when they correspond to adjacent tessellation tiles. In contrast, in the tessellations marked with (c), distances are computed according to the underlying geometry. The sets $D$ used in each pair are roughly the same size
(we have less control over $|D|$ in discrete tessellations). In the case of the product geometry $\bbH^2\times\bbR$, we compare geometric distances (c) to angular distance (a). The angular distance
$d_a(X,Y)$ is, intuitively, how small an object at $Y$ appears to an observer placed at $X$, assuming that the light travels along geodesics. The angular size of an object in distance $d$
is proportional to $1/d$ in the Euclidean case and $1/\exp(d)$ in the hyperbolic case; for anisotropic geometries, it may depend on the axis. More precisely, 
$d_a(X,Y)$ is proportional to $\lim_{r \ra 0} r^2/p(X,Y,r)$, where $p(X,Y,r)$ is the probability that a random geodesic starting in $X$ passes within a distance of at least $r$ from $Y$.
For technical reasons, distances are rounded to the nearest integer multiple of 1/20 absolute unit (for continuous distances) and multiplied by 20 (for discrete distances),
except for the sphere, where the unit is 1/200 of absolute unit. Thus, a diameter of 316 for a continuous tessellation is 15.8 absolute units; the diameter of 560
for a discrete tessellation is 28 steps, and the sphere has a diameter (i.e., half the circumference) $\pi$.

It was not certain if we benefit from those technical subtleties. As the data is not normally distributed
and the sample sizes are small (30 observations), we perfomed Wilcoxon tests (with Bonferroni correction for multiple comparisons).
Figure~\ref{fig:comparisons} visualizes the results of the procedure.
We notice that we generally do not benefit from discrete versions of hyperbolic tessellations, which is why we decided to exclude them from further analysis.
In the case of trees, we notice that the discrete version yields significantly better results for greedy success rates, so we keep that tessellation. 
Finally, we excluded the angular version of product geometry $\bbH^2 \times \bbR$ -- we did not notice systematic gains compared to the non-angular version.

\paragraph{Our results vs previous approaches}

We compare the performance of our embedder against
the previous embedders
on the CElegans, Cat2, Drosophila1, Human1, Human6, Macaque3, and Mouse3 connectomes.
For $\bbH^2$,
we have compared against the BFKL embedder \cite{tobias},
Mercator \cite{mercatorembedding} (fast and full version), 
2D Poincar\'e embeddings \cite{nickel} and 2D Lorentz embeddings \cite{nickel2}.
We ran each competing algorithm five times, found the best result
of these 25 runs, and compared them to our results. We have also performed a similar analysis for $\bbH^3$*
against 3D Poincar\'e and Mercator \cite{dmercator} embeddings.
Tables \ref{sota1}, \ref{sota2}, \ref{sota3} and \ref{sota4}
list our results for four measures (we omit log-likelihood since
not all embedders are based on MLE).
We use the benchmark from \cite{anonymous2024bridging}.

In most cases, our result turned out to give better results 
on mAP measures in all 30
runs, and in nearly all cases, we have received better results in
most of the runs. We have not managed to beat the state of the art
for Rat and Drosophila2, probably due to the large 
radius required for good embeddings of these connectomes.
The results are also very good on MR,
although we have not managed to beat Poincar\'e 2D embeddings
on CElegans and Lorentz 2D embeddings on Human1.
Mercator embeddings often turn out to be better on SC and IST measures.
Furthermore, our embeddings use a smaller radius
(7.7 for $\bbH^2$, 3.7 for $\bbH^3$), and use less time than Lorentz or Poincar\'e embeddings
(about 220 seconds per run on Mouse3 in $\bbH^3$).
A smaller radius means that our embeddings avoid
numerical precision issues that tend to be a serious issue in
hyperbolic embeddings \cite{tobias_alenex,reptradeoff,dhrgex},
are better able to use both the large-scale (tree-like)
and smaller-scale (Euclidean-like) nature of 
hyperbolic geometry (while large radius embeddings tend to be tree-like)
and make them more applicable for visualization.
(In large-radius visualizations, fewer nodes are visible.)

\begin{table}
\centering
\resizebox{0.75\columnwidth}{!}{%
\begin{tabular}{|lc|rlr|rrr|} \hline
connectome      &dim&  mAP & method         & rad  & time & ours  & better \\ \hline
CElegans        & 2 & 0.500 & Poincar\'e     &  7.2 &  278 & 0.540 & 30 \\ 
CElegans        & 3 & 0.583 & Poincar\'e     & 10.1 &  274 & 0.584 & 21 \\ 
Cat1            & 2 & 0.888 & Mercator (full)& 10.0 &    0 & 0.877 & 10 \\ 
Cat1            & 3 & 0.880 & Lorentz        &  9.2 &  104 & 0.934 & 30 \\ 
Cat2            & 2 & 0.804 & Mercator (full)& 12.2 &    0 & 0.825 & 29 \\ 
Cat2            & 3 & 0.831 & Poincar\'e     & 10.7 &  277 & 0.861 & 30 \\ 
Cat3            & 2 & 0.911 & Poincar\'e     &  9.7 &  291 & 0.916 & 23 \\ 
Cat3            & 3 & 0.911 & Lorentz        &  9.2 &   86 & 0.952 & 30 \\ 
Drosophila1     & 2 & 0.435 & Mercator (full)& 24.6 &    0 & 0.483 & 30 \\ 
Drosophila1     & 3 & 0.495 & Lorentz        & 10.5 &  248 & 0.512 & 29 \\ 
Drosophila2     & 2 & 0.475 & Mercator (full)& 34.4 &    0 & 0.346 & 0 \\ 
Drosophila2     & 3 & 0.393 & Poincar\'e     & 12.2 &  912 & 0.360 & 0 \\ 
Human1          & 2 & 0.654 & Lorentz        & 11.7 & 1554 & 0.675 & 29 \\ 
Human1          & 3 & 0.722 & Poincar\'e     &  9.4 &  827 & 0.799 & 30 \\ 
Human2          & 2 & 0.649 & Lorentz        &  9.8 &  601 & 0.671 & 29 \\ 
Human2          & 3 & 0.720 & Lorentz        &  9.3 &  596 & 0.804 & 30 \\ 
Human6          & 2 & 0.811 & Lorentz        & 11.0 &  128 & 0.841 & 30 \\ 
Human6          & 3 & 0.832 & Lorentz        & 11.5 &  128 & 0.915 & 30 \\ 
Human7          & 2 & 0.816 & Mercator (full)&  8.5 &    0 & 0.826 & 29 \\ 
Human7          & 3 & 0.836 & Poincar\'e     &  9.5 &  484 & 0.913 & 30 \\ 
Human8          & 2 & 0.845 & BFKL           &  4.0 &    2 & 0.871 & 30 \\ 
Human8          & 3 & 0.865 & Mercator       & 13.1 &  114 & 0.922 & 30 \\ 
Macaque1        & 2 & 0.914 & Mercator (full)& 17.8 &    0 & 0.932 & 26 \\ 
Macaque1        & 3 & 0.883 & Poincar\'e     & 12.2 &  150 & 0.957 & 30 \\ 
Macaque2        & 2 & 0.792 & Lorentz        & 10.5 &   80 & 0.821 & 20 \\ 
Macaque2        & 3 & 0.810 & Lorentz        & 10.8 &   80 & 0.865 & 30 \\ 
Macaque3        & 2 & 0.587 & Mercator (full)& 22.5 &    0 & 0.614 & 30 \\ 
Macaque3        & 3 & 0.605 & Poincar\'e     &  9.8 &  315 & 0.647 & 30 \\ 
Macaque4        & 2 & 0.984 & Mercator (full)& 14.0 &    0 & 0.992 & 29 \\ 
Macaque4        & 3 & 0.991 & Mercator       &  7.8 &   11 & 0.998 & 30 \\ 
Mouse2          & 2 & 0.604 & Mercator (full)& 28.8 &    0 & 0.625 & 30 \\ 
Mouse2          & 3 & 0.672 & Lorentz        & 12.7 & 5172 & 0.667 & 1 \\ 
Mouse3          & 2 & 0.585 & Mercator (full)& 29.9 &    0 & 0.612 & 30 \\ 
Mouse3          & 3 & 0.655 & Lorentz        & 12.4 & 6068 & 0.655 & 10 \\ 
Rat1            & 2 & 0.966 & Mercator (full)& 26.1 &    0 & 0.766 & 0 \\ 
Rat1            & 3 & 0.951 & Lorentz        & 14.5 & 1664 & 0.679 & 0 \\ 
Rat2            & 2 & 0.970 & Mercator (full)& 30.7 &    0 & 0.777 & 0 \\ 
Rat2            & 3 & 0.930 & Lorentz        & 14.5 & 1710 & 0.691 & 0 \\ 
Rat3            & 2 & 0.969 & Mercator (full)& 20.8 &    0 & 0.803 & 0 \\ 
Rat3            & 3 & 0.887 & Lorentz        & 14.5 & 1836 & 0.724 & 0 \\ 
ZebraFinch2     & 2 & 0.325 & Mercator (full)& 21.4 &    0 & 0.323 & 2 \\ 
ZebraFinch2     & 3 & 0.317 & Lorentz        &  6.1 & 1086 & 0.310 & 1 \\ \hline
\end{tabular}}
\caption{Our embeddings vs. state-of-the-art.
For each connectome and dimension, we list the best prior method and
its result, the radius of the embedding, time elapsed in seconds, the best result
of our method, and how many times (out of 30) our result was better.
MAP measure.
\label{sota1}}
\end{table}

\begin{table}
\centering
\resizebox{0.75\columnwidth}{!}{%
\begin{tabular}{|lc|rlr|rrr|} \hline
connectome      &dim& MeanRank & method         & rad  & time & ours  & better \\ \hline
CElegans        & 2 &  31.0 & Poincar\'e     &  7.2 &  267 &  30.1 & 1 \\ 
CElegans        & 3 &  26.6 & Poincar\'e     & 10.3 &  270 &  26.3 & 18 \\ 
Cat1            & 2 &   3.8 & Mercator (full)& 10.0 &    0 &   3.1 & 30 \\ 
Cat1            & 3 &   4.0 & Mercator       &  9.7 &   19 &   2.0 & 30 \\ 
Cat2            & 2 &   7.1 & Poincar\'e     & 12.2 &  565 &   5.6 & 30 \\ 
Cat2            & 3 &   6.2 & Poincar\'e     & 10.1 &  570 &   4.3 & 30 \\ 
Cat3            & 2 &   3.2 & Mercator (full)&  6.5 &    0 &   1.9 & 30 \\ 
Cat3            & 3 &   3.1 & Mercator       &  5.5 &   21 &   1.3 & 30 \\ 
Drosophila1     & 2 &  46.3 & Poincar\'e     &  8.2 &  232 &  45.0 & 28 \\ 
Drosophila1     & 3 &  38.7 & Lorentz        & 10.5 &  248 &  37.1 & 29 \\ 
Drosophila2     & 2 & 122.1 & Mercator (full)& 35.3 &    0 & 112.9 & 27 \\ 
Drosophila2     & 3 & 112.2 & Poincar\'e     & 11.8 & 1044 &  95.0 & 29 \\ 
Human1          & 2 &  38.6 & Lorentz        &  8.2 &  572 &  38.6 & 1 \\ 
Human1          & 3 &  24.1 & Mercator       & 14.9 &   26 &  17.8 & 28 \\ 
Human2          & 2 &  43.2 & Lorentz        &  9.9 &  602 &  40.1 & 7 \\ 
Human2          & 3 &  26.4 & Mercator       & 11.6 &   35 &  17.5 & 27 \\ 
Human6          & 2 &   5.8 & Mercator (full)&  9.9 &    0 &   4.6 & 20 \\ 
Human6          & 3 &   5.5 & Mercator       &  9.2 &   15 &   2.1 & 30 \\ 
Human7          & 2 &   6.1 & Mercator (full)&  8.5 &    0 &   5.1 & 30 \\ 
Human7          & 3 &   5.2 & Mercator       &  7.8 &   12 &   2.4 & 30 \\ 
Human8          & 2 &  13.3 & BFKL           &  4.0 &    2 &  10.6 & 30 \\ 
Human8          & 3 &  10.1 & Mercator       & 13.1 &  114 &   5.7 & 30 \\ 
Macaque1        & 2 &   3.3 & Poincar\'e     & 12.2 &  416 &   1.5 & 30 \\ 
Macaque1        & 3 &   3.2 & Poincar\'e     & 12.2 &  190 &   1.2 & 30 \\ 
Macaque2        & 2 &   5.5 & Lorentz        & 10.5 &   80 &   4.0 & 30 \\ 
Macaque2        & 3 &   5.0 & Lorentz        & 10.8 &   80 &   2.9 & 30 \\ 
Macaque3        & 2 &  24.9 & Poincar\'e     &  8.8 & 1392 &  23.4 & 30 \\ 
Macaque3        & 3 &  21.3 & Lorentz        &  9.7 &  253 &  17.6 & 30 \\ 
Macaque4        & 2 &   1.3 & Mercator (full)& 14.0 &    0 &   0.2 & 30 \\ 
Macaque4        & 3 &   1.2 & Mercator       &  7.8 &   11 &   0.0 & 30 \\ 
Mouse2          & 2 &  84.0 & Lorentz        & 10.3 & 5172 &  80.7 & 29 \\ 
Mouse2          & 3 &  71.3 & Poincar\'e     & 12.2 & 5670 &  67.2 & 30 \\ 
Mouse3          & 2 &  96.2 & Lorentz        & 10.4 & 6068 &  92.4 & 29 \\ 
Mouse3          & 3 &  83.8 & Lorentz        & 12.4 & 6068 &  78.5 & 29 \\ 
Rat1            & 2 &   4.9 & Mercator (full)& 23.5 &    0 &   7.1 & 0 \\ 
Rat1            & 3 &   7.0 & Lorentz        & 14.5 & 1647 &   8.7 & 0 \\ 
Rat2            & 2 &   4.4 & Mercator (full)& 19.1 &    0 &   6.1 & 0 \\ 
Rat2            & 3 &   7.3 & Lorentz        & 14.5 & 1710 &   8.0 & 0 \\ 
Rat3            & 2 &   5.4 & Mercator (full)& 20.8 &    0 &   6.5 & 0 \\ 
Rat3            & 3 &   7.7 & Poincar\'e     & 12.2 & 2407 &   7.8 & 4 \\ 
ZebraFinch2     & 2 & 121.6 & Mercator (full)& 21.9 &    0 & 121.1 & 27 \\ 
ZebraFinch2     & 3 & 115.9 & Poincar\'e     &  6.2 & 3494 & 117.1 & 10 \\ \hline

\end{tabular}}
\caption{Our embeddings vs. state-of-the-art.
For each connectome and dimension, we list the best prior method and
its result, the radius of the embedding, time elapsed in seconds, the best result
of our method, and how many times (out of 30) our result was better.
MeanRank measure.
\label{sota2}}
\end{table}

\begin{table}
\centering
\resizebox{0.75\columnwidth}{!}{%
\begin{tabular}{|lc|rlr|rrr|} \hline
connectome      &dim& success & method         & rad  & time & ours  & better \\ \hline
CElegans        & 2 & 0.903 & Poincar\'e     &  7.2 &  267 & 0.931 & 27 \\ 
CElegans        & 3 & 0.958 & Poincar\'e     & 10.1 &  274 & 0.930 & 0 \\ 
Cat1            & 2 & 1.000 & Mercator (full)&  9.9 &    0 & 1.000 & 0 \\ 
Cat1            & 3 & 1.000 & Lorentz        &  8.7 &  103 & 1.000 & 0 \\ 
Cat2            & 2 & 1.000 & Mercator (full)& 12.6 &    0 & 0.986 & 0 \\ 
Cat2            & 3 & 0.970 & Poincar\'e     & 10.7 &  277 & 0.980 & 29 \\ 
Cat3            & 2 & 1.000 & Poincar\'e     & 10.3 &   96 & 1.000 & 0 \\ 
Cat3            & 3 & 1.000 & Poincar\'e     & 10.3 &   96 & 1.000 & 0 \\ 
Drosophila1     & 2 & 0.783 & Mercator (full)& 24.6 &    0 & 0.847 & 29 \\ 
Drosophila1     & 3 & 0.844 & Poincar\'e     & 11.4 &  365 & 0.843 & 13 \\ 
Drosophila2     & 2 & 0.761 & Mercator (full)& 35.3 &    0 & 0.560 & 0 \\ 
Drosophila2     & 3 & 0.671 & Poincar\'e     & 12.2 &  912 & 0.555 & 0 \\ 
Human1          & 2 & 0.921 & Lorentz        &  9.0 &  563 & 0.929 & 10 \\ 
Human1          & 3 & 0.939 & Lorentz        &  8.8 &  566 & 0.958 & 23 \\ 
Human2          & 2 & 0.902 & Lorentz        &  9.9 &  602 & 0.924 & 13 \\ 
Human2          & 3 & 0.963 & Lorentz        &  8.9 & 1344 & 0.961 & 14 \\ 
Human6          & 2 & 0.996 & BFKL           &  5.4 &    0 & 0.995 & 3 \\ 
Human6          & 3 & 1.000 & Lorentz        & 10.7 &  127 & 0.979 & 0 \\ 
Human7          & 2 & 1.000 & Lorentz        &  9.5 &  118 & 0.941 & 0 \\ 
Human7          & 3 & 1.000 & Lorentz        &  8.9 &  114 & 0.990 & 0 \\ 
Human8          & 2 & 0.997 & Mercator (full)& 19.2 &    0 & 1.000 & 30 \\ 
Human8          & 3 & 0.996 & Poincar\'e     & 12.2 & 4856 & 0.997 & 30 \\ 
Macaque1        & 2 & 0.997 & Mercator (full)& 17.8 &    0 & 0.990 & 6 \\ 
Macaque1        & 3 & 0.980 & Mercator       & 16.4 &   41 & 1.000 & 29 \\ 
Macaque2        & 2 & 0.988 & Mercator (full)& 15.0 &    0 & 1.000 & 6 \\ 
Macaque2        & 3 & 0.985 & Poincar\'e     &  9.8 &  256 & 0.974 & 6 \\ 
Macaque3        & 2 & 0.924 & Mercator (full)& 22.5 &    0 & 0.944 & 28 \\ 
Macaque3        & 3 & 0.894 & Poincar\'e     &  9.8 &  315 & 0.919 & 30 \\ 
Macaque4        & 2 & 1.000 & Mercator (fast)& 16.1 &    0 & 1.000 & 0 \\ 
Macaque4        & 3 & 1.000 & Mercator       &  7.8 &   14 & 1.000 & 0 \\ 
Mouse2          & 2 & 0.967 & Mercator (full)& 28.8 &    0 & 0.968 & 29 \\ 
Mouse2          & 3 & 0.978 & Lorentz        & 13.2 & 5176 & 0.959 & 0 \\ 
Mouse3          & 2 & 0.962 & Mercator (full)& 34.5 &    0 & 0.967 & 30 \\ 
Mouse3          & 3 & 0.971 & Poincar\'e     & 12.2 & 8679 & 0.952 & 0 \\ 
Rat1            & 2 & 0.998 & Mercator (full)& 26.1 &    0 & 0.960 & 0 \\ 
Rat1            & 3 & 0.990 & Lorentz        & 14.5 & 1664 & 0.899 & 0 \\ 
Rat2            & 2 & 0.998 & Mercator (full)& 30.7 &    0 & 0.969 & 0 \\ 
Rat2            & 3 & 0.993 & Lorentz        & 14.5 & 1710 & 0.899 & 0 \\ 
Rat3            & 2 & 0.998 & Mercator (full)& 20.8 &    0 & 0.958 & 0 \\ 
Rat3            & 3 & 0.945 & Lorentz        & 14.5 & 1836 & 0.906 & 0 \\ 
ZebraFinch2     & 2 & 0.886 & Mercator (full)& 21.6 &    0 & 0.853 & 0 \\ 
ZebraFinch2     & 3 & 0.889 & Lorentz        &  6.3 & 1094 & 0.812 & 0 \\ \hline
\end{tabular}}
\caption{Our embeddings vs. state-of-the-art.
For each connectome and dimension, we list the best prior method and
its result, the radius of the embedding, time elapsed in seconds, the best result
of our method, and how many times (out of 30) our result was better.
Greedy routing success measure.
\label{sota3}}
\end{table}

\begin{table}
\centering
\resizebox{0.75\columnwidth}{!}{%
\begin{tabular}{|lc|rlr|rrr|} \hline
connectome      &dim& stretch & method         & rad  & time & ours  & better \\ \hline
CElegans        & 2 & 1.310 & Poincar\'e     &  7.2 &  278 & 1.254 & 30 \\ 
CElegans        & 3 & 1.216 & Poincar\'e     &  9.9 &  277 & 1.232 & 1 \\ 
Cat1            & 2 & 1.046 & Mercator (full)&  9.6 &    0 & 1.046 & 18 \\ 
Cat1            & 3 & 1.037 & Lorentz        &  8.7 &  103 & 1.021 & 30 \\ 
Cat2            & 2 & 1.053 & Mercator (full)& 12.6 &    0 & 1.058 & 7 \\ 
Cat2            & 3 & 1.062 & Poincar\'e     & 10.7 &  277 & 1.047 & 30 \\ 
Cat3            & 2 & 1.042 & Poincar\'e     &  9.7 &  291 & 1.039 & 23 \\ 
Cat3            & 3 & 1.042 & Lorentz        &  9.2 &   86 & 1.018 & 30 \\ 
Drosophila1     & 2 & 1.402 & Mercator (full)& 24.6 &    0 & 1.340 & 28 \\ 
Drosophila1     & 3 & 1.307 & Lorentz        & 11.3 &  243 & 1.328 & 3 \\ 
Drosophila2     & 2 & 1.202 & Mercator (full)& 35.3 &    0 & 1.964 & 0 \\ 
Drosophila2     & 3 & 1.394 & Poincar\'e     & 12.2 &  912 & 1.976 & 0 \\ 
Human1          & 2 & 1.299 & Lorentz        &  9.0 &  563 & 1.282 & 18 \\ 
Human1          & 3 & 1.231 & Lorentz        &  9.3 & 1554 & 1.176 & 27 \\ 
Human2          & 2 & 1.298 & Poincar\'e     & 11.3 & 3768 & 1.282 & 18 \\ 
Human2          & 3 & 1.203 & Poincar\'e     & 10.2 & 2977 & 1.177 & 25 \\ 
Human6          & 2 & 1.059 & BFKL           &  5.4 &    0 & 1.062 & 0 \\ 
Human6          & 3 & 1.061 & Lorentz        & 10.7 &  127 & 1.059 & 25 \\ 
Human7          & 2 & 1.085 & Mercator (full)&  8.7 &    0 & 1.103 & 13 \\ 
Human7          & 3 & 1.078 & Lorentz        &  9.2 &  115 & 1.056 & 29 \\ 
Human8          & 2 & 1.026 & Lorentz        & 14.5 &  783 & 1.020 & 30 \\ 
Human8          & 3 & 1.024 & Poincar\'e     & 12.2 & 4856 & 1.017 & 30 \\ 
Macaque1        & 2 & 1.025 & Mercator (fast)& 20.9 &    0 & 1.026 & 9 \\ 
Macaque1        & 3 & 1.040 & Mercator       & 16.4 &   41 & 1.019 & 30 \\ 
Macaque2        & 2 & 1.069 & Poincar\'e     & 12.2 &   88 & 1.056 & 5 \\ 
Macaque2        & 3 & 1.065 & Lorentz        & 10.8 &   80 & 1.068 & 16 \\ 
Macaque3        & 2 & 1.180 & Mercator (full)& 22.9 &    0 & 1.169 & 18 \\ 
Macaque3        & 3 & 1.194 & Poincar\'e     & 10.2 & 1496 & 1.171 & 28 \\ 
Macaque4        & 2 & 1.000 & Mercator (fast)& 16.1 &    0 & 1.000 & 0 \\ 
Macaque4        & 3 & 1.000 & Mercator       &  7.8 &   14 & 1.000 & 0 \\ 
Mouse2          & 2 & 1.102 & Mercator (full)& 28.8 &    0 & 1.140 & 0 \\ 
Mouse2          & 3 & 1.064 & Lorentz        & 13.2 & 5176 & 1.120 & 0 \\ 
Mouse3          & 2 & 1.108 & Mercator (full)& 29.8 &    0 & 1.156 & 0 \\ 
Mouse3          & 3 & 1.077 & Poincar\'e     & 12.2 & 9207 & 1.145 & 0 \\ 
Rat1            & 2 & 1.003 & Mercator (full)& 21.4 &    0 & 1.048 & 0 \\ 
Rat1            & 3 & 1.006 & Lorentz        & 14.5 & 1664 & 1.120 & 0 \\ 
Rat2            & 2 & 1.002 & Mercator (full)& 30.7 &    0 & 1.038 & 0 \\ 
Rat2            & 3 & 1.004 & Lorentz        & 14.5 & 1710 & 1.114 & 0 \\ 
Rat3            & 2 & 1.005 & Mercator (full)& 20.8 &    0 & 1.059 & 0 \\ 
Rat3            & 3 & 1.019 & Mercator       & 17.4 &  471 & 1.105 & 0 \\ 
ZebraFinch2     & 2 & 1.267 & Mercator (full)& 21.4 &    0 & 1.351 & 0 \\ 
ZebraFinch2     & 3 & 1.323 & Lorentz        &  6.1 & 1086 & 1.403 & 0 \\ \hline
\end{tabular}}
\caption{Our embeddings vs. state-of-the-art.
For each connectome and dimension, we list the best prior method and
its result, the radius of the embedding, time elapsed in seconds, the best result
of our method, and how many times (out of 30) our result was better. 
Greedy routing stretch measure.
\label{sota4}}
\end{table}

\section{Conclusions}

In this paper, we presented an experimental analysis of embeddings of 21 connectomes to various geometries (both three-dimesnional and two-dimensional). To our knowledge, we are the first to compare embeddings of connectomes to all Thurston geometries. Our findings unveil new prospects for connectome modeling, introducing a novel method based on Simulated Annealing. Our results demonstrate the efficacy of this approach -- it yields superior embeddings compared to the state-of-the-art.

Although previous studies suggested that one could find a universal winner geometry for the embeddings (usually pointing at two-dimensional hyperbolic geometry), our results reveals a nuanced scenario when considering the third dimension. We showed that the universal winner ceases to exist when we consider the third dimension. In particular, $\bbH^2$ embeddings tend to be worse than (non-Euclidean) 3D geometries, even if our $\bbH^2$ embeddings are good -- better than \cite{tobias,mercatorembedding,nickel,nickel2}. If we were to suggest a set of geometries to pay attention to when modeling connectomes, we would mention three-dimensional hyperbolic geometry, Solv, and product geometries. Surprisingly, three-dimensional Euclidean geometry is a suitable choice for Human connectomes. There is a correlation between the zone of the connectome (also its primary function) and the best choice for the embedding, e.g., nervous systems tend to be well modeled by trees. 

Some technicalities that do not matter (discrete versions or angularity). However, the size of the grid or the setup of the Simulated Annealing can affect the results. Remarkably, in some experiment, standard grid versions have given significantly better results than the so-called \emph{big} versions.
Since the difference between these two cases is that the \emph{big}-variant has a larger number of cells, this should not happen since
any embedding in the standard variant is also an embedding in the \emph{big}-variant. This is caused either by a failure to correctly
guess the optimal values of the parameters or possibly because Simulated Annealing requires more iterations to find good embeddings at larger distances.

Our results stem from an extensive simulation scheme with numerous robustness checks. While our results regarding log-likelihood, MAP, and MeanRank were similar and robust to the changes in the Simulated Annealing setup, we noticed that optimizing log-likelihood may affect the quality measured by greedy success rate and stretch. We suppect that one explanation is that these two sets of quality measures capture different aspects (functions) of the networks. However, finding out the relationships among connectomes or embeddings characteristics and quality measures is beyond the scope of this paper and will be the subject of future work.

\section*{Acknowledgments}
We wish to thank Piotr Pokarowski for his suggestions related to the presentation of the statistical analysis. We are also thankful to the anonymous referees for their suggestions. 
This work has been supported by the National Science Centre, Poland, grant UMO-2019/35/B/ST6/04456.

\def\ext{}
\bibliographystyle{alpha}
\bibliography{../master.bib}

\end{document}